\documentclass[12pt]{article}
\usepackage{graphicx}

\begin{document}

\title{Thermal effects in Jaynes-Cummings model derived with low-temperature expansion}

\author{Hiroo Azuma${}^{1,}$\thanks{On leave from
Information and Mathematical Science Laboratory Inc.,
Meikei Building, 1-5-21 Ohtsuka, Bunkyo-ku, Tokyo 112-0012, Japan.
Email: hiroo.azuma@m3.dion.ne.jp}
\ \ 
and
\ \ 
Masashi Ban${}^{2,}$\thanks{Email: m.ban@phys.ocha.ac.jp}
\\
\\
{\small ${}^{1}$Advanced Algorithm \& Systems Co., Ltd.,}\\
{\small 7F Ebisu-IS Building, 1-13-6 Ebisu, Shibuya-ku, Tokyo 150-0013, Japan}\\
{\small ${}^{2}$Graduate School of Humanities and Sciences, Ochanomizu University,}\\
{\small 2-1-1 Ohtsuka, Bunkyo-ku, Tokyo 112-8610, Japan}
}

\date{\today}

\maketitle

\begin{abstract}
In this paper, we investigate thermal effects of the Jaynes-Cummings model (JCM)
at finite temperature with a perturbative approach.
We assume a single two-level atom and a single cavity mode
to be initially in the thermal equilibrium state and the thermal coherent state, respectively,
at a certain finite low temperature.
Describing this system with Thermo Field Dynamics formalism,
we obtain a low-temperature expansion of the atomic population inversion
in a systematic manner.
Letting the system evolve in time with the JCM Hamiltonian,
we examine thermal effects of the collapse and the revival of the Rabi oscillations
by means of the third-order perturbation theory
under the low-temperature limit,
that is to say,
using the low-temperature expansion up to the third order terms.
From an intuitive discussion,
we can expect that the period of the revival of the Rabi oscillations
becomes longer
as the temperature rises.
Numerical results obtained with the perturbation theory
reproduce well this temperature dependence of the period.
\end{abstract}

\bigskip


\bigskip

\section{\label{section-introduction}Introduction}
The Jaynes-Cummings model (JCM) was originally
proposed for describing the spontaneous emission
in a semi-classical manner
by Jaynes and Cummings during the 1960s
\cite{Jaynes1963,Shore1993,Louisell1973,Schleich2001}.
The JCM consists of a single two-level atom
and a single cavity mode of the electromagnetic field.
The JCM interaction between the atom and the cavity mode is obtained
by the rotating wave approximation,
so that
each photon creation causes an atomic de-excitation
and
each photon annihilation causes an atomic excitation.
The JCM is a soluble quantum mechanical model.
Moreover, it is simple enough for expressing the basic and most important characteristics
of the matter-radiation interaction.
Because of these distinct advantages,
both theoretical and experimental researchers
in the field of quantum optics
have been studying the JCM eagerly for decades. 

If we initially prepare the atom in the ground state
and the cavity mode in the coherent state,
the JCM shows the periodic spontaneous collapse and the revival of the Rabi oscillations
during its time-evolution
\cite{Cummings1965,Eberly1980,Narozhny1981,Yoo1981,Yoo1985}.
This phenomenon was experimentally demonstrated in the 1980s
\cite{Rempe1987}.
We can regard this phenomenon as a direct evidence for discreteness of energy states of photons.
Thus, the JCM has a fully quantum property,
which cannot be explained by semi-classical physics.

In this paper,
we investigate the non-dissipative JCM at finite low temperature.
We assume that the atom and the cavity mode are initially
in the thermal equilibrium state and the thermal coherent state, respectively,
at a certain finite low temperature
$\beta[=1/(k_{\mbox{\scriptsize B}}T)]$.
Thus, we can describe an initial probability distribution of quantum states
of the system with the canonical ensemble.
Moreover,
we assume the time-evolution of the system is governed
by a unitary operator generated with the JCM Hamiltonian.
This implies that the system does not suffer from dissipation and its time-evolution is reversible.

The Thermo Field Dynamics (TFD) formalism was developed by Takahashi and Umezawa
for dealing with phenomena in isolated systems during the 1970s
\cite{Takahashi1975,Ojima1981,Umezawa1982,Umezawa1992}.
The TFD formalism has a wide range of applications in equilibrium situations of closed systems, as well.
In this paper,
we think about applications of the TFD
to non-dissipative closed systems.
The TFD formalism helps us to be more successful than conventional formalisms do as follows:
calculating an expectation value of a pure state created by the TFD mechanics,
we can obtain a thermal average in statical mechanics.
In return for this benefit,
the TFD formalism requires us to let the original Hilbert space double in size for the tensor product.
Then,
the TFD formalism induces thermal-like noises in two mode squeezed states
generated by the thermal Bogoliubov transformation.

Describing the JCM at finite low temperature with the TFD formalism,
we can write down a thermal average (an expectation value) of an observable
as a series expansion
containing powers of $\theta(\beta)$ in a systematic manner.
[The explicit form of the function $\theta(\beta)$ is given by
$\theta(\beta)=\mbox{arctanh}[\exp(-\beta\hbar\omega/2)]$,
where $\omega$ is a frequency of the cavity mode.
We note that $\theta(\beta)\to +0$ as $\beta\to +\infty$.]
We call this series the low-temperature expansion.
This prescription is a new key point of this paper as compared with the other past works.

Strictly speaking,
because we introduce a finite temperature into both the atom and the cavity field,
the perturbation theory has to include two parameters of the temperature.
In this paper,
we let $\Theta(\beta)$ and $\theta(\beta)$ denote the parameters of the temperature
for the atom and the cavity field, respectively.
[The explicit form of the function $\Theta(\beta)$ is given by
$\Theta(\beta)=\arctan[\exp(-\beta\hbar\omega_{0}/2)]$,
where $\omega_{0}$ is a transition frequency of the two-level atom.
We note that $\Theta(\beta)\to +0$ as $\beta\to +\infty$.]
However,
the Hilbert space of the atom is two-dimensional,
so that we can  solve the problem concerning the atom exactly.
On the other hand,
the dimension of the Hilbert space of the cavity field is infinite,
so that we cannot give a rigorous treatment for the problem
concerning the cavity field,
which interacts with the atom.
This is the reason why the low-temperature expansion contains
powers of $\theta(\beta)$, but not $\Theta(\beta)$.

As mentioned before,
in this paper,
we initially prepare the cavity mode in the thermal coherent state,
which is
proposed by Barnett, Knight, Mann and Revzen
as a natural extension of the zero-temperature ordinary coherent state according to the TFD
\cite{Barnett1985,Mann1989a}.
Then, we investigate the time-evolution of the atomic population inversion
using the low-temperature expansion up to the third order terms
in $\theta(\beta)$.
We regard this low-temperature expansion as the third order perturbation theory.

To take another approach,
we give an intuitive discussion and obtain the following expectation:
the period of the revival of the Rabi oscillations becomes longer as the temperature rises,
where the temperature is finite but low enough and varies in the neighborhood of the absolute zero.
Numerical results obtained by the third order perturbation theory
reproduce well this temperature dependence of the period.

Here, we write about related works.
The thermal JCM without dissipation is investigated
in Refs.~\cite{Arroyo-Correa1990,Liu1992,Chumakov1993,Klimov1999,Azuma2008,Azuma2010}.
In these references,
the cavity field is prepared initially in a thermal equilibrium state,
whose statistical behavior is given by the Bose-Einstein distribution.
Then,
the system consisting of
the atom and the cavity mode is assumed to evolve
with a unitary operator generated by the JCM Hamiltonian,
so that its time-evolution is reversible.
However,
in these works,
thermal coherent states are not considered to be the initial states of the cavity field.

The thermal JCM with dissipation is investigated
in Refs.~\cite{Eiselt1991,Daeubler1992,Murao1995,deFaria1999,Kuang1997}.
In Refs.~\cite{Eiselt1991,Daeubler1992,Murao1995,deFaria1999},
the cavity damping in the JCM is discussed.
In these works,
the equation of motion for the density operator $\rho$ includes
the term $\kappa L_{\mbox{\scriptsize ir}}(\rho)$,
which describes the irreversible motion caused by the cavity damping.
A typical form of $L_{\mbox{\scriptsize ir}}(\rho)$ is given by
$(2a\rho a^{\dagger}-\rho a^{\dagger}a-a^{\dagger}a\rho)$
in Refs.~\cite{Daeubler1992,deFaria1999},
where $a^{\dagger}$ and $a$ are the creation and annihilation operators of the cavity photons,
respectively.
On the other hand, in Ref.~\cite{Kuang1997},
the phase damping in the JCM is discussed.
In this work,
the equation of motion for the density operator $\rho$ includes the term
$(-\gamma[H,[H,\rho]])$,
which describes the phase damping.
To obtain this term,
we assume the system to interact with a heat-bath environment,
namely an infinite set of harmonic oscillators.
Considering the Liouville-von Neumann equation for the system and the heat bath,
and applying the Markovian approximation to it,
we can derive the term of the phase damping.

Some researchers attempt to extend the JCM Hamiltonian according to the TFD formalism
\cite{Barnett1985,Azuma2010,Fan2004}.
However,
in these works,
they introduce the temperature only into the cavity field
and let the atom maintain the temperature at absolute zero.
In this paper,
we explain that such extended JCM Hamiltonians cannot be genuine ones
based on the TFD formalism.

The organization of this paper is as follows:
In Sec.~\ref{section-review-JCM},
we give a brief review of the original JCM.
In Sec.~\ref{section-review-TFD},
we give a brief review of the TFD.
In Sec.~\ref{section-thermal-effects-period-qualitative-estimation},
we give an intuitive discussion about the thermal effects found
in the period of the revival of the Rabi oscillations.
In Sec.~\ref{section-formulation-perturbetion-theory},
we formulate the perturbation theory under the low-temperature limit.
In Sec.~\ref{section-comparison-of-TFD-and-Liouville-von-Neumann-equation},
we compare our TFD formalism applied to the JCM with the Liouville-von Neumann equation.
Although the equilibrium TFD is equivalent to the Liouville-von Neumann equation in principle,
we find that the TFD formalism is more effective and easier than the Liouville-von Neumann equation
for deriving the low-temperature expansion.
In Secs.~\ref{section-0th-order-correction-term},
\ref{section-1st-order-correction-term}
and
\ref{section-2nd-order-correction-term},
we calculate the zero-th, first and second order perturbation corrections, respectively.
In Sec.~\ref{section-numerical-calculations},
we give numerical results of calculations.
In Sec.~\ref{section-counter-rotating-terms},
we consider thermal effects of the counter-rotating terms.
In Sec.~\ref{section-discussion},
we give a brief discussion.
In Appendix~\ref{section-3rd-order-correction-term},
we give details of calculations of the third order perturbation correction.

The evaluation of the perturbation corrections described in
Secs.~\ref{section-0th-order-correction-term},
\ref{section-1st-order-correction-term},
\ref{section-2nd-order-correction-term}
and
Appendix~\ref{section-3rd-order-correction-term},
we make use of techniques for calculations
developed in Ref.~\cite{Azuma2010}.
Thus, we can regard this paper as a sequel of Ref.~\cite{Azuma2010}.

\section{\label{section-review-JCM}A brief review of the JCM}
In this section, we review the JCM briefly.
The Hamiltonian of the original
JCM is expressed in the form,
\begin{equation}
H
=
\frac{\hbar}{2}\omega_{0}\sigma_{z}
+
\hbar\omega a^{\dagger}a
+
\hbar\kappa(\sigma_{+}a+\sigma_{-}a^{\dagger}),
\label{JCM-Hamiltonian-0}
\end{equation}
where
$\hbar=h/2\pi$,
$\sigma_{\pm}=(1/2)(\sigma_{x}\pm i\sigma_{y})$
and
$[a,a^{\dagger}]=1$.
The Pauli matrices
($\sigma_{i}$ for $i=x,y,z$)
are operators acting on quantum states of the single atom in the cavity.
The creation and annihilation operators
$a^{\dagger}$
and
$a$
act on quantum states of the single cavity mode.
The transition frequency of the two-level atom is given by
$\omega_{0}$.
The frequency of the single cavity mode is given by
$\omega$.
From now on,
for simplicity,
we assume
$\kappa$
to be real.

We can divide the Hamiltonian $H$ given by Eq.~(\ref{JCM-Hamiltonian-0})
into two parts as follows:
\begin{eqnarray}
H&=&\hbar(C_{1}+C_{2}), \nonumber \\
C_{1}&=&\omega(\frac{1}{2}\sigma_{z}+a^{\dagger}a), \nonumber \\
C_{2}&=&\kappa(\sigma_{+}a+\sigma_{-}a^{\dagger})-\frac{\Delta\omega}{2}\sigma_{z},
\label{JCM-Hamiltonian-decomposition-0}
\end{eqnarray}
where
$\Delta\omega=\omega-\omega_{0}$.
We can confirm $[C_{1},C_{2}]=0$.
Moreover, we can diagonalize $C_{1}$ at ease.

Because of the above facts,
we take the following interaction picture for expressing the time-evolution of the quantum state:
First, we write down the state vector of the atom and the cavity photons in the Schr{\"o}dinger picture as
$|\psi_{\mbox{\scriptsize S}}(t)\rangle$.
Second,
assuming $|\psi_{\mbox{\scriptsize I}}(0)\rangle=|\psi_{\mbox{\scriptsize S}}(0)\rangle$,
we define the state vector in the interaction picture as
$|\psi_{\mbox{\scriptsize I}}(t)\rangle
=\exp(iC_{1}t)|\psi_{\mbox{\scriptsize S}}(t)\rangle$.
Thus,
we can describe the time-evolution
as
$|\psi_{\mbox{\scriptsize I}}(t)\rangle=U(t)|\psi_{\mbox{\scriptsize I}}(0)\rangle$,
where
$U(t)=\exp(-iC_{2}t)$.

We give eigenstates of the two-level atom and the cavity mode in the following forms:
First, we describe the ground state and the excited state of the atom as two-component vectors,
\begin{equation}
|g\rangle_{\mbox{\scriptsize A}}
=
|0\rangle_{\mbox{\scriptsize A}}
=
\left(
\begin{array}{c}
0 \\
1
\end{array}
\right),
\quad\quad
|e\rangle_{\mbox{\scriptsize A}}
=
|1\rangle_{\mbox{\scriptsize A}}
=
\left(
\begin{array}{c}
1 \\
0
\end{array}
\right),
\label{atom-basis-vectors}
\end{equation}
where
$|g\rangle_{\mbox{\scriptsize A}}(=|0\rangle_{\mbox{\scriptsize A}})$
and
$|e\rangle_{\mbox{\scriptsize A}}(=|1\rangle_{\mbox{\scriptsize A}})$
are eigenvectors of
$\sigma_{z}$,
and their corresponding eigenvalues are
$(-1)$
and
$1$,
respectively.
The index $\mbox{A}$ stands for the atom.
We can regard
$|i\rangle_{\mbox{\scriptsize A}}$ for $i\in\{0,1\}$
as the number states of the fermions.
Second,
we
write down the number states of the cavity photons
as
$|n\rangle_{\mbox{\scriptsize P}}
=(1/\sqrt{n!})(a^{\dagger})^{n}|0\rangle_{\mbox{\scriptsize P}}$ for $n=0,1,2,...$.
The index $\mbox{P}$ stands for the photons.

We describe the unitary operator for the time-evolution $U(t)$
as the $2\times 2$ matrix,
\begin{equation}
U(t)
=
\exp[-it
\left(
\begin{array}{cc}
-\Delta\omega/2 & \kappa a \\
\kappa a^{\dagger} & \Delta\omega/2
\end{array}
\right)
]
=
\left(
\begin{array}{cc}
u_{11} & u_{10} \\
u_{01} & u_{00}
\end{array}
\right),
\label{unitary-evolution-1}
\end{equation}
where
\begin{eqnarray}
u_{11}
&=&
\cos(\sqrt{a^{\dagger}a+c+1}|\kappa|t)
+
i
\frac{\Delta \omega}{|\Delta\omega|}
\sqrt{c}
\frac{\sin(\sqrt{a^{\dagger}a+c+1}|\kappa|t)}{\sqrt{a^{\dagger}a+c+1}}, \nonumber \\
u_{10}
&=&
-i
\frac{\kappa}{|\kappa|}
\frac{\sin(\sqrt{a^{\dagger}a+c+1}|\kappa|t)}{\sqrt{a^{\dagger}a+c+1}}
a, \nonumber \\
u_{01}
&=&
-i
\frac{\kappa}{|\kappa|}
\frac{\sin(\sqrt{a^{\dagger}a+c}|\kappa|t)}{\sqrt{a^{\dagger}a+c}}
a^{\dagger}, \nonumber \\
u_{00}
&=&
\cos(\sqrt{a^{\dagger}a+c}|\kappa|t)
-
i
\frac{\Delta \omega}{|\Delta\omega|}
\sqrt{c}
\frac{\sin(\sqrt{a^{\dagger}a+c}|\kappa|t)}{\sqrt{a^{\dagger}a+c}},
\label{unitary-evolution-2} 
\end{eqnarray}
and
\begin{equation}
c
=
(\frac{\Delta\omega}{2\kappa})^{2}.
\label{unitary-evolution-3}
\end{equation}
Because we take the basis vectors $\{|1\rangle_{\mbox{\scriptsize A}},|0\rangle_{\mbox{\scriptsize A}}\}$,
the indices $i,j\in\{1,0\}$ for $u_{ij}$
are arranged in descending order.
That is to say,
we take the index `$1$' for representing the first row and the first column of the $2\times2$ matrix $U(t)$,
and
we take the index `$0$' for representing the second row and the second column of the $2\times2$ matrix $U(t)$.

After these preparations,
the probability for detecting the ground state of the atom $|g\rangle_{\mbox{\scriptsize A}}$
at the time $t$
is given by
\begin{equation}
P_{g}(t)
=
\|_{\mbox{\scriptsize A}}\langle g|\psi_{\mbox{\scriptsize I}}(t)\rangle\|^{2}.
\end{equation}
Moreover,
the atomic population inversion is given by
\begin{equation}
\langle\sigma_{z}(t)\rangle
=
\langle\psi_{\mbox{\scriptsize I}}(t)|\sigma_{z}|\psi_{\mbox{\scriptsize I}}(t)\rangle
=
1-2P_{g}(t).
\label{definition-atomic-population-inversion}
\end{equation}

As a particular case, we consider
the initial state
to be in $|\psi_{\mbox{\scriptsize I}}(0)\rangle
=|g\rangle_{\mbox{\scriptsize A}}|\alpha\rangle_{\mbox{\scriptsize P}}$,
where $|\alpha\rangle_{\mbox{\scriptsize P}}$ represents the coherent state,
\begin{eqnarray}
|\alpha\rangle_{\mbox{\scriptsize P}}
&=&
\exp(\alpha a^{\dagger}-\alpha^{*}a)|0\rangle_{\mbox{\scriptsize P}} \nonumber \\
&=&
e^{-|\alpha|^{2}/2}
\sum_{n=0}^{\infty}
\frac{\alpha^{n}}{\sqrt{n!}}|n\rangle_{\mbox{\scriptsize P}},
\label{zero-temperature-coherent-state}
\end{eqnarray}
and $\alpha$ is an arbitrary complex number.
From now on, for simplicity,
we always let the parameter $\alpha$
characterizing the coherent state $|\alpha\rangle_{\mbox{\scriptsize P}}$
be real.
Then,
we obtain $P_{g}(t)$ in the form,
\begin{eqnarray}
P_{g}(t)
&=&
\left\|
\begin{array}{c}
{
(
\begin{array}{cc}
0 & 1
\end{array}
)
} \\
\quad
\end{array}
\left(
\begin{array}{cc}
u_{11} & u_{10} \\
u_{01} & u_{00}
\end{array}
\right)
\left(
\begin{array}{c}
0 \\
|\alpha\rangle_{\mbox{\scriptsize P}}
\end{array}
\right)
\right\|^{2} \nonumber \\
&=&
{}_{\mbox{\scriptsize P}}
\langle\alpha|u_{00}^{\dagger}u_{00}|\alpha\rangle_{\mbox{\scriptsize P}} \nonumber \\
&=&
e^{-\alpha^{2}}
\sum_{n=0}^{\infty}
\frac{\alpha^{2n}}{n!}
[\cos^{2}(\sqrt{n+c}|\kappa|t)
+
c
\frac{\sin^{2}(\sqrt{n+c}|\kappa|t)}{n+c}].
\label{Probability-collapse-and-revival-Rabi-oscillations-0}
\end{eqnarray}
The atomic population inversion given by Eqs.~(\ref{definition-atomic-population-inversion})
and (\ref{Probability-collapse-and-revival-Rabi-oscillations-0})
shows
the collapse and the revival of the Rabi oscillations.

Here, we examine the time scale of the initial collapse
and the period of the revival of the Rabi oscillations
\cite{Barnett1997}.
Learning from experience,
we know this phenomenon becomes more distinct as $\alpha^{2}$ increases.
Thus,
we assume $\alpha^{2}\gg 1$.
We rewrite the index of the summation $n$ as $n=\alpha^{2}+\delta n$
in Eq.~(\ref{Probability-collapse-and-revival-Rabi-oscillations-0}).
Then,
because of the Poisson distribution,
the major contribution for the summation
in Eq.~(\ref{Probability-collapse-and-revival-Rabi-oscillations-0})
comes from $|\delta n|<\alpha^{2}$.
Moreover,
for simplicity,
we assume $c\ll \alpha^{2}$,
and we neglect the term
$[c/(n+c)]\sin^{2}(\sqrt{n+c}|\kappa|t)$
in the right-hand side of Eq.~(\ref{Probability-collapse-and-revival-Rabi-oscillations-0}).

From the above discussions and Eqs.~(\ref{definition-atomic-population-inversion})
and (\ref{Probability-collapse-and-revival-Rabi-oscillations-0}),
writing
$\sqrt{n+c}
\simeq
(\alpha^{2}+n)/(2|\alpha|)$,
we obtain the following approximation:
\begin{eqnarray}
\langle\sigma_{z}(t)\rangle
&\simeq&
-
e^{-\alpha^{2}}
\sum_{n=0}^{\infty}
\frac{\alpha^{2n}}{n!}
\frac{1}{2}
[
e^{i(\alpha^{2}+n)|\kappa|t/|\alpha|}
+
e^{-i(\alpha^{2}+n)|\kappa|t/|\alpha|}
] \nonumber \\
&=&
-
\exp[\alpha^{2}(\cos\frac{|\kappa|t}{|\alpha|}-1)]
\cos(|\alpha||\kappa|t+\alpha^{2}\sin\frac{|\kappa|t}{|\alpha|}).
\label{atomic-population-inversion-approximation}
\end{eqnarray}
In Eq.~(\ref{atomic-population-inversion-approximation}),
$\exp[\alpha^{2}(\cos(|\kappa|t/|\alpha|)-1)]$
represents
the amplitude envelope of the wave,
and
$\cos[|\alpha||\kappa|t+\alpha^{2}\sin(|\kappa|t/|\alpha|)]$
represents
the Rabi oscillations.
Therefore,
we can estimate
the time scale of the initial collapse
and the period of the revival of the Rabi oscillations at
$|\kappa|^{-1}$
and
$2\pi|\alpha|/|\kappa|$ around,
respectively.
Moreover,
paying attention to
$\alpha^{2}\sin(|\kappa|t/|\alpha|)\sim|\alpha||\kappa|t$ for $\alpha^{2}\gg 1$,
we can estimate the period of the Rabi oscillations at about
$\pi/(|\alpha||\kappa|)$.

\section{\label{section-review-TFD}A brief review of the TFD}
In this section,
we give a brief review of the TFD developed by Takahashi and Umezawa
\cite{Takahashi1975,Ojima1981,Umezawa1982,Umezawa1992}.
The TFD is a method for describing the quantum mechanics at finite temperature.
Using this formalism,
we can describe
the statistical average of an observable at finite temperature
as a pure state expectation value.
Thus,
if we take the TFD formalism,
we do not need to deal with a mixed state,
which is a statistical ensemble of pure states at finite temperature.

In return for the above advantage,
the TFD lets us introduce 
the so-called tilde particles corresponding to the ordinary particles.
Then,
we understand that the ordinary particles and the tilde particles
represent the dynamical degree of freedom and the thermal degree of freedom,
respectively.
Thus, to construct the TFD formalism,
we introduce a fictitious Hilbert space $\tilde{\cal H}$
corresponding to an original Hilbert space ${\cal H}$
and handle quantum mechanics on ${\cal H}\otimes\tilde{\cal H}$.

In this section,
according to the TFD formalism,
we define the thermal vacua of bosons and fermions.
Moreover, we discuss the thermal coherent state proposed
by Barnett, Knight, Mann and Revzen
\cite{Barnett1985,Mann1989a}.
After these preparations,
we rewrite the Hamiltonian of the JCM according to the TFD.

First, we consider the TFD formalism for describing the system of the bosons.
We define
the ordinary Hilbert space
\begin{equation}
{\cal H}_{\mbox{\scriptsize B}}:
\{|0\rangle_{\mbox{\scriptsize B}},|1\rangle_{\mbox{\scriptsize B}},
|2\rangle_{\mbox{\scriptsize B}},...\},
\end{equation}
and
its corresponding tilde space,
\begin{equation}
\tilde{\cal H}_{\mbox{\scriptsize B}}:
\{|\tilde{0}\rangle_{\mbox{\scriptsize B}},|\tilde{1}\rangle_{\mbox{\scriptsize B}},
|\tilde{2}\rangle_{\mbox{\scriptsize B}},...\}.
\end{equation}
Then,
the TFD formalism for the bosons is defined on the following space:
\begin{equation}
\{|n\rangle_{\mbox{\scriptsize B}}\otimes |\tilde{m}\rangle_{\mbox{\scriptsize B}}
\in
{\cal H}_{\mbox{\scriptsize B}}\otimes\tilde{\cal H}_{\mbox{\scriptsize B}}:
n,m\in\{0,1,2,..\}\}.
\end{equation}
We write the creation and annihilation operators on the Hilbert space ${\cal H}_{\mbox{\scriptsize B}}$
as $a^{\dagger}$ and $a$, respectively.
Moreover,
we write the creation and annihilation operators on the Hilbert space $\tilde{\cal H}_{\mbox{\scriptsize B}}$
as $\tilde{a}^{\dagger}$ and $\tilde{a}$, respectively.
Then,
we assume the commutation relations,
\begin{equation}
[a,a^{\dagger}]=[\tilde{a},\tilde{a}^{\dagger}]=1,
\quad
[a,\tilde{a}]=[a,\tilde{a}^{\dagger}]=0.
\label{ordinary-tilde-commutation-relations}
\end{equation}

Next,
we introduce the temperature $\beta=1/(k_{\mbox{\scriptsize B}}T)$
as follows:
\begin{equation}
\hat{U}_{\mbox{\scriptsize B}}(\theta)=\exp[i\theta(\beta)\hat{G}_{\mbox{\scriptsize B}}],
\label{definition-hat-U-operator}
\end{equation}
\begin{equation}
\hat{G}_{\mbox{\scriptsize B}}=i(a\tilde{a}-\tilde{a}^{\dagger}a^{\dagger}),
\label{definition-hat-G-operator}
\end{equation}
\begin{eqnarray}
\cosh\theta(\beta)&=&[1-\exp(-\beta\epsilon)]^{-1/2}, \nonumber \\
\sinh\theta(\beta)&=&[\exp(\beta\epsilon)-1]^{-1/2},
\label{definition-theta-beta-1}
\end{eqnarray}
where $\epsilon=\hbar\omega$.
We note that the relations
$\hat{G}_{\mbox{\scriptsize B}}^{\dagger}=\hat{G}_{\mbox{\scriptsize B}}$,
$\hat{U}_{\mbox{\scriptsize B}}^{\dagger}(\theta)=\hat{U}_{\mbox{\scriptsize B}}^{-1}(\theta)$
and
$\hat{U}_{\mbox{\scriptsize B}}(-\theta)=\hat{U}_{\mbox{\scriptsize B}}^{\dagger}(\theta)$ hold.
In Eqs.~(\ref{definition-hat-U-operator}) and (\ref{definition-hat-G-operator}),
to emphasize that $\hat{G}_{\mbox{\scriptsize B}}$ and $\hat{U}_{\mbox{\scriptsize B}}(\theta)$
are operators acting on both
${\cal H}_{\mbox{\scriptsize B}}$
and
$\tilde{\cal H}_{\mbox{\scriptsize B}}$,
we put an accent (a hat) on them.
Moreover,
we pay attention to the fact that
$\theta(\beta)(\geq 0)$ is real
and
$\theta(\beta)\to +0$ as $\beta\to +\infty$
in Eq.~(\ref{definition-theta-beta-1}).
The index $\mbox{B}$ appearing
in $\hat{G}_{\mbox{\scriptsize B}}$ and $\hat{U}_{\mbox{\scriptsize B}}(\theta)$
stands for the boson.

Because of introducing the temperature,
the creation and annihilation operators
defined on ${\cal H}_{\mbox{\scriptsize B}}$ and $\tilde{\cal H}_{\mbox{\scriptsize B}}$
are transformed as follows:
\begin{eqnarray}
a
&\rightarrow&
a(\theta)
=\hat{U}_{\mbox{\scriptsize B}}(\theta)a\hat{U}_{\mbox{\scriptsize B}}^{\dagger}(\theta)
=
\cosh \theta(\beta)a
-
\sinh \theta(\beta)\tilde{a}^{\dagger}, \nonumber \\
\tilde{a}
&\rightarrow&
\tilde{a}(\theta)
=\hat{U}_{\mbox{\scriptsize B}}(\theta)\tilde{a}\hat{U}_{\mbox{\scriptsize B}}^{\dagger}(\theta)
=
\cosh \theta(\beta)\tilde{a}
-
\sinh \theta(\beta)a^{\dagger}.
\label{temperature-transformations-a-tilde-a}
\end{eqnarray}
The transformation given by Eq.~(\ref{temperature-transformations-a-tilde-a})
is called the Bogoliubov transformation for two mode squeezed states.
From Eqs.~(\ref{ordinary-tilde-commutation-relations})
and
(\ref{temperature-transformations-a-tilde-a}),
we can derive the commutation relations,
\begin{equation}
[a(\theta),a^{\dagger}(\theta)]=[\tilde{a}(\theta),\tilde{a}^{\dagger}(\theta)]=1,
\quad
[a(\theta),\tilde{a}(\theta)]=[a(\theta),\tilde{a}^{\dagger}(\theta)]=0.
\label{ordinary-tilde-commutation-relations-finite-temperature}
\end{equation}
Thus, we can regard $a(\theta)$ and $\tilde{a}(\theta)$
as quasi-particles at the temperature $\beta$.

Here, we define the zero-temperature vacuum as
\begin{eqnarray}
|0,\tilde{0}\rangle_{\mbox{\scriptsize B}}
=
|0\rangle_{\mbox{\scriptsize B}}
\otimes
|\tilde{0}\rangle_{\mbox{\scriptsize B}}
\in{\cal H}_{\mbox{\scriptsize B}}\otimes\tilde{\cal H}_{\mbox{\scriptsize B}}.
\end{eqnarray}
Furthermore,
referring to
Sec.~2.2.3 of Ref.~\cite{Umezawa1992},
we define the thermal vacuum in the form,
\begin{eqnarray}
|0(\theta)\rangle_{\mbox{\scriptsize B}}
&=&
\hat{U}_{\mbox{\scriptsize B}}(\theta)
|0,\tilde{0}\rangle_{\mbox{\scriptsize B}} \nonumber \\
&=&
\exp(-\ln \cosh\theta)\exp[(\tanh\theta)a^{\dagger}\tilde{a}^{\dagger}]
|0,\tilde{0}\rangle_{\mbox{\scriptsize B}}.
\label{definition-thermal-vacuum}
\end{eqnarray}
From Eqs.~(\ref{temperature-transformations-a-tilde-a}) and (\ref{definition-thermal-vacuum}),
we obtain
\begin{equation}
a(\theta)|0(\theta)\rangle_{\mbox{\scriptsize B}}
=
\tilde{a}(\theta)|0(\theta)\rangle_{\mbox{\scriptsize B}}
=
0.
\end{equation}
Thus,
we can consider the thermal vacuum
$|0(\theta)\rangle_{\mbox{\scriptsize B}}$
to be a vacuum for the quasi-particles, which are represented by $a(\theta)$ and $\tilde{a}(\theta)$.

Second, we consider the TFD formalism for describing the system of the fermions.
We define
the ordinary Hilbert space,
\begin{equation}
{\cal H}_{\mbox{\scriptsize F}}:
\{|0\rangle_{\mbox{\scriptsize F}},|1\rangle_{\mbox{\scriptsize F}}\},
\end{equation}
and its corresponding
tilde space,
\begin{equation}
\tilde{\cal H}_{\mbox{\scriptsize F}}:
\{|\tilde{0}\rangle_{\mbox{\scriptsize F}},|\tilde{1}\rangle_{\mbox{\scriptsize F}}\}.
\end{equation}
Then,
the TFD formalism for the fermions is defined on the following space:
\begin{equation}
\{|n\rangle_{\mbox{\scriptsize F}}\otimes |\tilde{m}\rangle_{\mbox{\scriptsize F}}
\in
{\cal H}_{\mbox{\scriptsize F}}\otimes\tilde{\cal H}_{\mbox{\scriptsize F}}:
n,m\in\{0,1\}\}.
\end{equation}
We write the creation and annihilation operators on the Hilbert space
${\cal H}_{\mbox{\scriptsize F}}$
as $c^{\dagger}$ and $c$, respectively.
Moreover, we write the creation and annihilation operators on the Hilbert space
$\tilde{\cal H}_{\mbox{\scriptsize F}}$ as
$\tilde{c}^{\dagger}$ and $\tilde{c}$,
respectively.
Then,
we assume the anti-commutation relations,
\begin{equation}
\{c,c^{\dagger}\}=\{\tilde{c},\tilde{c}^{\dagger}\}=1,
\quad
\{c,\tilde{c}\}=\{c,\tilde{c}^{\dagger}\}=0.
\label{ordinary-tilde-commutation-relations-fermion}
\end{equation}

Next,
we introduce the temperature $\beta$ as follows:
\begin{equation}
\hat{U}_{\mbox{\scriptsize F}}(\theta)=\exp[i\theta(\beta)\hat{G}_{\mbox{\scriptsize F}}],
\label{definition-hat-U-operator-fermion}
\end{equation}
\begin{equation}
\hat{G}_{\mbox{\scriptsize F}}=i(c\tilde{c}-\tilde{c}^{\dagger}c^{\dagger}),
\label{definition-hat-G-operator-fermion}
\end{equation}
\begin{eqnarray}
\cos\theta(\beta)&=&[1+\exp(-\beta\epsilon)]^{-1/2}, \nonumber \\
\sin\theta(\beta)&=&\exp(-\beta\epsilon/2)[1+\exp(-\beta\epsilon)]^{-1/2}.
\label{definition-theta-beta-2}
\end{eqnarray}
We note that the relations
$\hat{G}_{\mbox{\scriptsize F}}^{\dagger}=\hat{G}_{\mbox{\scriptsize F}}$,
$\hat{U}_{\mbox{\scriptsize F}}^{\dagger}(\theta)=\hat{U}_{\mbox{\scriptsize F}}^{-1}(\theta)$
and
$\hat{U}_{\mbox{\scriptsize F}}(-\theta)=\hat{U}_{\mbox{\scriptsize F}}^{\dagger}(\theta)$ hold.
Moreover, we pay attention to the fact that
$\theta(\beta)(\geq 0)$ is real and $\theta(\beta)\to +0$ as $\beta\to +\infty$
in Eq.~(\ref{definition-theta-beta-2}).
$\hat{U}_{\mbox{\scriptsize F}}(\theta)$ in Eq.~(\ref{definition-hat-U-operator-fermion})
and
$\hat{G}_{\mbox{\scriptsize F}}$ in Eq.~(\ref{definition-hat-G-operator-fermion})
are again of the same forms as
$\hat{U}_{\mbox{\scriptsize B}}(\theta)$ in Eq.~(\ref{definition-hat-U-operator})
and
$\hat{G}_{\mbox{\scriptsize B}}$ in Eq.~(\ref{definition-hat-G-operator}),
respectively.
The index $\mbox{F}$ appearing
in $\hat{U}_{\mbox{\scriptsize F}}(\theta)$ and $\hat{G}_{\mbox{\scriptsize F}}$
stands for the fermion.

Because of introducing the temperature,
the creation and annihilation operators defined on 
${\cal H}_{\mbox{\scriptsize F}}$
and
$\tilde{\cal H}_{\mbox{\scriptsize F}}$
are transformed as follows:
\begin{eqnarray}
c
&\rightarrow&
c(\theta)
=\hat{U}_{\mbox{\scriptsize F}}(\theta)c\hat{U}_{\mbox{\scriptsize F}}^{\dagger}(\theta)
=
\cos \theta(\beta)c
+
\sin \theta(\beta)\tilde{c}^{\dagger}, \nonumber \\
\tilde{c}
&\rightarrow&
\tilde{c}(\theta)
=\hat{U}_{\mbox{\scriptsize F}}(\theta)\tilde{c}\hat{U}_{\mbox{\scriptsize F}}^{\dagger}(\theta)
=
\cos \theta(\beta)\tilde{c}
-
\sin \theta(\beta)c^{\dagger}.
\label{temperature-transformations-c-tilde-c}
\end{eqnarray}
From Eqs.~(\ref{ordinary-tilde-commutation-relations-fermion})
and (\ref{temperature-transformations-c-tilde-c}),
we can derive the anti-commutation relations,
\begin{equation}
\{c(\theta),c^{\dagger}(\theta)\}=\{\tilde{c}(\theta),\tilde{c}^{\dagger}(\theta)\}=1,
\quad
\{c(\theta),\tilde{c}(\theta)\}=\{c(\theta),\tilde{c}^{\dagger}(\theta)\}=0.
\label{ordinary-tilde-commutation-relations-finite-temperature-fermion}
\end{equation}
Thus,
we can regard $c(\theta)$ and $\tilde{c}(\theta)$
as quasi-particles at the temperature $\beta$.

Here,
we
define the zero-temperature vacuum as
\begin{eqnarray}
|0,\tilde{0}\rangle_{\mbox{\scriptsize F}}
=
|0\rangle_{\mbox{\scriptsize F}}
\otimes
|\tilde{0}\rangle_{\mbox{\scriptsize F}}
\in{\cal H}_{\mbox{\scriptsize F}}\otimes\tilde{\cal H}_{\mbox{\scriptsize F}}.
\end{eqnarray}
Furthermore,
referring to Sec.~2.4.2 of Ref.~\cite{Umezawa1992},
we define the thermal vacuum in the form,
\begin{eqnarray}
|0(\theta)\rangle_{\mbox{\scriptsize F}}
&=&
\hat{U}_{\mbox{\scriptsize F}}(\theta)
|0,\tilde{0}\rangle_{\mbox{\scriptsize F}} \nonumber \\
&=&
[\cos\theta+(\sin\theta)c^{\dagger}\tilde{c}^{\dagger}]
|0,\tilde{0}\rangle_{\mbox{\scriptsize F}}.
\label{definition-thermal-vacuum-fermion}
\end{eqnarray}
From Eqs.~(\ref{temperature-transformations-c-tilde-c})
and
(\ref{definition-thermal-vacuum-fermion}),
we obtain
\begin{equation}
c(\theta)|0(\theta)\rangle_{\mbox{\scriptsize F}}
=
\tilde{c}(\theta)|0(\theta)\rangle_{\mbox{\scriptsize F}}
=
0.
\end{equation}
Thus,
we can consider the thermal vacuum
$|0(\theta)\rangle_{\mbox{\scriptsize F}}$
to be a vacuum for the quasi-particles,
which are represented by
$c(\theta)$ and $\tilde{c}(\theta)$.

Third, we consider the thermal coherent state.
After the above preparations,
Barnett, Knight, Mann and Revzen
define the thermal coherent state as follows
\cite{Barnett1985,Mann1989a}:
\begin{equation}
|\alpha,\tilde{\gamma};\theta\rangle_{\mbox{\scriptsize B}}
=
\exp[\alpha a^{\dagger}(\theta)
+\tilde{\gamma}^{*}\tilde{a}^{\dagger}(\theta)
-\alpha^{*}a(\theta)
-\tilde{\gamma}\tilde{a}(\theta)]
|0(\theta)\rangle_{\mbox{\scriptsize B}},
\label{definition-thermal-coherent-state-1}
\end{equation}
where $\alpha(\theta)$ and $\tilde{\alpha}(\theta)$
are given by Eq.~(\ref{temperature-transformations-a-tilde-a}).

However,
the TFD formalism requires
all state vectors to be invariant under the tilde conjugation,
which is given by
\begin{eqnarray}
(XY)\tilde{\;\;}&=&\tilde{X}\tilde{Y}, \nonumber \\
(\xi_{1}X+\xi_{2}Y)\tilde{\;\;}&=&\xi_{1}^{*}\tilde{X}+\xi_{2}^{*}\tilde{Y}, \nonumber \\
(X^{\dagger})\tilde{\;\;}&=&\tilde{X}^{\dagger}, \nonumber \\
(\tilde{X})\tilde{\;\;}&=&\sigma X,
\end{eqnarray}
where
\begin{equation}
\sigma
=
\left\{
\begin{array}{ll}
1 & \mbox{(boson)}\\
-1 & \mbox{(fermion)}
\end{array}
\right.,
\end{equation}
$X$ and $Y$
are arbitrary operators defined on
${\cal H}_{\mbox{\scriptsize B}}$
and
${\cal H}_{\mbox{\scriptsize F}}$,
and
$\xi_{1}$ and $\xi_{2}$ are arbitrary complex numbers.
For example,
the thermal vacua are obviously invariant under the tilde conjugation,
that is to say,
$|0(\theta)\rangle_{\mbox{\scriptsize B}}\tilde{\;\;}
=|0(\theta)\rangle_{\mbox{\scriptsize B}}$,
${}_{\mbox{\scriptsize B}}\langle 0(\theta)|\tilde{\;\;}
={}_{\mbox{\scriptsize B}}\langle 0(\theta)|$,
$|0(\theta)\rangle_{\mbox{\scriptsize F}}\tilde{\;\;}
=|0(\theta)\rangle_{\mbox{\scriptsize F}}$
and
${}_{\mbox{\scriptsize F}}\langle 0(\theta)|\tilde{\;\;}
={}_{\mbox{\scriptsize F}}\langle 0(\theta)|$.
Moreover,
we can show
$(\hat{G}_{\mbox{\scriptsize B}})\tilde{\;\;}=-\hat{G}_{\mbox{\scriptsize B}}$,
$(\hat{U}_{\mbox{\scriptsize B}}(\theta))\tilde{\;\;}=\hat{U}_{\mbox{\scriptsize B}}(\theta)$,
$(\hat{G}_{\mbox{\scriptsize F}})\tilde{\;\;}=-\hat{G}_{\mbox{\scriptsize F}}$
and
$(\hat{U}_{\mbox{\scriptsize F}}(\theta))\tilde{\;\;}=\hat{U}_{\mbox{\scriptsize F}}(\theta)$,
at ease.

In the TFD formalism,
all state vectors realized actually in the physical system
have to be invariant under the tilde conjugation.
Thus, not only the time-evolution but also all possible transitions of state vectors
have to be invariant
under the tilde conjugation.

Requiring $|\alpha,\tilde{\gamma};\theta\rangle_{\mbox{\scriptsize B}}$
given by Eq.~(\ref{definition-thermal-coherent-state-1})
to be invariant under the tilde conjugation,
we obtain
\begin{eqnarray}
|\alpha;\theta\rangle_{\mbox{\scriptsize B}}
&=&
|\alpha,\alpha^{*};\theta\rangle_{\mbox{\scriptsize B}} \nonumber \\
&=&
\exp[\alpha a^{\dagger}(\theta)
+\alpha\tilde{a}^{\dagger}(\theta)
-\alpha^{*}a(\theta)
-\alpha^{*}\tilde{a}(\theta)]
|0(\theta)\rangle_{\mbox{\scriptsize B}}.
\label{definition-thermal-coherent-state-2}
\end{eqnarray}
From now on,
we call this state the thermal coherent state
\cite{Mann1989b,Kireev1989,Mann1989c}.
Moreover, for simplicity,
we assume $\alpha$ characterizing $|\alpha;\theta\rangle_{\mbox{\scriptsize B}}$
to be always real.

In the following paragraphs,
we examine the physical meanings of the thermal vacua,
$|0(\theta)\rangle_{\mbox{\scriptsize B}}$
and
$|0(\theta)\rangle_{\mbox{\scriptsize F}}$.

First,
we clarify the physical meanings of the thermal vacuum for the bosons
$|0(\theta)\rangle_{\mbox{\scriptsize B}}$.
We begin by considering the density operator defined on
${\cal H}_{\mbox{\scriptsize B}}$,
\begin{equation}
\rho_{\mbox{\scriptsize B}}(\theta)
=
\mbox{Tr}_{\tilde{\cal H}}|0(\theta)\rangle_{\mbox{\scriptsize B}}
{}_{\mbox{\scriptsize B}}\langle 0(\theta)|.
\label{density-operator-boson-thermal-vacuum-0}
\end{equation}
Then,
we derive an explicit representation of
$\rho_{\mbox{\scriptsize B}}(\theta)$
as follows:
At first,
from Eq.~(\ref{definition-thermal-vacuum}) and
(\ref{density-operator-boson-thermal-vacuum-0}),
we obtain
\begin{eqnarray}
\rho_{\mbox{\scriptsize B}}(\theta)
&=&
\frac{1}{\cosh^{2}\theta}
\sum_{n=0}^{\infty}
\frac{1}{n!}
{}_{\mbox{\scriptsize B}}\langle\tilde{0}|\tilde{a}^{n}
\exp[(\tanh\theta)a^{\dagger}\tilde{a}^{\dagger}]
|0,\tilde{0}\rangle_{\mbox{\scriptsize B}} \nonumber \\
&&\times
{}_{\mbox{\scriptsize B}}\langle 0,\tilde{0}|
\exp[(\tanh\theta)\tilde{a}a](\tilde{a}^{\dagger})^{n}
|\tilde{0}\rangle_{\mbox{\scriptsize B}}.
\label{density-operator-thermal-vacuum-boson-1}
\end{eqnarray}
Next, we apply the following relation
to Eq.~(\ref{density-operator-thermal-vacuum-boson-1}):
\begin{equation}
\exp[-(\tanh\theta)a^{\dagger}\tilde{a}^{\dagger}]
\tilde{a}^{n}
\exp[(\tanh\theta)a^{\dagger}\tilde{a}^{\dagger}]
=
[\tilde{a}+(\tanh\theta)a^{\dagger}]^{n}.
\label{formula-a-tilde-a-condensation-1}
\end{equation}
Then using Eq.~(\ref{definition-theta-beta-1}),
we obtain
\begin{eqnarray}
\rho_{\mbox{\scriptsize B}}(\theta)
&=&
\frac{1}{\cosh^{2}\theta}
\sum_{n=0}^{\infty}
\tanh^{2n}\theta
|n\rangle_{\mbox{\scriptsize B}}{}_{\mbox{\scriptsize B}}\langle n| \nonumber \\
&=&
(1-e^{-\beta\epsilon})\sum_{n=0}^{\infty}e^{-n\beta\epsilon}
|n\rangle_{\mbox{\scriptsize B}}{}_{\mbox{\scriptsize B}}\langle n|.
\label{explicit-form-density-operator-thermal-vacuum-boson}
\end{eqnarray}

Looking at Eq.~(\ref{explicit-form-density-operator-thermal-vacuum-boson}),
we notice that
$\rho_{\mbox{\scriptsize B}}(\theta)$
is an ensemble of quantum states
$\{|n\rangle_{\mbox{\scriptsize B}}:n=0,1,2,...\}$,
into each of which $n$ bosons ($a$-particles) are put.
Moreover,
the statistical probability of $|n\rangle_{\mbox{\scriptsize B}}$
is given by $(1-e^{-\beta\epsilon})e^{-n\beta\epsilon}
=\mbox{Const.}\times e^{-\beta(n\epsilon)}$,
so that
$\rho_{\mbox{\scriptsize B}}(\theta)$
represents a canonical ensemble of the Bose-Einstein distribution in thermal equilibrium.
From these considerations,
we understand that
the $a$-particle represents the dynamical degree of freedom
and the $\tilde{a}$-particle represents the thermal degree of freedom
in
$|0(\theta)\rangle_{\mbox{\scriptsize B}}
\in
{\cal H}_{\mbox{\scriptsize B}}
\otimes
\tilde{\cal H}_{\mbox{\scriptsize B}}$.

Second,
we clarify the physical meanings of the thermal vacuum for fermions
$|0(\theta)\rangle_{\mbox{\scriptsize F}}$.
We begin by considering
the density operator defined on ${\cal H}_{\mbox{\scriptsize F}}$,
\begin{equation}
\rho_{\mbox{\scriptsize F}}(\theta)
=
\mbox{Tr}_{\tilde{\cal H}}|0(\theta)\rangle_{\mbox{\scriptsize F}}
{}_{\mbox{\scriptsize F}}\langle 0(\theta)|.
\label{density-operator-fermion-thermal-vacuum-0}
\end{equation}
From Eqs.~(\ref{definition-theta-beta-2}),
(\ref{definition-thermal-vacuum-fermion})
and
(\ref{density-operator-fermion-thermal-vacuum-0}),
we derive an explicit representation of
$\rho_{\mbox{\scriptsize F}}(\theta)$
in the form,
\begin{eqnarray}
\rho_{\mbox{\scriptsize F}}(\theta)
&=&
\sum_{n\in\{0,1\}}
{}_{\mbox{\scriptsize F}}\langle\tilde{n}|
[\cos\theta+(\sin\theta)c^{\dagger}\tilde{c}^{\dagger}]
|0,\tilde{0}\rangle_{\mbox{\scriptsize F}}
{}_{\mbox{\scriptsize F}}\langle 0,\tilde{0}|
[\cos\theta+(\sin\theta)\tilde{c}c]
|\tilde{n}\rangle_{\mbox{\scriptsize F}} \nonumber \\
&=&
\cos^{2}\theta|0\rangle_{\mbox{\scriptsize F}}{}_{\mbox{\scriptsize F}}\langle 0|
+
\sin^{2}\theta|1\rangle_{\mbox{\scriptsize F}}{}_{\mbox{\scriptsize F}}\langle 1| \nonumber \\
&=&
\frac{1}{1+e^{-\beta\epsilon}}
|0\rangle_{\mbox{\scriptsize F}}{}_{\mbox{\scriptsize F}}\langle 0|
+
\frac{e^{-\beta\epsilon}}{1+e^{-\beta\epsilon}}
|1\rangle_{\mbox{\scriptsize F}}{}_{\mbox{\scriptsize F}}\langle 1|.
\label{density-operator-thermal-vacuum-fermion-1}
\end{eqnarray}

Looking at Eq.~(\ref{density-operator-thermal-vacuum-fermion-1}),
we notice that
$\rho(\theta)_{\mbox{\scriptsize F}}$
is an ensemble of quantum states
$\{|n\rangle_{\mbox{\scriptsize F}}:n\in\{0,1\}\}$,
into each of which $n$ fermions ($c$-particles) are put.
Moreover,
the statistical probability of $|n\rangle_{\mbox{\scriptsize F}}$
is given by
$(1+e^{-\beta\epsilon})^{-1}e^{-n\beta\epsilon}
=\mbox{Const.}\times e^{-\beta(n\epsilon)}$,
so that $\rho(\theta)_{\mbox{\scriptsize F}}$
represents a canonical ensemble of the Fermi-Dirac distribution in thermal equilibrium.
From these considerations,
we understand that
the $c$-particle represents the dynamical degree of freedom
and
the $\tilde{c}$-particle represents the thermal degree of freedom
in
$|0(\theta)\rangle_{\mbox{\scriptsize F}}
\in
{\cal H}_{\mbox{\scriptsize F}}
\otimes
\tilde{\cal H}_{\mbox{\scriptsize F}}$.

For convenience of calculations
that appear in the remains of this paper,
using Eqs.~(\ref{ordinary-tilde-commutation-relations})
and (\ref{temperature-transformations-a-tilde-a}),
we rewrite $|\alpha;\theta\rangle_{\mbox{\scriptsize B}}$
given by Eq.~(\ref{definition-thermal-coherent-state-2})
as
\begin{eqnarray}
|\alpha;\theta\rangle_{\mbox{\scriptsize B}}
&=&
\exp[\hat{U}_{\mbox{\scriptsize B}}(\theta)
(\alpha a^{\dagger}
+\alpha\tilde{a}^{\dagger}
-\alpha a
-\alpha\tilde{a})
\hat{U}_{\mbox{\scriptsize B}}^{\dagger}(\theta)]
\hat{U}_{\mbox{\scriptsize B}}(\theta)
|0,\tilde{0}\rangle_{\mbox{\scriptsize B}} \nonumber \\
&=&
\hat{U}_{\mbox{\scriptsize B}}(\theta)
\exp[\alpha(a^{\dagger}-a)]
\exp[\alpha(\tilde{a}^{\dagger}-\tilde{a})]
|0,\tilde{0}\rangle_{\mbox{\scriptsize B}} \nonumber \\
&=&
\hat{U}_{\mbox{\scriptsize B}}(\theta)
|\alpha\rangle_{\mbox{\scriptsize B}}
|\tilde{\alpha}\rangle_{\mbox{\scriptsize B}},
\label{another-form-of-thermal-coherent-state-1}
\end{eqnarray}
where
$|\alpha\rangle_{\mbox{\scriptsize B}}$
and
$|\tilde{\alpha}\rangle_{\mbox{\scriptsize B}}$
are the coherent states at zero temperature
defined on
${\cal H}_{\mbox{\scriptsize B}}$
and
$\tilde{\cal H}_{\mbox{\scriptsize B}}$,
respectively.

Next, after the above preparations,
we discuss how to construct the finite-temperature JCM according to the TFD formalism
from the original JCM Hamiltonian.

For example,
we consider a system,
which consists of bosons $a$ and fermions $c$.
We assume that
the time-evolution of the system is reversible and it never causes dissipation.
This implies that
the Hamiltonian $H$,
which is the Hermitian operator corresponding to the total energy of the system
(the $a$-particles and the $c$-particles),
is equivalent to the generator of the unitary operator for the time-evolution.

Obeying the TFD formalism,
we introduce the $\tilde{a}$-particle corresponding to the $a$-particle
and
the $\tilde{c}$-particle corresponding to the $c$-particle
into the system.
Because $\tilde{a}$ and $\tilde{c}$ are fictitious particles
representing the thermal degree of freedom,
the Schr{\"o}dinger equation governing the time-evolution of the particles $a$ and $c$
never suffers form the thermal effects.
This observation suggests
that the Schr{\"o}dinger equation for $a$ and $c$
never changes in spite of introducing $\tilde{a}$ and $\tilde{c}$.
From these considerations,
we can conclude that
the particles ($a$ and $c$) are never coupled to the tilde particles ($\tilde{a}$ and $\tilde{c}$)
direct in the Hamiltonian.
Hence,
the total Hamiltonian based on the TFD formalism
has to be a sum
of the interaction terms of the particles ($a$ and $c$)
and the interaction terms of the tilde particles ($\tilde{a}$ and $\tilde{c}$).

Here, we describe the total Hamiltonian for the TFD formalism as $\hat{H}$.
Then,
letting the Schr{\"o}dinger equation for the whole system consisting of the particles
of $a$, $\tilde{a}$, $c$ and $\tilde{c}$
be invariant under the tilde conjugation,
we can express $\hat{H}$ in the form,
\begin{equation}
\hat{H}=H-\tilde{H},
\label{hat-Hamiltonian-0}
\end{equation}
where $\tilde{H}=(H)\tilde{\;\;}$
\cite{Takahashi1975,Ojima1981,Umezawa1982,Umezawa1992}.
If we give the Hamiltonian of the system $\hat{H}$ in the form of Eq.~(\ref{hat-Hamiltonian-0}),
the time-evolution of the whole system consisting of the particles
and the tilde particles
is reversible and it never causes dissipation.

In Eq.~(\ref{hat-Hamiltonian-0}),
we emphasize the following:
The Hamiltonian $H$ is constructed from $a$ and $c$.
Then,
because of $\tilde{H}=(H)\tilde{\;\;}$,
the Hamiltonian $\tilde{H}$ has to include both $\tilde{a}$ and $\tilde{c}$.
This implies that we have to introduce the temperature into
both the $a$-particles and the $c$-particles.

If the Hamiltonian of the total system is given by Eq.~(\ref{hat-Hamiltonian-0}),
the thermal vacua $|0(\theta)\rangle_{\mbox{\scriptsize B}}$ given by Eq.~(\ref{definition-thermal-vacuum})
and $|0(\theta)\rangle_{\mbox{\scriptsize F}}$ given by Eq.~(\ref{definition-thermal-vacuum-fermion})
are not dependent on time,
so that they are stationary states.
We can explain this fact as follows:
Turning our eyes towards Eq.~(\ref{definition-thermal-vacuum}),
we notice that
the condensation of $(a\tilde{a})$-pairs into $|0(\theta)\rangle_{\mbox{\scriptsize B}}$ occurs.
Thus, in the vacuum $|0(\theta)\rangle_{\mbox{\scriptsize B}}$,
the $a$-particle receives a phase factor $\exp[-i(H/\hbar)t]$
and the $\tilde{a}$-particle receives a phase factor $\exp[i(\tilde{H}/\hbar)t]$.
Then,
these phase factors cancel out their effects with each other,
and the ($a\tilde{a}$)-pair acts like a zero-energy boson.
Hence, we understand that the thermal vacuum
$|0(\theta)\rangle_{\mbox{\scriptsize B}}$
is not dependent on time.
We can apply a similar discussion to the thermal vacuum of  the fermions,
because we can observe the condensation of $(c\tilde{c})$-pairs
into $|0(\theta)\rangle_{\mbox{\scriptsize F}}$ in Eq.~(\ref{definition-thermal-vacuum-fermion}).

To obtain the JCM Hamiltonian
$\hat{H}$ which takes the form given by Eq.~(\ref{hat-Hamiltonian-0}),
we rewrite the original Hamiltonian $H$ of the JCM
given by Eq.~(\ref{JCM-Hamiltonian-0})
as
\begin{equation}
H
=
\frac{\hbar}{2}\omega_{0}(2c^{\dagger}c-1)
+
\hbar\omega a^{\dagger}a
+
\hbar\kappa(c^{\dagger}a+ca^{\dagger}).
\label{JCM-Hamiltonian-1}
\end{equation}
The Pauli matrices $\sigma_{z}$, $\sigma_{+}$ and $\sigma_{-}$
in the original Hamiltonian $H$ given by Eq.~(\ref{JCM-Hamiltonian-0})
are replaced with the creation and annihilation operators of the fermions
$(2c^{\dagger}c-1)$, $c^{\dagger}$ and $c$
in the rewritten Hamiltonian $H$ given by Eq.~(\ref{JCM-Hamiltonian-1}).

Extending the Hamiltonian $H$ given by Eq.~(\ref{JCM-Hamiltonian-1})
according to Eq.~(\ref{hat-Hamiltonian-0}),
we obtain
\begin{equation}
\hat{H}
=
(H_{0}+H_{\mbox{\scriptsize I}})
-
(\tilde{H}_{0}+\tilde{H}_{\mbox{\scriptsize I}}),
\label{extended-JCM-Hamiltonian-0}
\end{equation}
where
\begin{eqnarray}
H_{0}
&=&
\frac{\hbar}{2}\omega_{0}
(2c^{\dagger}c-1)
+
\hbar\omega a^{\dagger}a, \nonumber \\
H_{\mbox{\scriptsize I}}
&=&
\hbar\kappa(c^{\dagger}a+ca^{\dagger}), \nonumber \\
\tilde{H}_{0}
&=&
\frac{\hbar}{2}\omega_{0}
(2\tilde{c}^{\dagger}\tilde{c}-1)
+
\hbar\omega \tilde{a}^{\dagger}\tilde{a}, \nonumber \\
\tilde{H}_{\mbox{\scriptsize I}}
&=&
\hbar\kappa(\tilde{c}^{\dagger}\tilde{a}+\tilde{c}\tilde{a}^{\dagger}).
\label{extended-JCM-Hamiltonian-1}
\end{eqnarray}
In this paper,
we concentrate on examining the Hamiltonian $\hat{H}$
given by Eqs.~(\ref{extended-JCM-Hamiltonian-0}) and (\ref{extended-JCM-Hamiltonian-1}).

Here, we think about
some Hamiltonians,
which are proposed in the other works.
Especially, we examine whether or not we can regard them
as genuine JCM Hamiltonians based on the TFD formalism.

Barnett, Knight and Azuma consider the following Hamiltonian
in Refs.~\cite{Barnett1985,Azuma2010}:
\begin{equation}
\hat{H}_{\mbox{\scriptsize BKA}}
=
\frac{\hbar}{2}\omega_{0}\sigma_{z}
+
\hbar\omega(a^{\dagger}a-\tilde{a}^{\dagger}\tilde{a})
+
\hbar\kappa(\sigma_{+}a+\sigma_{-}a^{\dagger}).
\label{TFD-JCM-Hamiltonian-Azuma}
\end{equation}
However,
the Hamiltonian $\hat{H}_{\mbox{\scriptsize BKA}}$ given by Eq.~(\ref{TFD-JCM-Hamiltonian-Azuma})
does not include the tilde operators
corresponding to the atomic operators $\sigma_{z}$ and $\sigma_{\pm}$,
so that the states of the atom are always at zero temperature.
Because
the Hamiltonian $\hat{H}_{\mbox{\scriptsize BKA}}$ given by Eq.~(\ref{TFD-JCM-Hamiltonian-Azuma})
does not introduce the temperature into the atom,
we cannot regard it as a genuine Hamiltonian based on the TFD formalism.

Fan and Lu propose the following Hamiltonian in Ref.~\cite{Fan2004}:
\begin{eqnarray}
\hat{H}_{\mbox{\scriptsize FL}}
&=&
\frac{\hbar}{2}\omega_{0}\sigma_{z}
+
\hbar\omega(a^{\dagger}a-\tilde{a}^{\dagger}\tilde{a}) \nonumber \\
&&
+
\hbar\kappa
[
\sigma_{+}
\sqrt{
\frac{a-\tilde{a}^{\dagger}}{a^{\dagger}-\tilde{a}}
}
\sqrt{a^{\dagger}a-\tilde{a}^{\dagger}\tilde{a}}
+
\sqrt{a^{\dagger}a-\tilde{a}^{\dagger}\tilde{a}}
\sqrt{
\frac{a^{\dagger}-\tilde{a}}{a-\tilde{a}^{\dagger}}
}
\sigma_{-}
].
\label{Fan-Lu-Hamiltonian}
\end{eqnarray}
The reason why Fan and Lu construct the Hamiltonian $\hat{H}_{\mbox{\scriptsize FL}}$
given by Eq.~(\ref{Fan-Lu-Hamiltonian})
is as follows:
They find the commutation relations,
\begin{eqnarray}
{[}
a^{\dagger}a-\tilde{a}^{\dagger}\tilde{a},
\sqrt{\frac{a^{\dagger}-\tilde{a}}{a-\tilde{a}^{\dagger}}}
{]}
&=&
\sqrt{\frac{a^{\dagger}-\tilde{a}}{a-\tilde{a}^{\dagger}}}, \nonumber \\
{[}
a^{\dagger}a-\tilde{a}^{\dagger}\tilde{a},
\sqrt{\frac{a-\tilde{a}^{\dagger}}{a^{\dagger}-\tilde{a}}}
{]}
&=&
-
\sqrt{\frac{a-\tilde{a}^{\dagger}}{a^{\dagger}-\tilde{a}}}.
\end{eqnarray}
Thus,
if we regard
$(a^{\dagger}a-\tilde{a}^{\dagger}\tilde{a})$
as an extended number operator,
we can think
\\
$\sqrt{(a^{\dagger}-\tilde{a})/(a-\tilde{a}^{\dagger})}$
and
$\sqrt{(a-\tilde{a}^{\dagger})/(a^{\dagger}-\tilde{a})}$
to be extended creation and annihilation operators, respectively. 
From these suggestions,
Fan and Lu propose the Hamiltonian $\hat{H}_{\mbox{\scriptsize FL}}$
given in Eq.~(\ref{Fan-Lu-Hamiltonian}).
However,
Fan and Lu's Hamiltonian $\hat{H}_{\mbox{\scriptsize FL}}$
does not include the tilde operators
corresponding to the atomic operators $\sigma_{z}$ and $\sigma_{\pm}$.
Thus, it can not be regarded as a genuine Hamiltonian based on the TFD formalism.
Moreover,
in the Hamiltonian $\hat{H}_{\mbox{\scriptsize FL}}$,
the operators $a$ and $a^{\dagger}$
are coupled direct to
the tilde operators $\tilde{a}$ and $\tilde{a}^{\dagger}$.
Thus, Fan and Lu's system suffers from dissipation
during the time-evolution.
To examine their system,
we have to deal with non-equilibrium states.
Because it is beyond the purpose of this paper,
we do not involve ourselves in it.

From now on,
we examine the Hamiltonian $\hat{H}$
given by Eqs.~(\ref{extended-JCM-Hamiltonian-0}) and (\ref{extended-JCM-Hamiltonian-1}).
We can divide the Hamiltonian $\hat{H}$ into two parts as follows: 
\begin{eqnarray}
\hat{H}
&=&
\hbar(\hat{C}_{1}+\hat{C}_{2}), \nonumber \\
\hat{C}_{1}
&=&
\omega
[(c^{\dagger}c-\tilde{c}^{\dagger}\tilde{c})
+
(a^{\dagger}a-\tilde{a}^{\dagger}\tilde{a})], \nonumber \\
\hat{C}_{2}
&=&
\kappa(c^{\dagger}a+ca^{\dagger})
-
\kappa(\tilde{c}^{\dagger}\tilde{a}+\tilde{c}\tilde{a}^{\dagger})
-
\Delta\omega(c^{\dagger}c-\tilde{c}^{\dagger}\tilde{c}).
\label{decomposition-hat-Hamiltonian}
\end{eqnarray}
Then, we obtain a commutation relation
$[\hat{C}_{1},\hat{C}_{2}]=0$.
Moreover, we can diagonalize $\hat{C}_{1}$ at ease.
Thus,
we describe the time-evolution of the total system with the interaction picture
as follows:
First, we write
the state vector of the total system in the Schr{\"o}dinger picture
as $|\Psi_{\mbox{\scriptsize S}}(t)\rangle$.
Second,
assuming $|\Psi_{\mbox{\scriptsize I}}(0)\rangle
=|\Psi_{\mbox{\scriptsize S}}(0)\rangle$,
we define
the state vector of the total system in the interaction picture
as
$|\Psi_{\mbox{\scriptsize I}}(t)\rangle
=\exp(i\hat{C}_{1}t)|\Psi_{\mbox{\scriptsize S}}(t)\rangle$.
Hence,
we can describe the time-evolution
as
$|\Psi_{\mbox{\scriptsize I}}(t)\rangle
=\hat{U}(t)|\Psi_{\mbox{\scriptsize I}}(0)\rangle$,
where
$\hat{U}(t)=\exp(-i\hat{C}_{2}t)$.

The unitary operator for the time-evolution of $|\Psi_{\mbox{\scriptsize I}}(t)\rangle$
is given by
\begin{eqnarray}
\hat{U}(t)
&=&
\exp[-i\hat{C}_{2}t] \nonumber \\
&=&
U(t)\otimes \tilde{U}(t) \nonumber \\
&=&
\exp[-it
\left(
\begin{array}{cc}
-\Delta\omega/2 & \kappa a \\
\kappa a^{\dagger} & \Delta\omega/2
\end{array}
\right)
]
\otimes
\exp[it
\left(
\begin{array}{cc}
-\Delta\omega/2 & \kappa \tilde{a} \\
\kappa \tilde{a}^{\dagger} & \Delta\omega/2
\end{array}
\right)
].
\label{hat-time-evolution-unitary-operator-0}
\end{eqnarray}
In the right-hand side of Eq.~(\ref{hat-time-evolution-unitary-operator-0}),
the first $2\times 2$ matrix of the tensor product acts on ${\cal H}_{\mbox{\scriptsize F}}$
and
the second $2\times 2$ matrix of the tensor product acts on $\tilde{\cal H}_{\mbox{\scriptsize F}}$.
The $2\times 2$ matrix $U(t)$ appearing in Eq.~(\ref{hat-time-evolution-unitary-operator-0})
and
the unitary operator for the time-evolution defined in Eq.~(\ref{unitary-evolution-1})
are in the same form.
Thus,
the elements of the $2\times 2$ matrices $U(t)$ and $\tilde{U}(t)$
appearing in Eq.~(\ref{hat-time-evolution-unitary-operator-0})
are given by Eq.~(\ref{unitary-evolution-2}).

\section{\label{section-thermal-effects-period-qualitative-estimation}
Thermal effects of the period of the revival of the Rabi oscillations}
In this paper,
putting the single cavity mode and the atom
in the thermal coherent state $|\alpha;\theta\rangle_{\mbox{\scriptsize B}}$
and
the thermal vacuum $|0(\theta)\rangle_{\mbox{\scriptsize F}}$,
respectively,
at the time $t=0$,
we aim at examining the time-evolution of the JCM.
We assume that
the system consisting of
the single cavity mode and the atom
evolve in time without dissipation,
and it maintains the constant temperature $\beta$ all the time.
As shown in Sec.~\ref{section-review-JCM},
we can estimate that
the period of the revival of the Rabi oscillations at zero temperature
is around $2\pi|\alpha|/|\kappa|$.
In this section,
we discuss how the period changes at finite low temperature.
We evaluate the thermal effects of the period in an intuitive manner.

The parameter $|\alpha|$,
which characterizes the ordinary zero-temperature coherent state $|\alpha\rangle$,
is given by
\begin{equation}
|\alpha|
=
(\langle\alpha|a^{\dagger}a|\alpha\rangle)^{1/2}.
\end{equation}
Thus,
we can guess that the parameter $|\alpha|$ varies
with the thermal effects of the finite low temperature
as
\begin{equation}
|\alpha|
\rightarrow
(\langle\alpha;\theta|a^{\dagger}a|\alpha;\theta\rangle)^{1/2}.
\end{equation}

On the other hand,
using Eqs.~(\ref{ordinary-tilde-commutation-relations}),
(\ref{definition-hat-U-operator}),
(\ref{definition-hat-G-operator}),
(\ref{temperature-transformations-a-tilde-a})
and
(\ref{another-form-of-thermal-coherent-state-1}),
and paying attention to
$a|\alpha\rangle=\alpha|\alpha\rangle$
and
$\tilde{a}|\tilde{\alpha}\rangle=\alpha|\tilde{\alpha}\rangle$,
we obtain
\begin{eqnarray}
&&
\langle\alpha;\theta|a^{\dagger}a|\alpha;\theta\rangle \nonumber \\
&=&
\langle\alpha|\langle\tilde{\alpha}|
\hat{U}_{\mbox{\scriptsize B}}(-\theta)
a^{\dagger}a
\hat{U}_{\mbox{\scriptsize B}}^{\dagger}(-\theta)
|\alpha\rangle|\tilde{\alpha}\rangle \nonumber \\
&=&
\langle\alpha|\langle\tilde{\alpha}|
a^{\dagger}(-\theta)a(-\theta)
|\alpha\rangle|\tilde{\alpha}\rangle \nonumber \\
&=&
\alpha^{2}e^{2\theta}
+
(1/4)(e^{2\theta}+e^{-2\theta}-2).
\end{eqnarray}

From the above observations, we can expect the following:
assuming $\alpha^{2}\gg 1$ and $\theta\ll 1$,
the thermal effects let
the parameter characterizing the coherent state
change 
under the low-temperature limit
as
\begin{equation}
|\alpha|
\rightarrow
(\langle\alpha;\theta|a^{\dagger}a|\alpha;\theta\rangle)^{1/2}
\simeq
|\alpha|e^{\theta}.
\label{thermal-effect-of-alpha-coherent-state-0}
\end{equation}
Thus,
we can expect that
the period of the revival of the Rabi oscillations
at finite low temperature varies as
\begin{equation}
2\pi|\alpha|/|\kappa|
\rightarrow
2\pi|\alpha|e^{\theta}/|\kappa|.
\label{thermal-effect-period-rivival-Rabi-oscillations}
\end{equation}
This phenomenon is confirmed by numerical calculations in Sec.~\ref{section-numerical-calculations}.

The intuitive discussions given in this section is effective,
when we can specify the whole system with the constant temperature $\beta$.
If the system is in a non-equilibrium state and the temperature $\beta$ varies during its time-evolution,
we cannot apply the above intuitive discussions to the system,
so that Eq.~(\ref{thermal-effect-period-rivival-Rabi-oscillations}) does not hold.

\section{\label{section-formulation-perturbetion-theory}The formulation of the perturbation theory}
In this section,
we initially
put the states of the atom and the cavity field
in the thermal vacuum of the fermions $|0(\theta)\rangle_{\mbox{\scriptsize F}}$
and
the thermal coherent state $|\alpha;\beta\rangle_{\mbox{\scriptsize B}}$,
respectively.
Then,
we formulate the time-evolution of the JCM based on the TFD
discussed in Sec.~\ref{section-review-TFD}
as the perturbation theory under the low-temperature limit.
After formulating the perturbation theory here,
we estimate the zero-th, first, second and third order corrections
in Secs.~\ref{section-0th-order-correction-term},
\ref{section-1st-order-correction-term},
\ref{section-2nd-order-correction-term}
and
Appendix~\ref{section-3rd-order-correction-term}.
To evaluate these correction terms,
we make use of techniques for calculations developed in Ref.~\cite{Azuma2010}.

At first, we express the state of the system for $t=0$ in the form,
\begin{equation}
|\Psi_{\mbox{\scriptsize I}}(0)\rangle
=
|0(\Theta)\rangle_{\mbox{\scriptsize F}}
|\alpha;\theta\rangle_{\mbox{\scriptsize B}}.
\label{extended-JCM-initial-state}
\end{equation}
During the time-evolution of the system,
we assume the atom and the cavity mode do not suffer dissipation
and maintain the constant temperature $\beta$.
Thus,
from Eqs.~(\ref{definition-theta-beta-1})
and
(\ref{definition-theta-beta-2}),
the parameters of the temperature for the fermionic atom $\Theta(\beta)$
and
the bosonic cavity mode $\theta(\beta)$
are given in the following forms, respectively:
\begin{eqnarray}
\cos\Theta(\beta)
&=&
[1+\exp(-\beta\hbar\omega_{0})]^{-1/2}, \nonumber \\
\sin\Theta(\beta)
&=&
\exp(-\beta\hbar\omega_{0}/2)
[1+\exp(-\beta\hbar\omega_{0})]^{-1/2}, \nonumber \\
\cosh\theta(\beta)
&=&
[1-\exp(-\beta\hbar\omega)]^{-1/2}, \nonumber \\
\sinh\theta(\beta)
&=&
[\exp(\beta\hbar\omega)-1]^{-1/2},
\label{definition-temperature-JCM-0}
\end{eqnarray}
where
$\omega_{0}$ represents the transition frequency of the two-level atom
and
$\omega$ represents the frequency of the single cavity mode,
as defined in Eq.~(\ref{JCM-Hamiltonian-0}).
From Eqs.~(\ref{another-form-of-thermal-coherent-state-1})
and
(\ref{hat-time-evolution-unitary-operator-0}),
we obtain $|\Psi_{\mbox{\scriptsize I}}(t)\rangle$
as
\begin{eqnarray}
|\Psi_{\mbox{\scriptsize I}}(t)\rangle
&=&
\hat{U}(t)|\Psi_{\mbox{\scriptsize I}}(0)\rangle \nonumber \\
&=&
[U(t)\otimes\tilde{U}(t)]
|0(\Theta)\rangle_{\mbox{\scriptsize F}}
|\alpha;\theta\rangle_{\mbox{\scriptsize B}} \nonumber \\
&=&
[U(t)\otimes\tilde{U}(t)]
[
|0(\Theta)\rangle_{\mbox{\scriptsize F}}
\hat{U}_{\mbox{\scriptsize B}}(\theta)
|\alpha\rangle_{\mbox{\scriptsize B}}|\tilde{\alpha}\rangle_{\mbox{\scriptsize B}}
].
\label{state-vector-Psi_I_t-0}
\end{eqnarray}

From Eq.~(\ref{definition-thermal-vacuum-fermion}),
we can rewrite $|0(\Theta)\rangle_{\mbox{\scriptsize F}}$
in the form,
\begin{equation}
|0(\Theta)\rangle_{\mbox{\scriptsize F}}
=
\cos\Theta|0,\tilde{0}\rangle_{\mbox{\scriptsize F}}
+
\sin\Theta|1,\tilde{1}\rangle_{\mbox{\scriptsize F}}.
\label{thermal-vacuum-fermion-2}
\end{equation}
Taking $\{|i,\tilde{j}\rangle_{\mbox{\scriptsize F}}:i,j\in\{1,0\}\}$
for the basis vectors of the four-dimensional Hilbert space
${\cal H}_{\mbox{\scriptsize F}}\otimes\tilde{\cal H}_{\mbox{\scriptsize F}}$,
we can write down $|0(\Theta)\rangle_{\mbox{\scriptsize F}}$
as a four-component vector,
\begin{equation}
|0(\Theta)\rangle_{\mbox{\scriptsize F}}
=
\left(
\begin{array}{c}
\sin\Theta \\
0 \\
0 \\
\cos\Theta
\end{array}
\right),
\label{fermion-thermal-vacuum-4-component-vector-0}
\end{equation}
where
the components of the above vector are arranged
in the order of
$|1,\tilde{1}\rangle_{\mbox{\scriptsize F}}$,
$|1,\tilde{0}\rangle_{\mbox{\scriptsize F}}$,
$|0,\tilde{1}\rangle_{\mbox{\scriptsize F}}$
and
$|0,\tilde{0}\rangle_{\mbox{\scriptsize F}}$.

Thus,
writing $|\Psi_{\mbox{\scriptsize I}}(t)\rangle$
as the four-component vector,
we obtain
\begin{equation}
|\Psi_{\mbox{\scriptsize I}}(t)\rangle
=
[U(t)\otimes\tilde{U}(t)]
\left(
\begin{array}{c}
\sin\Theta
\hat{U}_{\mbox{\scriptsize B}}(\theta)
|\alpha\rangle_{\mbox{\scriptsize B}}|\tilde{\alpha}\rangle_{\mbox{\scriptsize B}} \\
0 \\
0 \\
\cos\Theta
\hat{U}_{\mbox{\scriptsize B}}(\theta)
|\alpha\rangle_{\mbox{\scriptsize B}}|\tilde{\alpha}\rangle_{\mbox{\scriptsize B}}
\end{array}
\right).
\end{equation}
Moreover,
expressing $U(t)\otimes\tilde{U}(t)$
in the form of the $4\times 4$ matrix,
\begin{equation}
U(t)\otimes\tilde{U}(t)
=
\left(
\begin{array}{cc}
u_{11}\tilde{U}(t) & u_{10}\tilde{U}(t) \\
u_{01}\tilde{U}(t) & u_{00}\tilde{U}(t)
\end{array}
\right),
\end{equation}
\begin{equation}
\tilde{U}(t)
=
\left(
\begin{array}{cc}
\tilde{u}_{11} & \tilde{u}_{10} \\
\tilde{u}_{01} & \tilde{u}_{00}
\end{array}
\right),
\label{definition-tilde-U-0}
\end{equation}
where
$\{u_{ij}:i,j\in\{1,0\}\}$
and
$\{\tilde{u}_{ij}:i,j\in\{1,0\}\}$
are given by Eq.~(\ref{unitary-evolution-2}),
we can write down the four-component vector
$|\Psi_{\mbox{\scriptsize I}}(t)\rangle$
as the following explicit form:
\begin{equation}
|\Psi_{\mbox{\scriptsize I}}(t)\rangle
=
\left(
\begin{array}{c}
\psi_{1\tilde{1}} \\
\psi_{1\tilde{0}} \\
\psi_{0\tilde{1}} \\
\psi_{0\tilde{0}}
\end{array}
\right),
\end{equation}
where
\begin{eqnarray}
\psi_{1\tilde{1}}
&=&
(\sin\Theta u_{11}\tilde{u}_{11}+\cos\Theta u_{10}\tilde{u}_{10})
\hat{U}_{\mbox{\scriptsize B}}(\theta)
|\alpha\rangle_{\mbox{\scriptsize B}}|\tilde{\alpha}\rangle_{\mbox{\scriptsize B}}, \nonumber \\
\psi_{1\tilde{0}}
&=&
(\sin\Theta u_{11}\tilde{u}_{01}+\cos\Theta u_{10}\tilde{u}_{00})
\hat{U}_{\mbox{\scriptsize B}}(\theta)
|\alpha\rangle_{\mbox{\scriptsize B}}|\tilde{\alpha}\rangle_{\mbox{\scriptsize B}}, \nonumber \\
\psi_{0\tilde{1}}
&=&
(\sin\Theta u_{01}\tilde{u}_{11}+\cos\Theta u_{00}\tilde{u}_{10})
\hat{U}_{\mbox{\scriptsize B}}(\theta)
|\alpha\rangle_{\mbox{\scriptsize B}}|\tilde{\alpha}\rangle_{\mbox{\scriptsize B}}, \nonumber \\
\psi_{0\tilde{0}}
&=&
(\sin\Theta u_{01}\tilde{u}_{01}+\cos\Theta u_{00}\tilde{u}_{00})
\hat{U}_{\mbox{\scriptsize B}}(\theta)
|\alpha\rangle_{\mbox{\scriptsize B}}|\tilde{\alpha}\rangle_{\mbox{\scriptsize B}}.
\end{eqnarray}

Hence,
the probability that we detect the ground state of the atom at zero temperature
in the state of the total system $|\Psi_{\mbox{\scriptsize I}}(t)\rangle$
is given by
\begin{eqnarray}
P_{g}(\Theta,\theta;t)
&=&
\|{}_{\mbox{\scriptsize F}}\langle 0,\tilde{0}|\Psi_{\mbox{\scriptsize I}}(t)\rangle\|^{2}
+
\|{}_{\mbox{\scriptsize F}}\langle 0,\tilde{1}|\Psi_{\mbox{\scriptsize I}}(t)\rangle\|^{2} \nonumber \\
&=&
\|\psi_{0\tilde{0}}\|^{2}+\|\psi_{0\tilde{1}}\|^{2} \nonumber \\
&=&
{}_{\mbox{\scriptsize B}}\langle\alpha|
{}_{\mbox{\scriptsize B}}\langle\tilde{\alpha}|
\hat{U}_{\mbox{\scriptsize B}}^{\dagger}(\theta)
[
\sin^{2}\Theta u_{01}^{\dagger}u_{01}\tilde{u}_{01}^{\dagger}\tilde{u}_{01} \nonumber \\
&&
+
\sin\Theta\cos\Theta
(u_{01}^{\dagger}u_{00}\tilde{u}_{01}^{\dagger}\tilde{u}_{00}
+
u_{00}^{\dagger}u_{01}\tilde{u}_{00}^{\dagger}\tilde{u}_{01}) \nonumber \\
&&
+
\cos^{2}\Theta u_{00}^{\dagger}u_{00}\tilde{u}_{00}^{\dagger}\tilde{u}_{00} \nonumber \\
&&
+
\sin^{2}\Theta u_{01}^{\dagger}u_{01}\tilde{u}_{11}^{\dagger}\tilde{u}_{11} \nonumber \\
&&
+
\sin\Theta\cos\Theta
(u_{01}^{\dagger}u_{00}\tilde{u}_{11}^{\dagger}\tilde{u}_{10}
+
u_{00}^{\dagger}u_{01}\tilde{u}_{10}^{\dagger}\tilde{u}_{11}) \nonumber \\
&&
+
\cos^{2}\Theta u_{00}^{\dagger}u_{00}\tilde{u}_{10}^{\dagger}\tilde{u}_{10}
]
\hat{U}_{\mbox{\scriptsize B}}(\theta)
|\alpha\rangle_{\mbox{\scriptsize B}}
|\tilde{\alpha}\rangle_{\mbox{\scriptsize B}}.
\label{Probability-atom-ground-state-theta-0}
\end{eqnarray}
Here,
we pay attention to the following fact:
Because $\tilde{U}(t)$ given by Eq.~(\ref{definition-tilde-U-0})
is a unitary matrix,
we obtain
\begin{eqnarray}
\tilde{u}_{11}^{\dagger}\tilde{u}_{11}+\tilde{u}_{01}^{\dagger}\tilde{u}_{01}
&=&
1, \nonumber \\
\tilde{u}_{10}^{\dagger}\tilde{u}_{10}+\tilde{u}_{00}^{\dagger}\tilde{u}_{00}
&=&
1, \nonumber \\
\tilde{u}_{11}^{\dagger}\tilde{u}_{10}+\tilde{u}_{01}^{\dagger}\tilde{u}_{00}
&=&
0, \nonumber \\
\tilde{u}_{10}^{\dagger}\tilde{u}_{11}+\tilde{u}_{00}^{\dagger}\tilde{u}_{01}
&=&
0.
\label{relations-unitary-tilde-U-components}
\end{eqnarray}
Substitution of Eq.~(\ref{relations-unitary-tilde-U-components})
into Eq.~(\ref{Probability-atom-ground-state-theta-0})
yields
\begin{eqnarray}
P_{g}(\Theta,\theta;t)
&=&
{}_{\mbox{\scriptsize B}}\langle\alpha|
{}_{\mbox{\scriptsize B}}\langle\tilde{\alpha}|
\hat{U}_{\mbox{\scriptsize B}}^{\dagger}(\theta)
[
\sin^{2}\Theta u_{01}^{\dagger}u_{01}
+
\cos^{2}\Theta u_{00}^{\dagger}u_{00}
]
\hat{U}_{\mbox{\scriptsize B}}(\theta)
|\alpha\rangle_{\mbox{\scriptsize B}}
|\tilde{\alpha}\rangle_{\mbox{\scriptsize B}} \nonumber \\
&=&
\cos^{2}\Theta
{}_{\mbox{\scriptsize B}}\langle\alpha|
{}_{\mbox{\scriptsize B}}\langle\tilde{\alpha}|
\hat{U}_{\mbox{\scriptsize B}}^{\dagger}(\theta)
g_{1}(a^{\dagger}a+c)
\hat{U}_{\mbox{\scriptsize B}}(\theta)
|\alpha\rangle_{\mbox{\scriptsize B}}
|\tilde{\alpha}\rangle_{\mbox{\scriptsize B}} \nonumber \\
&&
+
\sin^{2}\Theta
{}_{\mbox{\scriptsize B}}\langle\alpha|
{}_{\mbox{\scriptsize B}}\langle\tilde{\alpha}|
\hat{U}_{\mbox{\scriptsize B}}^{\dagger}(\theta)
g_{2}(a^{\dagger}a+c+1)
\hat{U}_{\mbox{\scriptsize B}}(\theta)
|\alpha\rangle_{\mbox{\scriptsize B}}
|\tilde{\alpha}\rangle_{\mbox{\scriptsize B}},
\label{Probability-atom-ground-state-theta-1}
\end{eqnarray}
where
\begin{eqnarray}
g_{1}(a^{\dagger}a+c)
&=&
u_{00}^{\dagger}u_{00} \nonumber \\
&=&
\cos^{2}(\sqrt{a^{\dagger}a+c}|\kappa|t)
+
c
\frac{\sin^{2}(\sqrt{a^{\dagger}a+c}|\kappa|t)}{a^{\dagger}a+c}, \nonumber \\
g_{2}(a^{\dagger}a+c+1)
&=&
u_{01}^{\dagger}u_{01} \nonumber \\
&=&
a
\frac{\sin^{2}(\sqrt{a^{\dagger}a+c}|\kappa|t)}{a^{\dagger}a+c}a^{\dagger} \nonumber \\
&=&
\frac{\sin^{2}(\sqrt{a^{\dagger}a+c+1}|\kappa|t)}{a^{\dagger}a+c+1}[(a^{\dagger}a+c+1)-c].
\label{definition-functions-g-1&2}
\end{eqnarray}

Using Eqs.~(\ref{definition-hat-U-operator})
and
(\ref{definition-hat-G-operator}),
we can rewrite $P_{g}(\Theta,\theta;t)$
given by Eqs.~(\ref{Probability-atom-ground-state-theta-1})
and
(\ref{definition-functions-g-1&2})
as
\begin{eqnarray}
P_{g}(\Theta,\theta;t)
&=&
\cos^{2}\Theta
\langle\alpha|
\langle\tilde{\alpha}|
\exp[\theta(a\tilde{a}-\tilde{a}^{\dagger}a^{\dagger})]
g_{1}(a^{\dagger}a+c)
\exp[-\theta(a\tilde{a}-\tilde{a}^{\dagger}a^{\dagger})]
|\alpha\rangle
|\tilde{\alpha}\rangle \nonumber \\
&&
+
\sin^{2}\Theta
\langle\alpha|
\langle\tilde{\alpha}|
\exp[\theta(a\tilde{a}-\tilde{a}^{\dagger}a^{\dagger})]
g_{2}(a^{\dagger}a+c+1) \nonumber \\
&&\times
\exp[-\theta(a\tilde{a}-\tilde{a}^{\dagger}a^{\dagger})]
|\alpha\rangle
|\tilde{\alpha}\rangle.
\label{Probability-atom-ground-state-theta-2}
\end{eqnarray}
Thus,
we obtain the perturbative expansion of $P_{g}(\Theta,\theta;t)$
in the small parameter $\theta(\beta)$
as
\begin{equation}
P_{g}(\Theta,\theta;t)
=
\sum_{n=0}^{\infty}
\frac{\theta(\beta)^{n}}{n!}
P^{(n)}_{g}(\Theta,\theta;t),
\label{Perturbation-theory-formula-0}
\end{equation}
where
\begin{equation}
P^{(n)}_{g}(\Theta,\theta;t)
=
\cos^{2}\Theta P^{(n)}_{g,1}(t)
+
\sin^{2}\Theta P^{(n)}_{g,2}(t),
\label{Perturbation-theory-formula-1}
\end{equation}
\begin{eqnarray}
P^{(0)}_{g,1}(t)
&=&
\langle\alpha|
\langle\tilde{\alpha}|
g_{1}(a^{\dagger}a+c)
|\alpha\rangle
|\tilde{\alpha}\rangle, \nonumber \\
P^{(0)}_{g,2}(t)
&=&
\langle\alpha|
\langle\tilde{\alpha}|
g_{2}(a^{\dagger}a+c+1)
|\alpha\rangle
|\tilde{\alpha}\rangle, \nonumber \\
P^{(1)}_{g,1}(t)
&=&
\langle\alpha|
\langle\tilde{\alpha}|
[a\tilde{a}-\tilde{a}^{\dagger}a^{\dagger},g_{1}(a^{\dagger}a+c)]
|\alpha\rangle
|\tilde{\alpha}\rangle, \nonumber \\
P^{(1)}_{g,2}(t)
&=&
\langle\alpha|
\langle\tilde{\alpha}|
[a\tilde{a}-\tilde{a}^{\dagger}a^{\dagger},g_{2}(a^{\dagger}a+c+1)]
|\alpha\rangle
|\tilde{\alpha}\rangle, \nonumber \\
P^{(2)}_{g,1}(t)
&=&
\langle\alpha|
\langle\tilde{\alpha}|
[a\tilde{a}-\tilde{a}^{\dagger}a^{\dagger},
[a\tilde{a}-\tilde{a}^{\dagger}a^{\dagger},g_{1}(a^{\dagger}a+c)]]
|\alpha\rangle
|\tilde{\alpha}\rangle, \nonumber \\
P^{(2)}_{g,2}(t)
&=&
\langle\alpha|
\langle\tilde{\alpha}|
[a\tilde{a}-\tilde{a}^{\dagger}a^{\dagger},
[a\tilde{a}-\tilde{a}^{\dagger}a^{\dagger},g_{2}(a^{\dagger}a+c+1)]]
|\alpha\rangle
|\tilde{\alpha}\rangle, \nonumber \\
&&...,
\label{Perturbation-theory-formula-2}
\end{eqnarray}
\begin{eqnarray}
P^{(n)}_{g,1}(t)
&=&
\langle\alpha|
\langle\tilde{\alpha}|
\underbrace{
[a\tilde{a}-\tilde{a}^{\dagger}a^{\dagger},
...,
[a\tilde{a}-\tilde{a}^{\dagger}a^{\dagger},
g_{1}(a^{\dagger}a+c)]...]
}_{\mbox{\scriptsize $n$-fold bracket}}
|\alpha\rangle
|\tilde{\alpha}\rangle, \nonumber \\
P^{(n)}_{g,2}(t)
&=&
\langle\alpha|
\langle\tilde{\alpha}|
\underbrace{
[a\tilde{a}-\tilde{a}^{\dagger}a^{\dagger},
...,
[a\tilde{a}-\tilde{a}^{\dagger}a^{\dagger},
g_{2}(a^{\dagger}a+c+1)]...]
}_{\mbox{\scriptsize $n$-fold bracket}}
|\alpha\rangle
|\tilde{\alpha}\rangle \nonumber \\
&&
\mbox{for $n=1,2,3,...$}.
\label{Perturbation-theory-formula-3}
\end{eqnarray}

Here, we pay attention to the following fact:
The perturbative expansion given by
Eqs.~(\ref{Perturbation-theory-formula-0}),
(\ref{Perturbation-theory-formula-1}),
(\ref{Perturbation-theory-formula-2})
and
(\ref{Perturbation-theory-formula-3})
is a power series in the small parameter $\theta(\beta)$.
On the other hand,
all the terms of the parameter $\Theta(\beta)$ included
in the perturbative expansion,
namely
$\cos^{2}\Theta$ and $\sin^{2}\Theta$,
are expressed as explicit rigorous forms.
Thus,
we can strictly compute the functions of $\Theta(\beta)$ at ease
in Eqs.~(\ref{Perturbation-theory-formula-0})
and (\ref{Perturbation-theory-formula-1}),
so that
we do not need to worry about perturbative corrections of the parameter $\Theta(\beta)$.
In this paper,
we consider the power series in the small parameter $\theta(\beta)$
to be the perturbative expansion under the low-temperature limit.
In contrast,
we do not regard $\Theta(\beta)$ as the parameter for the perturbation.

Furthermore,
the following trick lets
actual computations of correction terms,
that is to say,
$P^{(n)}_{g}(\Theta,\theta;t)$ for $n=1,2,3,...$,
be tractable.
We can write down the functions $g_{1}(x)$ and $g_{2}(x)$
defined in Eq.~(\ref{definition-functions-g-1&2})
as
\begin{eqnarray}
g_{1}(x)
&=&
\cos^{2}(\sqrt{x}|\kappa|t)
+
c
\frac{\sin^{2}(\sqrt{x}|\kappa|t)}{x}, \nonumber \\
g_{2}(x)
&=&
\frac{\sin^{2}(\sqrt{x}|\kappa|t)}{x}(x-c),
\label{definition-function-g-1&2}
\end{eqnarray}
so that we can rewrite each of them as the Taylor series at $x=0$,
\begin{equation}
g_{i}(x)
=
\sum_{m=0}^{\infty}g^{(m)}_{i}x^{m}
\quad
\mbox{for $-\infty<x<+\infty$},
\label{g-function-Taylor-series-0}
\end{equation}
where
\begin{equation}
g^{(m)}_{i}
=
\frac{1}{m!}\frac{d^{m}}{dx^{m}}g_{i}(x)
\bigg|_{x=0}
\quad
\mbox{for $i\in\{1,2\}$}.
\label{g-function-Taylor-series-1}
\end{equation}
Thus,
we can rewrite Eq.~(\ref{Perturbation-theory-formula-3}) as
\begin{eqnarray}
P^{(n)}_{g,1}(t)
&=&
\sum_{m=0}^{\infty}
g^{(m)}_{1}
\langle\alpha|
\langle\tilde{\alpha}|
\underbrace{
[a\tilde{a}-\tilde{a}^{\dagger}a^{\dagger},
...,
[a\tilde{a}-\tilde{a}^{\dagger}a^{\dagger},
(a^{\dagger}a+c)^{m}]...]
}_{\mbox{\scriptsize $n$-fold bracket}}
|\alpha\rangle
|\tilde{\alpha}\rangle, \nonumber \\
P^{(n)}_{g,2}(t)
&=&
\sum_{m=0}^{\infty}
g^{(m)}_{2}
\langle\alpha|
\langle\tilde{\alpha}|
\underbrace{
[a\tilde{a}-\tilde{a}^{\dagger}a^{\dagger},
...,
[a\tilde{a}-\tilde{a}^{\dagger}a^{\dagger},
(a^{\dagger}a+c+1)^{m}]...]
}_{\mbox{\scriptsize $n$-fold bracket}}
|\alpha\rangle
|\tilde{\alpha}\rangle \nonumber \\
&&\mbox{for $n=1,2,3,...$.}
\label{Perturbation-theory-formula-4}
\end{eqnarray}

In Secs.~\ref{section-0th-order-correction-term},
\ref{section-1st-order-correction-term},
\ref{section-2nd-order-correction-term}
and
Appendix~\ref{section-3rd-order-correction-term},
using the perturbative expansion given by
Eqs.~(\ref{Perturbation-theory-formula-0}),
(\ref{Perturbation-theory-formula-1}),
(\ref{Perturbation-theory-formula-2}),
(\ref{Perturbation-theory-formula-3}),
(\ref{definition-function-g-1&2}),
(\ref{g-function-Taylor-series-0}),
(\ref{g-function-Taylor-series-1})
and
(\ref{Perturbation-theory-formula-4}),
we compute
$P_{g}(\Theta,\theta;t)$.

\section{\label{section-comparison-of-TFD-and-Liouville-von-Neumann-equation}
Comparison of the TFD formalism and the
\\
Liouville-von Neumann equation}
In the previous sections,
we discuss a method for examining the time-evolution
caused by the Hamiltonian $\hat{H}$
defined in Eqs.~(\ref{extended-JCM-Hamiltonian-0}) and (\ref{extended-JCM-Hamiltonian-1})
with the initial state
$|\Psi(0)\rangle
=|0(\Theta)\rangle_{\mbox{\scriptsize F}}|\alpha;\theta\rangle_{\mbox{\scriptsize B}}$
given by Eq.~(\ref{extended-JCM-initial-state})
according to the TFD formalism.
This method is equivalent to
solving the following Liouville-von Neumann equation:
\begin{equation}
\frac{\partial}{\partial t}
\rho(t)
=
-\frac{i}{\hbar}
[H,\rho(t)],
\label{Liouville-von-Neumann-equation-0}
\end{equation}
where
\begin{eqnarray}
\rho(0)
&=&
\mbox{Tr}_{\tilde{\cal H}}
[
|\Psi(0)\rangle\langle\Psi(0)|
] \nonumber \\
&=&
\mbox{Tr}_{\tilde{\cal H}}
[
|0(\Theta)\rangle_{\mbox{\scriptsize F}}{}_{\mbox{\scriptsize F}}\langle 0(\Theta)|
\otimes
|\alpha;\theta\rangle_{\mbox{\scriptsize B}}{}_{\mbox{\scriptsize B}}\langle\alpha;\theta|
],
\label{initial-density-operator-0}
\end{eqnarray}
and the Hamiltonian $H$ appearing in Eq.~(\ref{Liouville-von-Neumann-equation-0})
is given by Eqs.~(\ref{JCM-Hamiltonian-0}) and (\ref{JCM-Hamiltonian-1}).

Both the Hamiltonian $\hat{H}$ based on the TFD formalism
defined by Eqs.~(\ref{extended-JCM-Hamiltonian-0}) and (\ref{extended-JCM-Hamiltonian-1})
and
the Liouville-von Neumann equation
given by Eq.~(\ref{Liouville-von-Neumann-equation-0})
represent that
the total system evolves in time
with maintaining the constant temperature,
so that it never suffers from dissipation
and
its time-evolution is reversible.
Thus, we understand
that
the Hermitian operator corresponding to the energy of the total system
is equivalent
to the generator of the unitary operator for the time-evolution.

Here,
thinking about the Liouville-von Neumann equation given by Eq.~(\ref{Liouville-von-Neumann-equation-0}),
we divide $H$ into the two parts $C_{1}$ and $C_{2}$
as shown in Eq.~(\ref{JCM-Hamiltonian-decomposition-0})
and take the interaction picture.
Assuming $\rho_{\mbox{\scriptsize I}}(0)=\rho(0)$,
we introduce the density operator described in the interaction picture
as
\begin{equation}
\rho_{\mbox{\scriptsize I}}(t)
=
e^{iC_{1}t}\rho(t)e^{-iC_{1}t}.
\end{equation}
Then,
using the commutation relation $[C_{1},C_{2}]=0$,
we obtain
\begin{equation}
\frac{\partial}{\partial t}
\rho_{\mbox{\scriptsize I}}(t)
=
-i
[C_{2},\rho_{\mbox{\scriptsize I}}(t)].
\label{Liouville-von-Neumann-equation-1}
\end{equation}

From Eq.~(\ref{Liouville-von-Neumann-equation-1}),
we notice that we can rewrite
$\rho_{\mbox{\scriptsize I}}(t)$
as
\begin{eqnarray}
\rho_{\mbox{\scriptsize I}}(t)
&=&
e^{-iC_{2}t}\rho_{\mbox{\scriptsize I}}(0)e^{iC_{2}t} \nonumber \\
&=&
\mbox{Tr}_{\tilde{\cal H}}
\Bigl[
e^{-iC_{2}t}
[
|0(\Theta)\rangle_{\mbox{\scriptsize F}}{}_{\mbox{\scriptsize F}}\langle 0(\Theta)|
\otimes
|\alpha;\theta\rangle_{\mbox{\scriptsize B}}{}_{\mbox{\scriptsize B}}\langle\alpha;\theta|
]e^{iC_{2}t}
\Bigr].
\label{Liouville-von-Neumann-equation-2}
\end{eqnarray}
Moreover,
the probability that
we detect the ground state of the atom at zero temperature
is given by
\begin{equation}
P_{g}(\Theta,\theta;t)
=
{}_{\mbox{\scriptsize F}}\langle 0|
\mbox{Tr}_{\mbox{\scriptsize B}}[
\rho_{\mbox{\scriptsize I}}(t)
]
|0\rangle_{\mbox{\scriptsize F}}.
\label{Liouville-von-Neumann-equation-3}
\end{equation}
The physical meaning of Eqs.~(\ref{Liouville-von-Neumann-equation-2})
and (\ref{Liouville-von-Neumann-equation-3})
is equivalent to
the discussion developed
from Eq.~(\ref{extended-JCM-initial-state})
until Eq.~(\ref{Probability-atom-ground-state-theta-2})
in Sec.~\ref{section-formulation-perturbetion-theory}.

Thus,
comparing Eqs.~(\ref{Probability-atom-ground-state-theta-2})
and
(\ref{Liouville-von-Neumann-equation-3}),
we cannot find distinct differences
between the TFD formalism and the Liouville-von Neumann equation.
However,
if we take the TFD formalism,
we can express a physical quantity
as a power series in $\theta(\beta)$
such as Eqs.~(\ref{Perturbation-theory-formula-0}),
(\ref{Perturbation-theory-formula-1}),
(\ref{Perturbation-theory-formula-2})
and
(\ref{Perturbation-theory-formula-3}).
Because of this advantage,
the TFD formalism is superior than the Liouville-von Neumann equation
for computing physical quantities actually.
The reason why we take the TFD formalism in this paper
for describing the JCM at finite temperature is the fact mentioned above.
And this prescription is a new key point of this paper as compared with the other past works.

In fact,
if we rewrite Eqs.~(\ref{Liouville-von-Neumann-equation-2})
and
(\ref{Liouville-von-Neumann-equation-3})
as a low-temperature expansion without using the TFD formalism,
we have to carry out the following calculations:
\begin{eqnarray}
&&
\mbox{Tr}_{\tilde{\cal H}}
[|\alpha;\theta\rangle_{\mbox{\scriptsize B}}{}_{\mbox{\scriptsize B}}\langle\alpha;\theta|] \nonumber \\
&=&
\mbox{Tr}_{\tilde{\cal H}}
[
\hat{U}_{\mbox{\scriptsize B}}(\theta)
|\alpha\rangle_{\mbox{\scriptsize B}}
|\tilde{\alpha}\rangle_{\mbox{\scriptsize B}}
{}_{\mbox{\scriptsize B}}\langle\alpha|
{}_{\mbox{\scriptsize B}}\langle\tilde{\alpha}|
\hat{U}_{\mbox{\scriptsize B}}^{\dagger}(\theta)
] \nonumber \\
&=&
\mbox{Tr}_{\tilde{\cal H}}
\Bigl(
\exp[-\theta(a\tilde{a}-\tilde{a}^{\dagger}a^{\dagger})]
|\alpha\rangle_{\mbox{\scriptsize B}}
|\tilde{\alpha}\rangle_{\mbox{\scriptsize B}}
{}_{\mbox{\scriptsize B}}\langle\alpha|
{}_{\mbox{\scriptsize B}}\langle\tilde{\alpha}|
\exp[\theta(a\tilde{a}-\tilde{a}^{\dagger}a^{\dagger})]
\Bigr) \nonumber \\
&=&
\mbox{Tr}_{\tilde{\cal H}}
\Bigl(
|\alpha\rangle_{\mbox{\scriptsize B}}
|\tilde{\alpha}\rangle_{\mbox{\scriptsize B}}
{}_{\mbox{\scriptsize B}}\langle\alpha|
{}_{\mbox{\scriptsize B}}\langle\tilde{\alpha}|
-\theta
[a\tilde{a}-\tilde{a}^{\dagger}a^{\dagger},
|\alpha\rangle_{\mbox{\scriptsize B}}
|\tilde{\alpha}\rangle_{\mbox{\scriptsize B}}
{}_{\mbox{\scriptsize B}}\langle\alpha|
{}_{\mbox{\scriptsize B}}\langle\tilde{\alpha}|] \nonumber \\
&&
+\frac{\theta^{2}}{2!}
[a\tilde{a}-\tilde{a}^{\dagger}a^{\dagger},
[a\tilde{a}-\tilde{a}^{\dagger}a^{\dagger},
|\alpha\rangle_{\mbox{\scriptsize B}}
|\tilde{\alpha}\rangle_{\mbox{\scriptsize B}}
{}_{\mbox{\scriptsize B}}\langle\alpha|
{}_{\mbox{\scriptsize B}}\langle\tilde{\alpha}|]]
+...
\Bigr).
\end{eqnarray}
The above calculations are essentially equivalent to
Eqs.~(\ref{Perturbation-theory-formula-0}),
(\ref{Perturbation-theory-formula-1}),
(\ref{Perturbation-theory-formula-2})
and
(\ref{Perturbation-theory-formula-3}).
However,
the perturbation theory via
the TFD formalism provides us
a clearer insight and a more accurate understanding
than the Liouville-von Neumann equation does.

\section{\label{section-0th-order-correction-term}The zero-th order correction}
From Eqs.~(\ref{definition-functions-g-1&2}),
(\ref{Perturbation-theory-formula-1})
and
(\ref{Perturbation-theory-formula-2}),
we can write down the zero-th order correction as
\begin{equation}
P^{(0)}_{g}(\Theta,\theta;t)
=
\cos^{2}\Theta P^{(0)}_{g,1}(t)
+
\sin^{2}\Theta P^{(0)}_{g,2}(t),
\label{0th-order-perturbation-0}
\end{equation}
\begin{eqnarray}
P^{(0)}_{g,1}(t)
&=&
\langle\alpha|
g_{1}(a^{\dagger}a+c)
|\alpha\rangle \nonumber \\
&=&
e^{-\alpha^{2}}
\sum_{n=0}^{\infty}
\frac{\alpha^{2n}}{n!}
g_{1}(n+c) \nonumber \\
&=&
e^{-\alpha^{2}}
\sum_{n=0}^{\infty}
\frac{\alpha^{2n}}{n!}
[
\cos^{2}(\sqrt{n+c}|\kappa|t)
+
c
\frac{\sin^{2}(\sqrt{n+c}|\kappa|t)}{n+c}
], \nonumber \\
P^{(0)}_{g,2}(t)
&=&
\langle\alpha|
g_{2}(a^{\dagger}a+c+1)
|\alpha\rangle \nonumber \\
&=&
e^{-\alpha^{2}}
\sum_{n=0}^{\infty}
\frac{\alpha^{2n}}{n!}
g_{2}(n+c+1) \nonumber \\
&=&
e^{-\alpha^{2}}
\sum_{n=0}^{\infty}
\frac{\alpha^{2n}}{n!}
\frac{\sin^{2}(\sqrt{n+c+1}|\kappa|t)}{n+c+1}(n+1).
\label{0th-order-perturbation-term}
\end{eqnarray}
Referring to Eq.~(\ref{Probability-collapse-and-revival-Rabi-oscillations-0}),
we note that
$P^{(0)}_{g,1}(t)=P_{g}(t)$
and
$P^{(0)}_{g}(0,0;t)=P_{g}(t)$.

For the convenience of calculations
carried out in the remains of this paper,
we define the following functions,
each of which is represented as an infinite series:
\begin{eqnarray}
Q_{1}^{(l)}(t)
&=&
e^{-\alpha^{2}}
\sum_{n=0}^{\infty}
\frac{\alpha^{2n}}{n!}
g_{1}(n+c+l) \nonumber \\
&=&
e^{-\alpha^{2}}
\sum_{n=0}^{\infty}
\frac{\alpha^{2n}}{n!}
[
\cos^{2}(\sqrt{n+c+l}|\kappa|t)
+
c
\frac{\sin^{2}(\sqrt{n+c+l}|\kappa|t)}{n+c+l}
], \nonumber \\
Q_{2}^{(l)}(t)
&=&
e^{-\alpha^{2}}
\sum_{n=0}^{\infty}
\frac{\alpha^{2n}}{n!}
g_{2}(n+c+1+l) \nonumber \\
&=&
e^{-\alpha^{2}}
\sum_{n=0}^{\infty}
\frac{\alpha^{2n}}{n!}
\frac{\sin^{2}(\sqrt{n+c+1+l}|\kappa|t)}{n+c+1+l}(n+1+l).
\end{eqnarray}
From the above definitions,
we obtain the zero-th order correction terms as
\begin{eqnarray}
P^{(0)}_{g,1}(t)
&=&
Q_{1}^{(0)}(t), \nonumber \\
P^{(0)}_{g,2}(t)
&=&
Q_{2}^{(0)}(t).
\label{0th-order-perturbation-term-1}
\end{eqnarray}

\section{\label{section-1st-order-correction-term}The first order correction}
From Eqs.~(\ref{Perturbation-theory-formula-1})
and
(\ref{Perturbation-theory-formula-4}),
we can write down the first order correction as
\begin{equation}
P^{(1)}_{g}(\Theta,\theta;t)
=
\cos^{2}\Theta P^{(1)}_{g,1}(t)
+
\sin^{2}\Theta P^{(1)}_{g,2}(t),
\label{1st-order-perturbation-0}
\end{equation}
\begin{eqnarray}
P^{(1)}_{g,1}(t)
&=&
\sum_{n=0}^{\infty}
g^{(n)}_{1}
\langle\alpha|
\langle\tilde{\alpha}|
[a\tilde{a}-a^{\dagger}\tilde{a}^{\dagger},
(a^{\dagger}a+c)^{n}]
|\alpha\rangle
|\tilde{\alpha}\rangle, \nonumber \\
P^{(1)}_{g,2}(t)
&=&
\sum_{n=0}^{\infty}
g^{(n)}_{2}
\langle\alpha|
\langle\tilde{\alpha}|
[a\tilde{a}-a^{\dagger}\tilde{a}^{\dagger},
(a^{\dagger}a+c+1)^{n}]
|\alpha\rangle
|\tilde{\alpha}\rangle.
\label{first-order-perturbation-terms-0}
\end{eqnarray}
From Eq.~(\ref{first-order-perturbation-terms-0}),
we notice that we have to calculate the commutation relations,
\begin{equation}
[a\tilde{a}-a^{\dagger}\tilde{a}^{\dagger},
(a^{\dagger}a+c)^{n}]
\quad\mbox{for $n=0,1,2,...$}.
\end{equation}

At first,
we define the following three operators:
\begin{equation}
\hat{A}
=
a^{\dagger}\tilde{a}^{\dagger}-a\tilde{a},
\quad
\hat{B}
=
a^{\dagger}a+c,
\quad
\hat{C}
=
a^{\dagger}\tilde{a}^{\dagger}+a\tilde{a}.
\label{definition-hatA-hatB-hatC-1}
\end{equation}
[We pay attention to the fact that
the operator $\hat{C}$ defined in Eq.~(\ref{definition-hatA-hatB-hatC-1})
is different from
$C_{1}$, $C_{2}$, $\hat{C}_{1}$ and $\hat{C}_{2}$
given by Eqs.~(\ref{JCM-Hamiltonian-decomposition-0})
and
(\ref{decomposition-hat-Hamiltonian}).]
Then,
we obtain the commutation relations,
\begin{eqnarray}
[\hat{A},\hat{B}]
&=&
-\hat{C}, \nonumber \\
{[}\hat{A},\hat{B}^{2}{]}
&=&
\hat{A}-2\hat{B}\hat{C}, \nonumber \\
{[}\hat{A},\hat{B}^{3}{]}
&=&
-\hat{C}+3\hat{B}\hat{A}-3\hat{B}^{2}\hat{C}, \nonumber \\
{[}\hat{A},\hat{B}^{4}{]}
&=&
\hat{A}-4\hat{B}\hat{C}+6\hat{B}^{2}\hat{A}-4\hat{B}^{3}\hat{C}, \nonumber \\
&&....
\label{commutation-relations-A-Bn-1}
\end{eqnarray}

Next,
we define the following two operators:
\begin{equation}
\hat{\mu}=a^{\dagger}\tilde{a}^{\dagger},
\quad
\hat{\nu}=a\tilde{a},
\label{definition-operators-mu-nu-1}
\end{equation}
and
we obtain
\begin{equation}
\hat{A}=\hat{\mu}-\hat{\nu},
\quad
\hat{C}=\hat{\mu}+\hat{\nu}.
\label{definition-operators-mu-nu-2}
\end{equation}
Using the operators $\hat{\mu}$ and $\hat{\nu}$,
we can rewrite Eq.~(\ref{commutation-relations-A-Bn-1})
as the general form,
\begin{equation}
[\hat{A},\hat{B}^{n}]
=
(\hat{B}-1)^{n}\hat{\mu}-(\hat{B}+1)^{n}\hat{\nu}-\hat{B}^{n}\hat{A}
\quad\mbox{for $n=1,2,3,...$}.
\label{commutation-relations-A-Bn-3}
\end{equation}
We can prove Eq.~(\ref{commutation-relations-A-Bn-3})
with the mathematical induction as follows:
First, we can confirm that Eq.~(\ref{commutation-relations-A-Bn-3}) holds for $n=1$, at ease.
Second, we assume Eq.~(\ref{commutation-relations-A-Bn-3}) holds for some unspecified number $n(\geq 1)$.
Third, we compute the commutation relation,
\begin{eqnarray}
[\hat{A},\hat{B}^{n+1}]
&=&
[\hat{A},\hat{B}^{n}]\hat{B}+\hat{B}^{n}[\hat{A},\hat{B}] \nonumber \\
&=&
\Bigl((\hat{B}-1)^{n}\hat{\mu}-(\hat{B}+1)^{n}\hat{\nu}-\hat{B}^{n}\hat{A}\Bigr)\hat{B}
+\hat{B}^{n}[\hat{A},\hat{B}] \nonumber \\
&=&
(\hat{B}-1)^{n+1}\hat{\mu}-(\hat{B}+1)^{n+1}\hat{\nu}-\hat{B}^{n+1}\hat{A},
\end{eqnarray}
where we use $\hat{\mu}\hat{B}=(\hat{B}-1)\hat{\mu}$ and $\hat{\nu}\hat{B}=(\hat{B}+1)\hat{\nu}$.

Thus,
we obtain
\begin{eqnarray}
&&
{[}
a\tilde{a}-a^{\dagger}\tilde{a}^{\dagger}, (a^{\dagger}a+c)^{n}
{]} \nonumber \\
&=&
-(a^{\dagger}a+c-1)^{n}\tilde{a}^{\dagger}a^{\dagger}
+(a^{\dagger}a+c+1)^{n}a\tilde{a}
-(a^{\dagger}a+c)^{n}(a\tilde{a}-a^{\dagger}\tilde{a}^{\dagger}) \nonumber \\
&=&
-a^{\dagger}\tilde{a}^{\dagger}(a^{\dagger}a+c)^{n}
+(a^{\dagger}a+c+1)^{n}a\tilde{a}
-(a^{\dagger}a+c)^{n}a\tilde{a} \nonumber \\
&&
+a^{\dagger}\tilde{a}^{\dagger}(a^{\dagger}a+c+1)^{n}.
\label{first-order-perturbation-commutation-relations-a}
\end{eqnarray}
Moreover,
replacing $c$ in Eq.~(\ref{first-order-perturbation-commutation-relations-a})
with $(c+1)$,
we obtain the commutation relation,
\begin{eqnarray}
&&
{[}
a\tilde{a}-a^{\dagger}\tilde{a}^{\dagger}, (a^{\dagger}a+c+1)^{n}
{]} \nonumber \\
&=&
-a^{\dagger}\tilde{a}^{\dagger}(a^{\dagger}a+c+1)^{n}
+(a^{\dagger}a+c+2)^{n}a\tilde{a} \nonumber \\
&&
-(a^{\dagger}a+c+1)^{n}a\tilde{a}
+a^{\dagger}\tilde{a}^{\dagger}(a^{\dagger}a+c+2)^{n}.
\label{first-order-perturbation-commutation-relations-b}
\end{eqnarray}

Hence,
using the relations $a|\alpha\rangle=\alpha|\alpha\rangle$
and $\tilde{a}|\tilde{\alpha}\rangle=\alpha|\tilde{\alpha}\rangle$,
substitution of Eq.~(\ref{first-order-perturbation-commutation-relations-a})
into Eq.~(\ref{first-order-perturbation-terms-0})
yields
\begin{eqnarray}
P^{(1)}_{g,1}(t)
&=&
\sum_{n=0}^{\infty}
g^{(n)}_{1}
\langle\alpha|
\langle\tilde{\alpha}|
[
-a^{\dagger}\tilde{a}^{\dagger}(a^{\dagger}a+c)^{n}
+(a^{\dagger}a+c+1)^{n}a\tilde{a} \nonumber \\
&&
-(a^{\dagger}a+c)^{n}a\tilde{a}
+a^{\dagger}\tilde{a}^{\dagger}(a^{\dagger}a+c+1)^{n}
]
|\alpha\rangle
|\tilde{\alpha}\rangle, \nonumber \\
&=&
\langle\alpha|
\langle\tilde{\alpha}|
[
-a^{\dagger}\tilde{a}^{\dagger}g_{1}(a^{\dagger}a+c)
+g_{1}(a^{\dagger}a+c+1)a\tilde{a} \nonumber \\
&&
-g_{1}(a^{\dagger}a+c)a\tilde{a}
+a^{\dagger}\tilde{a}^{\dagger}g_{1}(a^{\dagger}a+c+1)
]
|\alpha\rangle
|\tilde{\alpha}\rangle, \nonumber \\
&=&
-2\alpha^{2}
\langle\alpha|
\langle\tilde{\alpha}|
[g_{1}(a^{\dagger}a+c)-g_{1}(a^{\dagger}a+c+1)]
|\alpha\rangle
|\tilde{\alpha}\rangle \nonumber \\
&=&
-2\alpha^{2}
e^{-\alpha^{2}}\sum_{n=0}^{\infty}
\frac{\alpha^{2n}}{n!}[g_{1}(n+c)-g_{1}(n+c+1)] \nonumber \\
&=&
-2\alpha^{2}[Q_{1}^{(0)}(t)-Q_{1}^{(1)}(t)].
\label{1st-order-perturbation-term-g1}
\end{eqnarray}
Similarly,
substitution of Eq.~(\ref{first-order-perturbation-commutation-relations-b})
into
Eq.~(\ref{first-order-perturbation-terms-0})
yields
\begin{eqnarray}
P^{(1)}_{g,2}(t)
&=&
-2\alpha^{2}
e^{-\alpha^{2}}\sum_{n=0}^{\infty}
\frac{\alpha^{2n}}{n!}[g_{2}(n+c+1)-g_{2}(n+c+2)] \nonumber \\
&=&
-2\alpha^{2}[Q_{2}^{(0)}(t)-Q_{2}^{(1)}(t)].
\label{1st-order-perturbation-term-g2}
\end{eqnarray}

\section{\label{section-2nd-order-correction-term}The second order correction}
From Eq.~(\ref{Perturbation-theory-formula-4}),
we can write down the second order terms as
\begin{eqnarray}
P^{(2)}_{g,1}(t)
&=&
\sum_{n=0}^{\infty}
g^{(n)}_{1}
\langle\alpha|
\langle\tilde{\alpha}|
[a\tilde{a}-a^{\dagger}\tilde{a}^{\dagger},
[a\tilde{a}-a^{\dagger}\tilde{a}^{\dagger},
(a^{\dagger}a+c)^{n}]]
|\alpha\rangle
|\tilde{\alpha}\rangle, \nonumber \\
P^{(2)}_{g,2}(t)
&=&
\sum_{n=0}^{\infty}
g^{(n)}_{2}
\langle\alpha|
\langle\tilde{\alpha}|
[a\tilde{a}-a^{\dagger}\tilde{a}^{\dagger},
[a\tilde{a}-a^{\dagger}\tilde{a}^{\dagger},
(a^{\dagger}a+c+1)^{n}]]
|\alpha\rangle
|\tilde{\alpha}\rangle.
\label{second-order-perturbation-terms-0}
\end{eqnarray}
Looking at Eqs.~(\ref{definition-hatA-hatB-hatC-1})
and
(\ref{second-order-perturbation-terms-0}),
we notice that we need to calculate the commutation relation
$[\hat{A},[\hat{A},\hat{B}^{n}]]$.

From now on,
referring to Eq.~(\ref{commutation-relations-A-Bn-3}),
we divide the commutation relation
$[\hat{A},[\hat{A},\hat{B}^{n}]]$
into the following two parts
and examine each of them:
\begin{equation}
[\hat{A},[\hat{A},\hat{B}^{n}]]
=
\hat{R}_{n}+\hat{S}_{n}
\quad\mbox{for $n=1,2,3,...$},
\label{commutation-relation-Rn-Sn}
\end{equation}
where
\begin{eqnarray}
\hat{R}_{n}
&=&
[\hat{A},(\hat{B}-1)^{n}]\hat{\mu}
-
[\hat{A},(\hat{B}+1)^{n}]\hat{\nu}
-
[\hat{A},\hat{B}^{n}]\hat{A}, \nonumber \\
\hat{S}_{n}
&=&
(\hat{B}-1)^{n}[\hat{A},\hat{\mu}]
-
(\hat{B}+1)^{n}[\hat{A},\hat{\nu}].
\label{defintion-Rn-Sn-i}
\end{eqnarray}
According to Eq.~(\ref{commutation-relation-Rn-Sn}),
we divide one of the second order terms given by Eq.~(\ref{second-order-perturbation-terms-0})
into two parts as
\begin{equation}
P^{(2)}_{g,1}(t)
=
\sum_{n=0}^{\infty}
g^{(n)}_{1}
\langle \alpha|\langle\tilde{\alpha}|
\hat{R}_{n}
|\alpha\rangle|\tilde{\alpha}\rangle
+
\sum_{n=0}^{\infty}
g^{(n)}_{1}
\langle \alpha|\langle\tilde{\alpha}|
\hat{S}_{n}
|\alpha\rangle|\tilde{\alpha}\rangle.
\label{second-order-correction-i}
\end{equation}
Here,
we show $P^{(2)}_{g,1}(t)$
as a concrete example in Eq.~(\ref{second-order-correction-i}).
We understand obviously
that we can compute $P^{(2)}_{g,2}(t)$
after the manner of $P^{(2)}_{g,1}(t)$.
Thus,
for simplicity,
we concentrate on evaluating $P^{(2)}_{g,1}(t)$ in the following paragraphs.

At first,
we examine the part which includes $\{\hat{R}_{n}\}$ in Eq.~(\ref{second-order-correction-i}).
From Eq.~(\ref{defintion-Rn-Sn-i}),
we obtain
\begin{eqnarray}
\hat{R}_{0}
&=&
0, \nonumber \\
\hat{R}_{1}
&=&
0, \nonumber \\
\hat{R}_{2}
&=&
-2[\hat{A},\hat{B}]\hat{C}, \nonumber \\
\hat{R}_{3}
&=&
-3[\hat{A},\hat{B}^{2}]\hat{C}
+3[\hat{A},\hat{B}]\hat{A}, \nonumber \\
\hat{R}_{4}
&=&
-4[\hat{A},\hat{B}^{3}]\hat{C}
+6[\hat{A},\hat{B}^{2}]\hat{A}
-4[\hat{A},\hat{B}]\hat{C}, \nonumber \\
\hat{R}_{5}
&=&
-5[\hat{A},\hat{B}^{4}]\hat{C}
+10[\hat{A},\hat{B}^{3}]\hat{A}
-10[\hat{A},\hat{B}^{2}]\hat{C}
+5[\hat{A},\hat{B}]\hat{A}, \nonumber \\
\hat{R}_{6}
&=&
-6[\hat{A},\hat{B}^{5}]\hat{C}
+15[\hat{A},\hat{B}^{4}]\hat{A}
-20[\hat{A},\hat{B}^{3}]\hat{C}
+15[\hat{A},\hat{B}^{2}]\hat{A}
-6[\hat{A},\hat{B}]\hat{C}, \nonumber \\
&&....
\label{defintion-Rn-Sn-ii}
\end{eqnarray}
Thus,
using Eq.~(\ref{commutation-relations-A-Bn-3}) and the following formula:
\begin{eqnarray}
\sum_{n=0}^{\infty}
\left(
\begin{array}{c}
n+m+1 \\
m
\end{array}
\right)
g^{(n+m+1)}_{1}
x^{n+1}
&=&
\frac{1}{m!}\frac{d^{m}}{dx^{m}}g_{1}(x)-g^{(m)}_{1} \nonumber \\
&&
\quad
\mbox{for $m=1,2,3,...$},
\label{formula-g-polynomial-1}
\end{eqnarray}
where $g^{(m)}_{1}$ is defined
in Eqs.~(\ref{g-function-Taylor-series-0}) and (\ref{g-function-Taylor-series-1}),
we can rewrite
the part including $\{\hat{R}_{n}\}$
in the second order term of Eq.~(\ref{second-order-correction-i})
as
\begin{eqnarray}
&&
\langle \alpha|\langle\tilde{\alpha}|
\Bigr(
-
\sum_{n=0}^{\infty}
\left(
\begin{array}{c}
n+2 \\
1
\end{array}
\right)
g^{(n+2)}_{1}[\hat{A},\hat{B}^{n+1}]\hat{C} \nonumber \\
&&
+
\sum_{n=0}^{\infty}
\left(
\begin{array}{c}
n+3 \\
2
\end{array}
\right)
g^{(n+3)}_{1}[\hat{A},\hat{B}^{n+1}]\hat{A} \nonumber \\
&&
-
\sum_{n=0}^{\infty}
\left(
\begin{array}{c}
n+4 \\
3
\end{array}
\right)
g^{(n+4)}_{1}[\hat{A},\hat{B}^{n+1}]\hat{C}
+
...
\Bigr)
|\alpha\rangle|\tilde{\alpha}\rangle \nonumber \\
&=&
\langle \alpha|\langle\tilde{\alpha}|
\Bigr(
-\hat{F}_{1}\hat{C}
+\frac{1}{2!}\hat{F}_{2}\hat{A}
-\frac{1}{3!}\hat{F}_{3}\hat{C}
+...
\Bigr)
|\alpha\rangle|\tilde{\alpha}\rangle \nonumber \\
&=&
\langle \alpha|\langle\tilde{\alpha}|
\Bigr(
\sum_{n=0}^{\infty}
\frac{(-1)^{n}}{n!}\hat{F}_{n}\hat{\mu}
-
\sum_{n=0}^{\infty}
\frac{1}{n!}\hat{F}_{n}\hat{\nu}
\Bigr)
|\alpha\rangle|\tilde{\alpha}\rangle \nonumber \\
&&
-
\langle \alpha|\langle\tilde{\alpha}|
\Bigr(
g_{1}(\hat{B}-1)\hat{\mu}-g_{1}(\hat{B}+1)\hat{\nu}-g_{1}(\hat{B})\hat{A}
\Bigr)\hat{A}
|\alpha\rangle|\tilde{\alpha}\rangle,
\label{second-order-correction-R-part-i}
\end{eqnarray}
where
\begin{equation}
\hat{F}_{n}
=
\frac{d^{n}}{dx^{n}}g_{1}(x)\bigg|_{x=\hat{B}-1}\hat{\mu}
-
\frac{d^{n}}{dx^{n}}g_{1}(x)\bigg|_{x=\hat{B}+1}\hat{\nu}
-
\frac{d^{n}}{dx^{n}}g_{1}(x)\bigg|_{x=\hat{B}}\hat{A}.
\label{definition-operator-Fn}
\end{equation}
In the derivation of Eq.~(\ref{second-order-correction-R-part-i}),
we use Eq.~(\ref{commutation-relations-A-Bn-3})
in an effective manner.
The form of $\hat{F}_{n}$ in Eq.~(\ref{definition-operator-Fn})
reflects Eq.~(\ref{commutation-relations-A-Bn-3}).

Using the operators $e^{\pm d/dx}$,
we can rewrite the first term of the right-hand side of Eq.~(\ref{second-order-correction-R-part-i})
as
\begin{eqnarray}
&&
\langle \alpha|\langle\tilde{\alpha}|
\Bigr(
[
e^{-d/dx}g_{1}(\hat{B}-1)\hat{\mu}
-
e^{-d/dx}g_{1}(\hat{B}+1)\hat{\nu}
-
e^{-d/dx}g_{1}(\hat{B})\hat{A}
]\hat{\mu} \nonumber \\
&&
-
[
e^{d/dx}g_{1}(\hat{B}-1)\hat{\mu}
-
e^{d/dx}g_{1}(\hat{B}+1)\hat{\nu}
-
e^{d/dx}g_{1}(\hat{B})\hat{A}
]\hat{\nu}
\Bigr)
|\alpha\rangle|\tilde{\alpha}\rangle.
\label{second-order-correction-R-part-ii}
\end{eqnarray}
Then,
we apply the following technique to Eq.~(\ref{second-order-correction-R-part-ii}):
\begin{equation}
e^{\pm d/dx}g_{1}(\hat{X})
=
\sum_{n=0}^{\infty}
\frac{(\pm 1)^{n}}{n!}
\frac{d^{n}}{dx^{n}}g_{1}(x)\bigg|_{x=\hat{X}}
=
g_{1}(\hat{X}\pm 1),
\end{equation}
where $\hat{X}$ is an arbitrary operator.
Thus,
we can rewrite Eq.~(\ref{second-order-correction-R-part-ii})
as
\begin{eqnarray}
&&
\langle \alpha|\langle\tilde{\alpha}|
\Bigr(
[
g_{1}(\hat{B}-2)\hat{\mu}
-
g_{1}(\hat{B})\hat{\nu}
-
g_{1}(\hat{B}-1)\hat{A}
]\hat{\mu} \nonumber \\
&&
-
[
g_{1}(\hat{B})\hat{\mu}
-
g_{1}(\hat{B}+2)\hat{\nu}
-
g_{1}(\hat{B}+1)\hat{A}
]\hat{\nu}
\Bigr)
|\alpha\rangle|\tilde{\alpha}\rangle.
\label{second-order-correction-R-part-iii}
\end{eqnarray}

Next,
we examine the part including $\{\hat{S}_{n}\}$
in Eq.~(\ref{second-order-correction-i}).
From Eqs.~(\ref{definition-operators-mu-nu-1})
and
(\ref{definition-operators-mu-nu-2}),
we obtain
\begin{equation}
[\hat{A},\hat{\mu}]=[\hat{A},\hat{\nu}]=[\hat{\mu},\hat{\nu}]=-\hat{D},
\label{commutation-relationd-related-D}
\end{equation}
where
\begin{equation}
\hat{D}=a^{\dagger}a+\tilde{a}^{\dagger}\tilde{a}+1.
\end{equation}
In the remains of this section and Appendix~\ref{section-3rd-order-correction-term},
we use Eq.~(\ref{commutation-relationd-related-D}) and the following commutation relations often
without notice:
\begin{equation}
[\hat{A},\hat{C}]=-2\hat{D},
\quad\quad
[\hat{A},\hat{D}]=-2\hat{C}.
\end{equation}
Then,
we can rewrite $\hat{S}_{n}$ given by Eq.~(\ref{defintion-Rn-Sn-i}) as
\begin{equation}
\hat{S}_{n}
=
-
\Bigl(
(\hat{B}-1)^{n}-(\hat{B}+1)^{n}
\Bigr)
\hat{D}.
\label{formula-Sn-1}
\end{equation}
[In Eq.~(57) of Ref.~\cite{Azuma2010},
a calculation concerning
$\hat{S}_{n}$
is wrong.]
Thus, we can write down the part including $\{\hat{S}_{n}\}$
in the second order term given by Eq.~(\ref{second-order-correction-i})
as
\begin{equation}
-
\langle \alpha|\langle\tilde{\alpha}|
\Bigr(
g_{1}(\hat{B}-1)-g_{1}(\hat{B}+1)
\Bigr)
\hat{D}
|\alpha\rangle|\tilde{\alpha}\rangle.
\label{second-order-correction-S-part-i}
\end{equation}

Putting together Eqs.~(\ref{second-order-correction-i}),
(\ref{second-order-correction-R-part-i}),
(\ref{second-order-correction-R-part-iii})
and
(\ref{second-order-correction-S-part-i}),
we can write down the whole of the second order term
as
\begin{eqnarray}
&&
P^{(2)}_{g,1}(t) \nonumber \\
&=&
\langle \alpha|\langle\tilde{\alpha}|
[
g_{1}(\hat{B}-2)\hat{\mu}^{2}
+
g_{1}(\hat{B}-1)(-\hat{D}-\hat{\mu}\hat{A}-\hat{A}\hat{\mu}) \nonumber \\
&&
+
g_{1}(\hat{B})(\hat{A}^{2}-\hat{\nu}\hat{\mu}-\hat{\mu}\hat{\nu})
+
g_{1}(\hat{B}+1)(\hat{D}+\hat{\nu}\hat{A}+\hat{A}\hat{\nu}) \nonumber \\
&&
+
g_{1}(\hat{B}+2)\hat{\nu}^{2}]
|\alpha\rangle|\tilde{\alpha}\rangle.
\label{second-order-correction-ii}
\end{eqnarray}

Here,
to compute $P^{(2)}_{g,1}(t)$ given by Eq.~(\ref{second-order-correction-ii}),
we arrange $\hat{\mu}$ in the left side of the product of operators
and $\hat{\nu}$ in the right side of the product of operators.
For the arrangement of operators,
we carry out the calculations,
\begin{eqnarray}
(\hat{B}-2)^{n}\hat{\mu}^{2}
&=&
\hat{\mu}^{2}
\hat{B}^{n}, \nonumber \\
(\hat{B}-1)^{n}(-\hat{D}-\hat{\mu}\hat{A}-\hat{A}\hat{\mu})
&=&
-2\hat{\mu}^{2}(\hat{B}+1)^{n}
+2\hat{\mu}\hat{B}^{n}\hat{\nu}, \nonumber \\
\hat{B}^{n}(\hat{A}^{2}-\hat{\nu}\hat{\mu}-\hat{\mu}\hat{\nu})
&=&
\hat{\mu}^{2}(\hat{B}+2)^{n}
-4\hat{\mu}(\hat{B}+1)^{n}\hat{\nu}
-2\hat{B}^{n}\hat{D}+\hat{B}^{n}\hat{\nu}^{2}, \nonumber \\
(\hat{B}+1)^{n}(\hat{D}+\hat{\nu}\hat{A}+\hat{A}\hat{\nu})
&=&
2\hat{\mu}(\hat{B}+2)^{n}\hat{\nu}
-2(\hat{B}+1)^{n}\hat{\nu}^{2}
+2(\hat{B}+1)^{n}\hat{D} \nonumber \\
&&
\mbox{for $n=1,2,3,...$.}
\end{eqnarray}
Substitution of the above relations into Eq.~(\ref{second-order-correction-ii})
yields
\begin{eqnarray}
P^{(2)}_{g,1}(t)
&=&
\langle\alpha|\langle\tilde{\alpha}|
[
\hat{\mu}^{2}g_{1}(\hat{B})
-2\hat{\mu}^{2}g_{1}(\hat{B}+1)
+2\hat{\mu}g_{1}(\hat{B})\hat{\nu}
+\hat{\mu}^{2}g_{1}(\hat{B}+2) \nonumber \\
&&
-4\hat{\mu}g_{1}(\hat{B}+1)\hat{\nu}
-2g_{1}(\hat{B})\hat{D}
+g_{1}(\hat{B})\hat{\nu}^{2}
+2\hat{\mu}g_{1}(\hat{B}+2)\hat{\nu} \nonumber \\
&&
-2g_{1}(\hat{B}+1)\hat{\nu}^{2}
+2g_{1}(\hat{B}+1)\hat{D}
+g_{1}(\hat{B}+2)\hat{\nu}^{2}
]
|\alpha\rangle|\tilde{\alpha}\rangle \nonumber \\
&=&
4\alpha^{4}
\langle\alpha|\langle\tilde{\alpha}|
[
g_{1}(\hat{B})
-2g_{1}(\hat{B}+1)
+g_{1}(\hat{B}+2)
]
|\alpha\rangle|\tilde{\alpha}\rangle \nonumber \\
&&
-2
\langle\alpha|\langle\tilde{\alpha}|
g_{1}(\hat{B})\hat{D}
|\alpha\rangle|\tilde{\alpha}\rangle
+2
\langle\alpha|\langle\tilde{\alpha}|
g_{1}(\hat{B}+1)\hat{D}
|\alpha\rangle|\tilde{\alpha}\rangle.
\end{eqnarray}
Moreover,
preparing the following formula:
\begin{eqnarray}
&&
\langle\alpha|\langle\tilde{\alpha}|
g_{1}(\hat{B})\hat{D}
|\alpha\rangle|\tilde{\alpha}\rangle \nonumber \\
&=&
\langle\alpha|\langle\tilde{\alpha}|
e^{-\alpha^{2}}
\sum_{n=0}^{\infty}
\sum_{m=0}^{\infty}
\frac{\alpha^{n+m}}{\sqrt{n!m!}}
(n+m+1)g_{1}(n+c)
|n\rangle|\tilde{m}\rangle \nonumber \\
&=&
e^{-\alpha^{2}}
\sum_{n=0}^{\infty}
\frac{\alpha^{2n}}{n!}
g_{1}(n+c)
+
e^{-\alpha^{2}}
\sum_{n=0}^{\infty}
\frac{\alpha^{2n}}{n!}
n
g_{1}(n+c) \nonumber \\
&&
+
e^{-2\alpha^{2}}
\sum_{n=0}^{\infty}
\sum_{m=0}^{\infty}
\frac{\alpha^{2(n+m)}}{n!m!}
m
g_{1}(n+c), \nonumber \\
&=&
(1+\alpha^{2})
e^{-\alpha^{2}}
\sum_{n=0}^{\infty}
\frac{\alpha^{2n}}{n!}
g_{1}(n+c)
+
\alpha^{2}
e^{-\alpha^{2}}
\sum_{n=0}^{\infty}
\frac{\alpha^{2n}}{n!}
g_{1}(n+c+1),
\label{g-B-D-foemula}
\end{eqnarray}
we arrive at the final representation of $P^{(2)}_{g,1}(t)$ as
\begin{eqnarray}
P^{(2)}_{g,1}(t)
&=&
4\alpha^{4}
e^{-\alpha^{2}}
\sum_{n=0}^{\infty}
\frac{\alpha^{2n}}{n!}
[g_{1}(n+c)-2g_{1}(n+c+1)+g_{1}(n+c+2)] \nonumber \\
&&
-2
(1+\alpha^{2})
e^{-\alpha^{2}}
\sum_{n=0}^{\infty}
\frac{\alpha^{2n}}{n!}
g_{1}(n+c)
-2
\alpha^{2}
e^{-\alpha^{2}}
\sum_{n=0}^{\infty}
\frac{\alpha^{2n}}{n!}
g_{1}(n+c+1) \nonumber \\
&&
+2
(1+\alpha^{2})
e^{-\alpha^{2}}
\sum_{n=0}^{\infty}
\frac{\alpha^{2n}}{n!}
g_{1}(n+c+1)
+2
\alpha^{2}
e^{-\alpha^{2}}
\sum_{n=0}^{\infty}
\frac{\alpha^{2n}}{n!}
g_{1}(n+c+2) \nonumber \\
&=&
2(2\alpha^{2}+1)(\alpha+1)(\alpha-1)
Q_{1}^{(0)}(t) \nonumber \\
&&
-2(2\alpha^{2}+1)(2\alpha^{2}-1)
Q_{1}^{(1)}(t) \nonumber \\
&&
+2\alpha^{2}(2\alpha^{2}+1)
Q_{1}^{(2)}(t).
\label{2nd-order-perturbation-term-g1}
\end{eqnarray}

Similarly,
we obtain
$P^{(2)}_{g,2}(t)$
as
\begin{eqnarray}
P^{(2)}_{g,2}(t)
&=&
2(2\alpha^{2}+1)(\alpha+1)(\alpha-1)
Q_{2}^{(0)}(t) \nonumber \\
&&
-2(2\alpha^{2}+1)(2\alpha^{2}-1)
Q_{2}^{(1)}(t) \nonumber \\
&&
+2\alpha^{2}(2\alpha^{2}+1)
Q_{2}^{(2)}(t).
\label{2nd-order-perturbation-term-g2}
\end{eqnarray}

\section{\label{section-numerical-calculations}The numerical calculations}
In this section,
we show numerical results for the atomic population inversion
obtained with the third order perturbation theory under the low-temperature limit.
In Secs.~\ref{section-0th-order-correction-term},
\ref{section-1st-order-correction-term},
\ref{section-2nd-order-correction-term}
and Appendix~\ref{section-3rd-order-correction-term},
we obtain
$\{P_{g,1}^{(n)}(t),P_{g,2}^{(n)}(t):n\in\{0,1,2,3\}\}$
in the form of Eqs.~(\ref{0th-order-perturbation-term-1}),
(\ref{1st-order-perturbation-term-g1}),
(\ref{1st-order-perturbation-term-g2}),
(\ref{2nd-order-perturbation-term-g1}),
(\ref{2nd-order-perturbation-term-g2}),
(\ref{3rd-order-perturbation-term-g1})
and
(\ref{3rd-order-perturbation-term-g2}).
Thus,
from Eqs.~(\ref{definition-atomic-population-inversion}),
(\ref{Perturbation-theory-formula-0})
and
(\ref{Perturbation-theory-formula-1}),
we can calculate
$\langle\sigma_{z}(t)\rangle$
as the third order perturbation theory,
\begin{equation}
\langle\sigma_{z}(t)\rangle
=
1-2
[
\cos^{2}\Theta(\beta)
\sum_{n=0}^{3}
\frac{\theta(\beta)^{n}}{n!}
P_{g,1}^{(n)}(t)
+
\sin^{2}\Theta(\beta)
\sum_{n=0}^{3}
\frac{\theta(\beta)^{n}}{n!}
P_{g,2}^{(n)}(t)
].
\label{sigma_z_t_upto-3rd-order-perturbation}
\end{equation}

\begin{figure}
\begin{center}
\mbox{\scalebox{1.0}[1.0]{\includegraphics{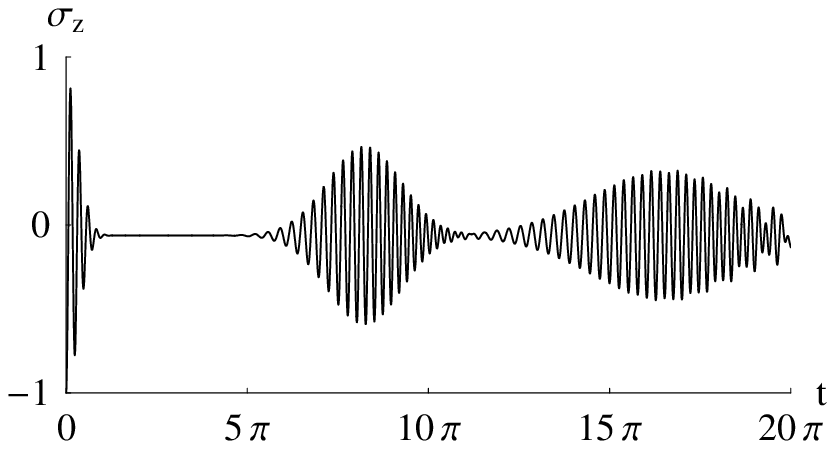}}}
\vspace*{8pt}
\caption{The atomic population inversion $\langle\sigma_{z}(t)\rangle$
as a function of the time $t$
obtained from numerical calculations of Eq.~(\ref{sigma_z_t_upto-3rd-order-perturbation})
with
$\alpha=4$, $c=1$, $\kappa=1$ and $\Theta(\beta)=\theta(\beta)=0$.
Looking at the graph,
we estimate the time scale of the initial collapse
and
the period of the revival
of the Rabi oscillations 
at unity and $8\pi$ around, respectively.}
\label{Figure01}
\end{center}
\vspace*{8pt}
\begin{center}
\mbox{\scalebox{1.0}[1.0]{\includegraphics{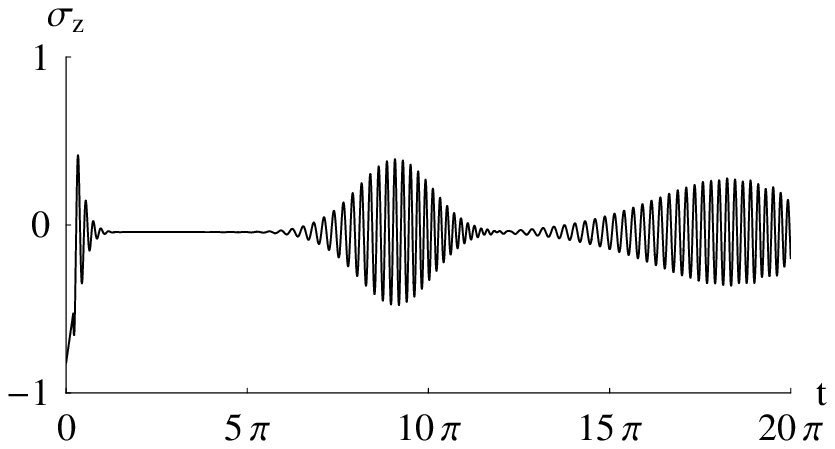}}}
\vspace*{8pt}
\caption{The atomic population inversion $\langle\sigma_{z}(t)\rangle$
as a function of the time $t$
obtained from numerical calculations of Eq.~(\ref{sigma_z_t_upto-3rd-order-perturbation})
with
$\alpha=4$, $c=1$, $\kappa=1$,
$\omega_{0}=2$, $\omega=4$,
$\theta(\beta)=\pi/32$
and
$\Theta(\beta)=\arctan[\tanh^{1/2}(\pi/32)]$.
Looking at the graph,
we estimate the time scale of the initial collapse
and
the period of the revival
of the Rabi oscillations 
at unity and $8\pi e^{\pi/32}\simeq (8.83)\pi$ around, respectively.}
\label{Figure02}
\end{center}
\end{figure}

\begin{figure}
\begin{center}
\mbox{\scalebox{1.0}[1.0]{\includegraphics{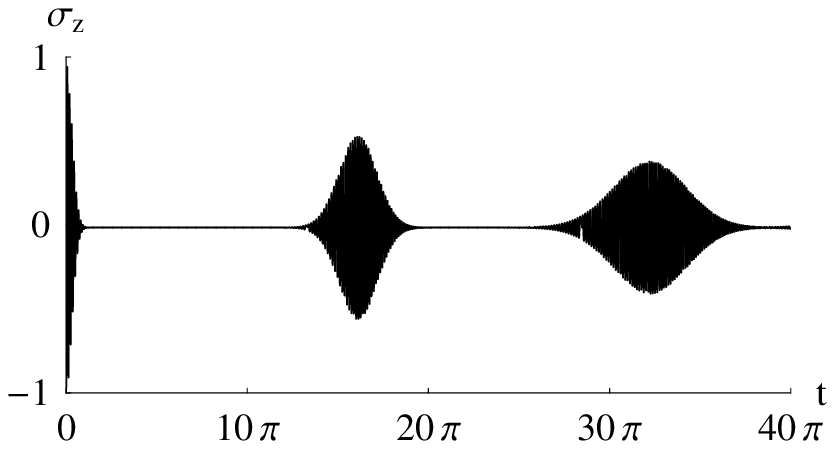}}}
\vspace*{8pt}
\caption{The atomic population inversion $\langle\sigma_{z}(t)\rangle$
as a function of the time $t$
obtained from numerical calculations of Eq.~(\ref{sigma_z_t_upto-3rd-order-perturbation})
with
$\alpha=8$, $c=1$, $\kappa=1$ and $\Theta(\beta)=\theta(\beta)=0$.
Looking at the graph,
we estimate the time scale of the initial collapse
and
the period of the revival
of the Rabi oscillations 
at unity and $16\pi$ around, respectively.}
\label{Figure03}
\end{center}
\vspace*{8pt}
\begin{center}
\mbox{\scalebox{1.0}[1.0]{\includegraphics{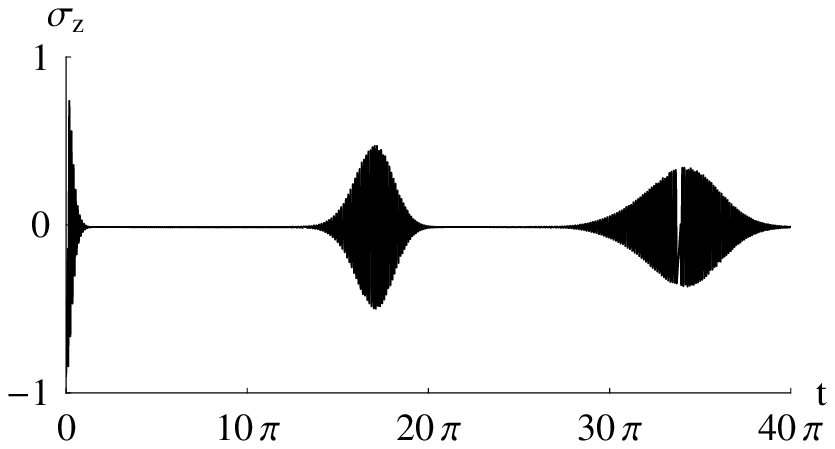}}}
\vspace*{8pt}
\caption{The atomic population inversion $\langle\sigma_{z}(t)\rangle$
as a function of the time $t$
obtained from numerical calculations of Eq.~(\ref{sigma_z_t_upto-3rd-order-perturbation})
with
$\alpha=8$, $c=1$, $\kappa=1$,
$\omega_{0}=2$, $\omega=4$,
$\theta(\beta)=\pi/60$
and
$\Theta(\beta)=\arctan[\tanh^{1/2}(\pi/60)]$.
Looking at the graph,
we estimate the time scale of the initial collapse
and
the period of the revival
of the Rabi oscillations 
at unity and $16\pi e^{\pi/60}\simeq (16.9)\pi$ around, respectively.}
\label{Figure04}
\end{center}
\end{figure}

Figure~\ref{Figure01} shows
the atomic population inversion $\langle\sigma_{z}(t)\rangle$
given by Eq.~(\ref{sigma_z_t_upto-3rd-order-perturbation})
as a function of the time $t$
with
$\alpha=4$, $c=1$, $\kappa=1$ and $\Theta(\beta)=\theta(\beta)=0$.
Figure~\ref{Figure03} shows
the atomic population inversion $\langle\sigma_{z}(t)\rangle$
given by Eq.~(\ref{sigma_z_t_upto-3rd-order-perturbation})
as a function of the time $t$
with
$\alpha=8$, $c=1$, $\kappa=1$ and $\Theta(\beta)=\theta(\beta)=0$.
Carrying out numerical calculations for Figs.~\ref{Figure01} and \ref{Figure03},
we replace the summation $\sum_{n=0}^{\infty}$ in $Q_{1}^{(0)}(t)$
with $\sum_{n=0}^{100}$,
so that we compute the sum of first one hundred and one terms in the series.
(In this section,
whenever we carry out numerical calculations of $Q_{1}^{(l)}(t)$ and $Q_{2}^{(l)}(t)$,
we replace their summation $\sum_{n=0}^{\infty}$ with $\sum_{n=0}^{100}$.)
In Figs.~\ref{Figure01} and \ref{Figure03},
we assume the system to be at zero temperature.
Thus, the graphs in Figs.~\ref{Figure01} and \ref{Figure03}
do not suffer from thermal effects. 
We can observe the collapse and the revival of the Rabi oscillations
obviously in these graphs.

Figure~\ref{Figure02} shows
the atomic population inversion $\langle\sigma_{z}(t)\rangle$
given by Eq.~(\ref{sigma_z_t_upto-3rd-order-perturbation})
as a function of the time $t$
with
$\alpha=4$, $c=1$, $\kappa=1$, $\theta(\beta)=\pi/32$, $\omega_{0}=2$ and $\omega=4$.
From Eq.~(\ref{definition-temperature-JCM-0}),
we obtain
\begin{eqnarray}
\theta(\beta)
&=&
\mbox{arctanh}(e^{-\beta\hbar\omega/2}), \nonumber \\
\Theta(\beta)
&=&
\arctan(e^{-\beta\hbar\omega_{0}/2}),
\label{definitions-theta-Theta-1}
\end{eqnarray}
so that the relation
$\exp(-2\beta\hbar)=\tanh[\theta(\beta)]=\tanh(\pi/32)$
holds.
Thus,
we can derive the following relation:
\begin{equation}
\Theta(\beta)
=\arctan(e^{-\beta\hbar})
=\arctan[\tanh^{1/2}(\theta(\beta))]
=\arctan[\tanh^{1/2}(\pi/32)].
\label{definitions-theta-Theta-2}
\end{equation}
Because the system of Fig.~\ref{Figure02} evolves in time with maintaining constant low temperature,
its time-evolution is under the thermal effects.
Comparing the graphs shown in Figs.~\ref{Figure01} and \ref{Figure02},
we notice that the period of Fig.~\ref{Figure02} is longer than the period of Fig.~\ref{Figure01}.
Thus,
we can suppose that the thermal effects let the period of the revival of the Rabi oscillations
become longer.

Figure~\ref{Figure04} shows
the atomic population inversion $\langle\sigma_{z}(t)\rangle$
given by Eq.~(\ref{sigma_z_t_upto-3rd-order-perturbation})
as a function of the time $t$
with
$\alpha=8$, $c=1$, $\kappa=1$, $\theta(\beta)=\pi/60$, $\omega_{0}=2$ and $\omega=4$.
Then,
in a similar manner for obtaining Eqs.~(\ref{definitions-theta-Theta-1})
and
(\ref{definitions-theta-Theta-2}),
we achieve
$\Theta(\beta)
=\arctan[\tanh^{1/2}(\theta(\beta))]
=\arctan[\tanh^{1/2}(\pi/60)]$.
Comparing the graphs shown in Figs.~\ref{Figure03} and \ref{Figure04},
we notice that the period of Fig.~\ref{Figure04} is longer than the period of Fig.~\ref{Figure03},
so that
we can suppose that the thermal effects let the period of the revival of the Rabi oscillations
become longer.

\begin{table}
\caption{The ranges of numerical values of the perturbation corrections with
$\alpha=4$, $c=1$, $\kappa=1$, $0\leq t\leq 20\pi$ and $\theta(\beta)=\pi/32$.
The estimations of the minimum and the maximum in every row of the table are based on values
of each correction term,
which we obtain numerically at equally spaced intervals $\Delta t=20\pi\times 10^{-4}$
during $0\leq t\leq 20\pi$.}
\label{Table01}
\begin{center}
\begin{tabular}{|c|c|c|}
\hline
correction term & min & max \\
\hline
$\theta(\beta)P_{g,1}^{(1)}(t)$          & $-0.227$\hphantom{0} & $0.199$\hphantom{0} \\
$\theta(\beta)P_{g,2}^{(1)}(t)$          & $-0.208$\hphantom{0} & $0.225$\hphantom{0} \\
$(1/2)\theta(\beta)^{2}P_{g,1}^{(2)}(t)$ & $-0.108$\hphantom{0} & $0.102$\hphantom{0} \\
$(1/2)\theta(\beta)^{2}P_{g,2}^{(2)}(t)$ & $-0.0996$            & $0.109$\hphantom{0} \\
$(1/6)\theta(\beta)^{3}P_{g,1}^{(3)}(t)$ & $-0.0441$            & $0.0483$ \\
$(1/6)\theta(\beta)^{3}P_{g,2}^{(3)}(t)$ & $-0.0478$            & $0.0467$ \\
\hline
\end{tabular}
\end{center}
\end{table}

\begin{table}
\caption{The ranges of numerical values of the perturbation corrections with
$\alpha=8$, $c=1$, $\kappa=1$, $0\leq t\leq 40\pi$ and $\theta(\beta)=\pi/60$.
The estimations of the minimum and the maximum in every row of the table are based on values
of each correction term,
which we obtain numerically at equally spaced intervals $\Delta t=40\pi\times 10^{-4}$
during $0\leq t\leq 40\pi$.}
\label{Table02}
\begin{center}
\begin{tabular}{|c|c|c|}
\hline
correction term & min & max \\
\hline
$\theta(\beta)P_{g,1}^{(1)}(t)$          & $-0.246$\hphantom{0} & $0.247$\hphantom{0} \\
$\theta(\beta)P_{g,2}^{(1)}(t)$          & $-0.248$\hphantom{0} & $0.245$\hphantom{0} \\
$(1/2)\theta(\beta)^{2}P_{g,1}^{(2)}(t)$ & $-0.125$\hphantom{0} & $0.128$\hphantom{0} \\
$(1/2)\theta(\beta)^{2}P_{g,2}^{(2)}(t)$ & $-0.127$\hphantom{0} & $0.126$\hphantom{0} \\
$(1/6)\theta(\beta)^{3}P_{g,1}^{(3)}(t)$ & $-0.0566$            & $0.0570$ \\
$(1/6)\theta(\beta)^{3}P_{g,2}^{(3)}(t)$ & $-0.0565$            & $0.0567$ \\
\hline
\end{tabular}
\end{center}
\end{table}

When we take $\alpha=4$, $c=1$, $\kappa=1$, $0\leq t\leq 20\pi$ and $\theta(\beta)=\pi/32$,
a numerical value of each order perturbation correction varies as shown in Table~\ref{Table01}.
On the other hand,
when we take $\alpha=8$, $c=1$, $\kappa=1$, $0\leq t\leq 40\pi$ and $\theta(\beta)=\pi/60$,
a numerical value of each order perturbation correction varies as shown in Table~\ref{Table02}.
Turning our eyes towards Table~\ref{Table01},
we observe that the contribution of the third order correction is nearly equal to
a half of the contribution of the second order correction
in the perturbative expansion.
From Table~\ref{Table01},
we consider the perturbative expansion to be reliable for $\theta(\beta)=\pi/32$.
Thus,
taking $\alpha=4$, $c=1$ and $\kappa=1$,
we can conclude that the third order perturbation theory is effective
for the parameter $0\leq \theta(\beta) \leq \pi/32$.
We notice that a similar thing happens in Table~\ref{Table02}, as well.
Thus,
taking $\alpha=8$, $c=1$ and $\kappa=1$,
we can conclude that the third order perturbation theory is effective
for the parameter $0\leq \theta(\beta) \leq \pi/60$.

\begin{figure}
\begin{center}
\includegraphics[scale=1.0]{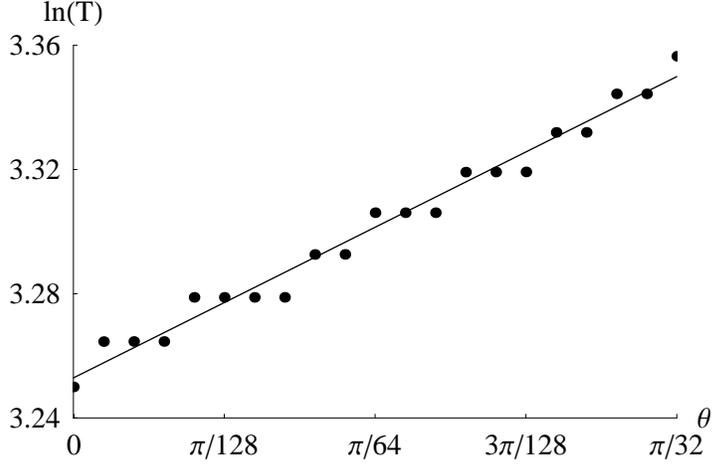}
\vspace*{8pt}
\caption{The period of the revival of the Rabi oscillations $T(\theta)$
plotted
as a function of the parameter of the temperature $\theta(\beta)$.
The points are
obtained from numerical calculations of the third order perturbation theory
with taking
$\alpha=4$, $c=1$, $\kappa=1$,
$\omega_{0}=2$, $\omega=4$
and
$0\leq \theta(\beta)\leq\pi/32$.
In the graph,
the vertical axis is scaled logarithmically as $\ln[T(\theta)]$
and the horizontal axis is scaled linearly as $\theta(\beta)$.
Fitting the points with the linear function according to the least-squares method,
we obtain $\ln[T(\theta)]=3.25+(0.988)\theta$,
which is drawn in the graph.}
\label{Figure05}
\end{center}
\end{figure}

\begin{figure}
\begin{center}
\includegraphics[scale=1.0]{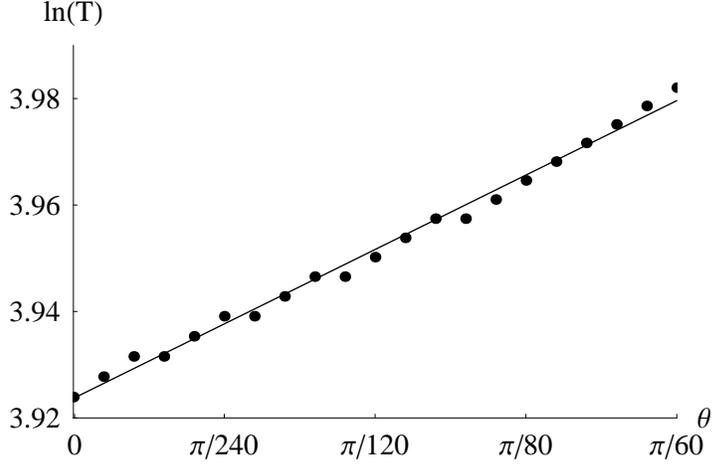}
\vspace*{8pt}
\caption{The period of the revival of the Rabi oscillations $T(\theta)$
plotted
as a function of the parameter of the temperature $\theta(\beta)$.
The points are
obtained from numerical calculations of the third order perturbation theory
with taking
$\alpha=8$, $c=1$, $\kappa=1$,
$\omega_{0}=2$, $\omega=4$
and
$0\leq \theta(\beta)\leq\pi/60$.
In the graph,
the vertical axis is scaled logarithmically as $\ln[T(\theta)]$
and the horizontal axis is scaled linearly as $\theta(\beta)$.
Fitting the points with the linear function according to the least-squares method,
we obtain $\ln[T(\theta)]=3.92+(1.07)\theta$,
which is drawn in the graph.}
\label{Figure06}
\end{center}
\end{figure}

In Figs.~\ref{Figure05} and \ref{Figure06},
we plot
the period of the revival of the Rabi oscillations $T(\theta)$
as a function of the parameter of the temperature $\theta(\beta)$.
The points in Fig.~\ref{Figure05} are
obtained from numerical calculations of the third order perturbation theory
with taking
$\alpha=4$, $c=1$, $\kappa=1$,
$\omega_{0}=2$, $\omega=4$
and
$0\leq \theta(\beta)\leq\pi/32$.
The points in Fig.~\ref{Figure06} are obtained similarly with taking
$\alpha=8$, $c=1$, $\kappa=1$,
$\omega_{0}=2$, $\omega=4$
and
$0\leq \theta(\beta)\leq\pi/60$.
In Fig.~\ref{Figure05},
we compute the period $T(\theta)$ numerically as follows:
First,
we calculate $\langle\sigma_{z}(t)\rangle$
given by Eq.~(\ref{sigma_z_t_upto-3rd-order-perturbation})
for a certain $\theta(\beta)$.
For every point of Fig.~\ref{Figure05}, taking the interval of the time $\Delta t=5\pi\times 10^{-4}$,
we obtain $\langle\sigma_{z}(t)\rangle$ at each time step
during $15\pi/2\leq t\leq 10\pi$.
We write the time at which $\langle\sigma_{z}(t)\rangle$ takes the maximum value
as $t_{\mbox{\scriptsize max}}$
and
write the time at which $\langle\sigma_{z}(t)\rangle$ takes the minimum value
as $t_{\mbox{\scriptsize min}}$.
Second, we obtain the period $T(\theta)$
with taking $T=(t_{\mbox{\scriptsize max}}+t_{\mbox{\scriptsize min}})/2$.
[For example,
taking $\theta=\pi/32$,
we obtain
$t_{\mbox{\scriptsize max}}\approx 28.52$
for $\langle\sigma_{z}(t_{\mbox{\scriptsize max}})\rangle\approx 0.3923$
and
$t_{\mbox{\scriptsize min}}\approx 28.86$
for $\langle\sigma_{z}(t_{\mbox{\scriptsize min}})\rangle\approx -0.4763$.
Thus,
we obtain $T(\pi/32)\approx 28.69$.]
The points of $T(\theta)$ in Fig.~\ref{Figure06} are
obtained in a similar manner with taking the interval of the time
$\Delta t=10\pi\times 10^{-4}$
and carrying out calculations of $\langle\sigma_{z}(t)\rangle$
at each time step during $15\pi\leq t\leq 20\pi$.

In the graphs of Figs.~\ref{Figure05} and \ref{Figure06},
the vertical axes are scaled logarithmically as $\ln[T(\theta)]$
and the horizontal axes are scaled linearly as $\theta(\beta)$.
In the graph of Fig.~\ref{Figure05},
the points form groups consisting of twos, threes and fours,
so that they appear in the shape of the stairs
as $\ln[T(\theta)]$ increases gradually.
The reason why the points appear in the shape of the stairs is as follows:
The atomic population inversion $\langle\sigma_{z}(t)\rangle$
is a bunch of the Rabi oscillations whose period is $\pi/(|\alpha||\kappa|)\simeq\pi/4$ around.
(We obtain this approximation in Sec.~\ref{section-review-JCM}.)
At the same time,
it shows the revival of the amplitude envelope with the period
$T(\theta)\simeq 2\pi|\alpha|e^{\theta}/|\kappa|=8\pi e^{\theta}$ around.
Thus,
calculating $t_{\mbox{\scriptsize max}}$ and $t_{\mbox{\scriptsize min}}$
numerically,
the rapid Rabi oscillations give us the smallest interval measurable as about
$\pi/8$, which is the half of the period of the Rabi oscillations.
This resolution of the time lets the points in Fig.~\ref{Figure05}
form the shape of the stairs.

Contrastingly,
the points in Fig.~\ref{Figure06} do not appear in the distinct shape of the stairs.
This is because the resolution of Fig.~\ref{Figure06} is finer than that of Fig.~\ref{Figure05}.
Indeed, in Fig.~\ref{Figure06},
the period of the Rabi oscillations is given by $\pi/8$ around,
so that the resolution of $T(\theta)$ is nearly equal to $\pi/16$.

Fitting the points in Fig.~\ref{Figure05}
with the linear function according to the least-squares method,
we obtain
\begin{equation}
\ln{[}T(\theta){]}
=
3.25+(0.988)\theta.
\end{equation}
We can interpret the above result as
\begin{equation}
T(\theta)
\simeq
e^{3.25}\times e^{(0.988)\theta}
\simeq
8\pi e^{\theta},
\end{equation}
which reminds us of Eq.~(\ref{thermal-effect-period-rivival-Rabi-oscillations}).
On the other hand,
fitting the points in Fig.~\ref{Figure06}
with the linear function according to the least-squares method,
we obtain
\begin{equation}
\ln{[}T(\theta){]}
=
3.92+(1.07)\theta.
\end{equation}
We can interpret the above result as
\begin{equation}
T(\theta)
\simeq
e^{3.92}\times e^{(1.07)\theta}
\simeq
16\pi e^{\theta},
\end{equation}
which also reminds us of Eq.~(\ref{thermal-effect-period-rivival-Rabi-oscillations}).

\section{\label{section-counter-rotating-terms}Thermal effects of the counter-rotating terms}
In this section,
we address thermal effects of the counter-rotating terms.
Because this topic is difficult and includes subtle problems,
we treat it with an intuitive manner.

First of all,
we have to go back to the derivation of the JCM.
At the beginning,
we consider the Hamiltonian for a magnetic dipole in a magnetic field,
and we obtain
\begin{equation}
H
=
\frac{\hbar}{2}\omega_{0}\sigma_{z}
+
\hbar\omega a^{\dagger}a
+
\hbar\kappa(\sigma_{-}+\sigma_{+})
(a+a^{\dagger}).
\label{dipole-field-Hamiltonian-0}
\end{equation}
Assuming near resonance $\omega\simeq\omega_{0}$,
the interaction terms $\sigma_{+}a$ and $\sigma_{-}a^{\dagger}$ are
practically independent of the time $t$,
while the terms $\sigma_{-}a$ and $\sigma_{+}a^{\dagger}$ vary
rapidly at frequencies $\pm(\omega_{0}+\omega)$.
Then,
applying the rotating wave approximation to Eq.~(\ref{dipole-field-Hamiltonian-0})
and removing the term $\hbar\kappa(\sigma_{+}a^{\dagger}+\sigma_{-}a)$,
we obtain the Hamiltonian of the JCM written down as Eq.~(\ref{JCM-Hamiltonian-0}).

As mentioned above,
the rotating wave approximation is used often in the field of the quantum optics.
However,
it is shown that the rotating wave approximation cannot always be a good treatment,
and sometimes it causes serious defects.
Ford {\it et al}. examine the Hamiltonian for an oscillator of the frequency $\omega_{0}$
interacting with a reservoir and its rotating wave approximation
\cite{Ford1988,Ford1997}.
The Hamiltonian of the original model is given by
\begin{equation}
H
=
\hbar\omega_{0}a^{\dagger}a
+
\sum_{j}\hbar\omega_{j} b_{j}^{\dagger}b_{j}
+
(a+a^{\dagger})\sum_{j}\lambda_{j}(b_{j}+b_{j}^{\dagger}),
\label{oscillator-reservoir-Hamiltonian-0}
\end{equation}
where $[a,a^{\dagger}]=1$, $[b_{j},b_{j}^{\dagger}]=1$ $\forall j$,
and its rotating wave approximation is given by
\begin{equation}
H_{\mbox{\scriptsize RWA}}
=
\hbar\omega_{0}a^{\dagger}a
+
\sum_{j}\hbar\omega_{j} b_{j}^{\dagger}b_{j}
+
\sum_{j}\lambda_{j}(a^{\dagger}b_{j}+ab_{j}^{\dagger}).
\label{oscillator-reservoir-Hamiltonian-RWA-1}
\end{equation}
Then,
the Hamiltonian $H_{\mbox{\scriptsize RWA}}$ defined
in Eq.~(\ref{oscillator-reservoir-Hamiltonian-RWA-1})
causes the following problem:
The expectation value (the energy) of $H_{\mbox{\scriptsize RWA}}$ has no lower bound,
so that we cannot specify the ground state.
Thus, we have to think the system described with $H_{\mbox{\scriptsize RWA}}$ to be unphysical.

As explained above, the rotating wave approximation sometimes manifests anomalous aspects.
Someone might complain that the rotating wave approximation brings us the JCM
that is an exactly soluble quantum mechanical model
for arbitrary $\Delta \omega$ and $\kappa$.
However, the JCM also has a defect,
which we cannot neglect.
Here, we think around the eigenvalues and the eigenvectors of the Hamiltonian of the JCM
given by Eq.~(\ref{JCM-Hamiltonian-0}).
They are written down as follows
\cite{Louisell1973}:
\begin{eqnarray}
E_{n,1}
&=&
\hbar[\omega(n+\frac{1}{2})+\lambda_{n}], \nonumber \\
E_{n,2}
&=&
\hbar[\omega(n+\frac{1}{2})-\lambda_{n}]
\quad\quad
\mbox{for $n=0,1,2,...$}, \nonumber \\
E_{0,0}
&=&
-\frac{\hbar}{2}\omega_{0},
\label{eigenvalues-JCM}
\end{eqnarray}
\begin{eqnarray}
|\varphi(n,1)\rangle
&=&
\cos\theta_{n}|n+1\rangle_{\mbox{\scriptsize P}}|g\rangle_{\mbox{\scriptsize A}}
+
\sin\theta_{n}|n\rangle_{\mbox{\scriptsize P}}|e\rangle_{\mbox{\scriptsize A}}, \nonumber \\
|\varphi(n,2)\rangle
&=&
-\sin\theta_{n}|n+1\rangle_{\mbox{\scriptsize P}}|g\rangle_{\mbox{\scriptsize A}}
+
\cos\theta_{n}|n\rangle_{\mbox{\scriptsize P}}|e\rangle_{\mbox{\scriptsize A}}
\quad\quad
\mbox{for $n=0,1,2,...$}, \nonumber \\
|0,0\rangle
&=&
|0\rangle_{\mbox{\scriptsize P}}|g\rangle_{\mbox{\scriptsize A}},
\label{eigenvectors-JCM}
\end{eqnarray}
where
\begin{equation}
\lambda_{n}=\sqrt{(\frac{\Delta\omega}{2})^{2}+\kappa^{2}(n+1)},
\end{equation}
$\Delta\omega=\omega-\omega_{0}$,
and
\begin{equation}
\tan\theta_{n}=\frac{\kappa\sqrt{n+1}}{(\Delta\omega/2)+\lambda_{n}}.
\end{equation}

\begin{figure}
\begin{center}
\mbox{\scalebox{1.0}[1.0]{\includegraphics{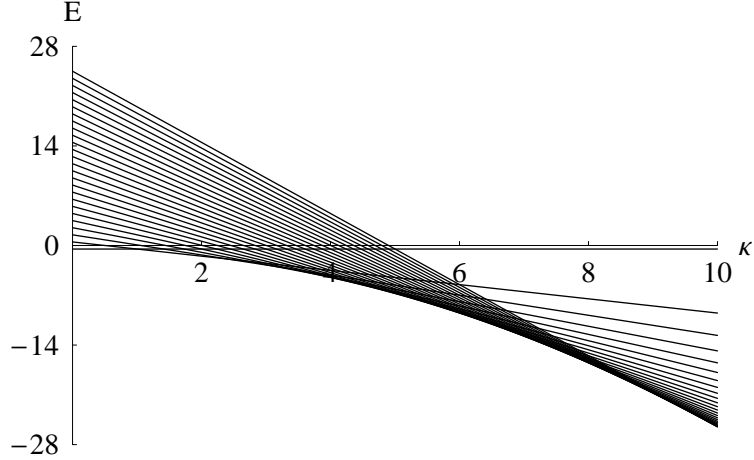}}}
\vspace*{8pt}
\caption{$E_{0,0}$ and $\{E_{n,2}:n=0,1,2,...,24\}$
as functions of $\kappa$,
where $0\leq \kappa\leq 10$.
Plotting them as graphs,
we assume $\omega_{0}=\omega=1$ and $\hbar=1$.
Because $\{E_{n,1}\}$ never can be the ground-state energy,
we do not plot them in this figure.
Looking these graphs,
we notice the following facts.
When $\kappa=0$, the ground-state energy is equal to $E_{0,0}$.
On the other hand, when $\kappa=10$, the ground-state energy is given by $E_{24,2}$.
In fact, these graphs show that the ground-state energy changes from $E_{0,0}$
to $E_{n,2}$ for $n\gg 1$ gradually as $\kappa$ becomes larger.}
\label{Figure07}
\end{center}
\end{figure}

\begin{figure}
\begin{center}
\mbox{\scalebox{1.0}[1.0]{\includegraphics{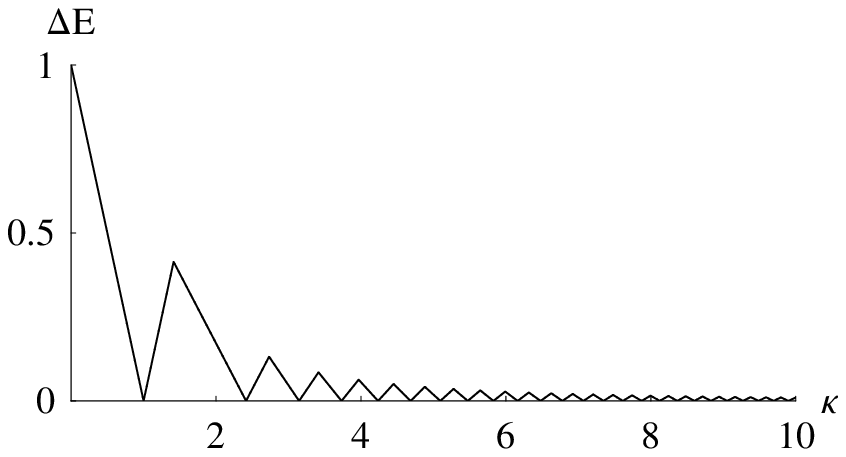}}}
\vspace*{8pt}
\caption{An excitation energy $\Delta E$,
which is required to promote the JCM system from the ground state to the first excited state,
as an function of $\kappa$,
where $0\leq\kappa\leq 10$.
Looking at this graph,
we notice that $\Delta E$ oscillates and its amplitude decreases rapidly.
When $\Delta E=0$, the JCM system has the degenerate ground states.}
\label{Figure08}
\end{center}
\vspace*{8pt}
\begin{center}
\mbox{\scalebox{1.0}[1.0]{\includegraphics{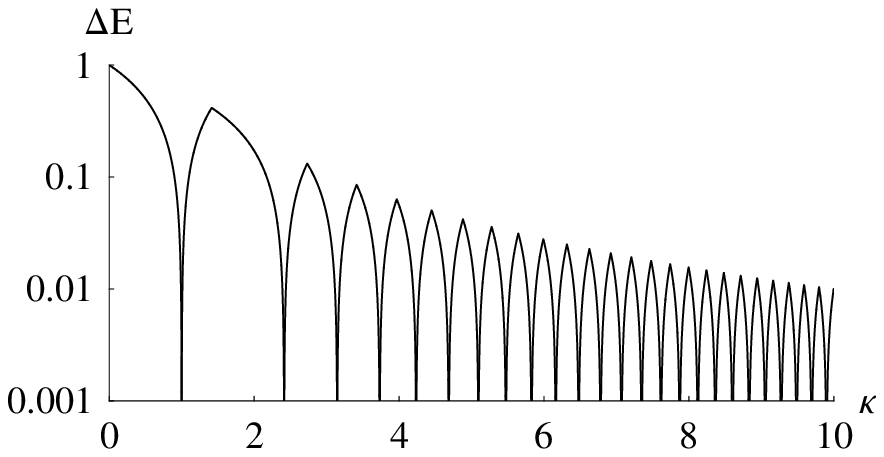}}}
\vspace*{8pt}
\caption{An excitation energy $\Delta E$,
which is required to promote the JCM system from the ground state to the first excited state,
as an function of $\kappa$,
where $0\leq\kappa\leq 10$.
The graph uses the logarithmic scale on the vertical axis
and the linear scale on the horizontal axis.
In Fig.~\ref{Figure08}, we show that the system has the degenerate ground states
at certain values of the parameter $\kappa$.
Because $\ln\Delta E\to -\infty$ as $\Delta E\to 0$,
we cannot plot small $\Delta E$, which is nearly equal to zero, in the graph.
Looking at this graph,
we notice that the amplitude of $\Delta E$,
which oscillates in the parameter $\kappa$,
decreases exponentially.}
\label{Figure09}
\end{center}
\end{figure}

Looking at Eq.~(\ref{eigenvalues-JCM}),
we notice that the ground state changes from $|0,0\rangle$
to $|\varphi(n,2)\rangle$ for $n\gg 1$ gradually
as $|\kappa|$ becomes larger.
To confirm it numerically,
we plot $E_{0,0}$ and $\{E_{n,2}:n=0,1,2,...,24\}$
as functions of $\kappa$ in Fig.~\ref{Figure07}.
At the same time,
an excitation energy, which is required to promote the JCM system from the ground state
to the first excited state,
becomes smaller rapidly as $|\kappa|\to\infty$.
To confirm it numerically,
we plot the excitation energy as a function of $\kappa$ in Figs.~\ref{Figure08} and \ref{Figure09}.

From the analyses performed in Figs.~\ref{Figure07}, \ref{Figure08} and \ref{Figure09},
we can conclude as follows:
If we take a large value of $|\kappa|$, the ground state of the JCM contains many photons.
Then,
the excitation energy takes a small value.
These properties of the ground state of the JCM relate to the uncertainty principle
$\Delta N \Delta \phi\geq (1/2)$.
Because $\Delta E$ decreases exponentially
as $|\kappa|$ becomes larger as shown in Figs.~\ref{Figure08} and \ref{Figure09},
the system of the JCM is able to jump from the ground state to excited states
at ease for $|\kappa|\gg 1$.
Thus, the fluctuation of the number of photons $\Delta N$ becomes very larger.
Hence, according to the uncertainty principle
$\Delta N \Delta \phi\geq (1/2)$,
the system of the JCM around the ground state acquires
very small fluctuation of the phase of each photon,
so that $\Delta \phi\to 0$.

However, the ground state that contains a large number of photons with small fluctuation of the phase
seems not to be practical.
We may realize the ground state for $\Delta N\gg 1$ and $\Delta\phi\simeq 0$
in the laboratory
by using a two-level atom in the cavity field,
which is induced by a very strong laser beam.

From the viewpoint explained above,
we cannot regard the JCM derived with the rotating wave approximation as a proper model
in the field of the quantum optics.
Hence, the JCM is valid and has physical meanings
if and only if a near resonance $\omega\simeq\omega_{0}$ is assumed and $|\kappa|$ is small enough.

To overcome the defects of the rotating wave approximation in the JCM,
some researchers try to extend and generalize the JCM.
Ng {\it et al}. investigate the two-photon JCM
and the intensity-dependent JCM with the counter-rotating terms
\cite{Ng1999,Ng2000}.
In these models,
the nonlinearity of the interaction between the two-level atom and the cavity field is emphasized.

In general,
it is very difficult and complicated to evaluate the contributions of the counter-rotating terms
in the JCM.
Feranchuk {\it et al}. study the Schr{\"o}dinger equation,
whose Hamiltonian is given by Eq.~(\ref{dipole-field-Hamiltonian-0}),
numerically
\cite{Feranchuk1996}.

Especially,
Phoenix presents several perturbative approaches to investigate this problem.
Here, we review one of his perturbation methods,
which is called short time expansion of the inversion.
First,
we begin with the Heisenberg picture of $\sigma_{z}$,
\begin{equation}
\sigma_{z}(t)
=
\exp(\frac{i}{\hbar}Ht)\sigma_{z}\exp(-\frac{i}{\hbar}Ht),
\label{Heisenberg-picture-sigma-z-0}
\end{equation}
where the Hamiltonian is given by Eq.~(\ref{dipole-field-Hamiltonian-0}).
Moreover,
we assume $\omega=\omega_{0}$,
so that we consider the optical resonance.

Second,
we expand Eq.~(\ref{Heisenberg-picture-sigma-z-0}) in a power series in $t$ and neglect third-order terms
\cite{Phoenix1989}.
So that,
we obtain
\begin{eqnarray}
\sigma_{z}(t)
&=&
\sigma_{z}(0)
+
\frac{it}{\hbar}[H,\sigma_{z}(0)]
-
\frac{t^{2}}{2\hbar^{2}}(H^{2}\sigma_{z}(0)+\sigma_{z}(0)H^{2})
+
\frac{t^{2}}{\hbar^{2}}H\sigma_{z}(0)H \nonumber \\
&&
+
{\cal O}(t^{3}) \nonumber \\
&=&
1-2(\kappa t)^{2}(a+a^{\dagger})^{2}
+
{\cal O}(t^{3}).
\label{short-time-expansion-inversion-0}
\end{eqnarray}
Third,
we assume the initial state as $|\alpha\rangle_{\mbox{\scriptsize P}}|e\rangle_{\mbox{\scriptsize A}}$,
where $\alpha=\sqrt{\bar{n}}e^{i\phi}$,
and substitute it into Eq.~(\ref{short-time-expansion-inversion-0}).
Finally,
we obtain
\begin{eqnarray}
\langle \sigma_{z}(t)\rangle_{\alpha}
&=&
{}_{\mbox{\scriptsize A}}\langle e|_{\mbox{\scriptsize P}}\langle\alpha|
\sigma_{z}(t)
|\alpha\rangle_{\mbox{\scriptsize P}}|e\rangle_{\mbox{\scriptsize A}} \nonumber \\
&=&
1-2(\kappa t)^{2}(4\bar{n}\cos^{2}\phi+1)+{\cal O}(t^{3}).
\label{short-time-expansion-inversion-coherent-state-1}
\end{eqnarray}
Because Eq.~(\ref{short-time-expansion-inversion-coherent-state-1}) is valid
for $0\leq t\ll 1$,
we can expect it to describe the initial collapse of the Rabi oscillations.

We note that Eq.~(\ref{short-time-expansion-inversion-coherent-state-1})
depends on the phase $\phi$.
This characteristic can always be found
in any perturbative expansion of $\langle\sigma_{z}(t)\rangle_{\alpha}$.
(This fact is indicated by Phoenix first.)

In Sec.~\ref{section-thermal-effects-period-qualitative-estimation},
we give the intuitive discussions about the thermal effects of the JCM,
and we obtain Eq.~(\ref{thermal-effect-of-alpha-coherent-state-0})
under the low-temperature limit.
Hence, we can add the thermal effects to Eq.~(\ref{short-time-expansion-inversion-coherent-state-1})
as
\begin{equation}
\langle\sigma_{z}(t)\rangle_{\alpha}
=
1-2(\kappa t)^{2}(4\bar{n}e^{2\theta}\cos^{2}\phi+1)+{\cal O}(t^{3}),
\label{short-time-expansion-inversion-coherent-state-2}
\end{equation}
where $\theta$ is given by Eq.~(\ref{definition-theta-beta-1}).
To examine whether or not Eq.~(\ref{short-time-expansion-inversion-coherent-state-2}) holds
remains to be solved in the future.

In this section,
we argue only the short time expansion of $\langle\sigma_{z}(t)\rangle_{\alpha}$
with the counter-rotating terms.
We point out that to examine long time behaviour of the JCM with counter-rotating terms
is very difficult even if we use perturbative techniques.

\section{\label{section-discussion}Discussion}
Turning our eyes towards the graphs shown
in Figs.~\ref{Figure05} and \ref{Figure06},
we observe that the period of the revival of the Rabi oscillations becomes longer
as the temperature rises.
This phenomenon is predicted form an intuitive discussion
in Sec.~\ref{section-thermal-effects-period-qualitative-estimation}.
In Sec.~\ref{section-numerical-calculations},
we confirm this phenomenon (or this expectation)
with numerical calculations based on the third order low-temperature expansion.

Why is the low-temperature expansion in $\theta(\beta)$ given
by Sec.~\ref{section-formulation-perturbetion-theory}
effective for a perturbation theory?
The reason why is as follows:
The thermal coherent state is defined
in Eqs.~(\ref{definition-thermal-coherent-state-2})
and
(\ref{another-form-of-thermal-coherent-state-1}).
This definition is suitable for the low-temperature expansion
because of the Baker-Hausdorff theorem
\cite{Louisell1973}.

\appendix

\section{\label{section-3rd-order-correction-term}The third order correction}
In this section,
we give details of calculations of
$P^{(3)}_{g,1}(t)$ and $P^{(3)}_{g,2}(t)$ defined in Eq.~(\ref{Perturbation-theory-formula-4}).
From Eq.~(\ref{Perturbation-theory-formula-4}),
we can write down the third order correction terms as
\begin{eqnarray}
P^{(3)}_{g,1}(t)
&=&
\sum_{n=0}^{\infty}
g^{(n)}_{1}
\langle\alpha|
\langle\tilde{\alpha}|
[a\tilde{a}-a^{\dagger}\tilde{a}^{\dagger},
[a\tilde{a}-a^{\dagger}\tilde{a}^{\dagger}, \nonumber \\
&&
[a\tilde{a}-a^{\dagger}\tilde{a}^{\dagger},
(a^{\dagger}a+c)^{n}]]]
|\alpha\rangle
|\tilde{\alpha}\rangle, \nonumber \\
P^{(3)}_{g,2}(t)
&=&
\sum_{n=0}^{\infty}
g^{(n)}_{2}
\langle\alpha|
\langle\tilde{\alpha}|
[a\tilde{a}-a^{\dagger}\tilde{a}^{\dagger},
[a\tilde{a}-a^{\dagger}\tilde{a}^{\dagger}, \nonumber \\
&&
[a\tilde{a}-a^{\dagger}\tilde{a}^{\dagger},
(a^{\dagger}a+c+1)^{n}]]]
|\alpha\rangle
|\tilde{\alpha}\rangle.
\label{third-order-perturbation-terms-0}
\end{eqnarray}
As mentioned in Secs.~\ref{section-0th-order-correction-term},
\ref{section-1st-order-correction-term}
and
\ref{section-2nd-order-correction-term},
we can obtain $P^{(3)}_{g,2}(t)$ after the manner of $P^{(3)}_{g,1}(t)$.
Thus, from now on,
we compute
$P^{(3)}_{g,1}(t)$.
For the convenience of calculations carried out in the remains of this section,
according to Eq.~(\ref{commutation-relation-Rn-Sn}),
we divide $P^{(3)}_{g,1}(t)$ into the following two parts:
\begin{eqnarray}
P^{(3)}_{g,1}(t)
&=&
-
\sum_{n=0}^{\infty}
g^{(n)}_{1}
\langle \alpha|\langle\tilde{\alpha}|
[\hat{A},
[\hat{A},
[\hat{A},
\hat{B}^{n}]]]
|\alpha\rangle|\tilde{\alpha}\rangle \nonumber \\
&=&
-
\sum_{n=0}^{\infty}
g^{(n)}_{1}
\langle \alpha|\langle\tilde{\alpha}|
[\hat{A},\hat{R}_{n}]
|\alpha\rangle|\tilde{\alpha}\rangle
-
\sum_{n=0}^{\infty}
g^{(n)}_{1}
\langle \alpha|\langle\tilde{\alpha}|
[\hat{A},\hat{S}_{n}]
|\alpha\rangle|\tilde{\alpha}\rangle.
\label{definition-3rd-order-correction-terms-1}
\end{eqnarray}

First,
we examine the part which includes
$\{[\hat{A},\hat{R}_{n}]\}$ in Eq.~(\ref{definition-3rd-order-correction-terms-1}).
From Eqs.~(\ref{defintion-Rn-Sn-i})
and
(\ref{defintion-Rn-Sn-ii}),
after slightly tough calculations,
we obtain
\begin{eqnarray}
[\hat{A},\hat{R}_{0}]
&=&
0, \nonumber \\
{[}\hat{A},\hat{R}_{1}{]}
&=&
0, \nonumber \\
{[}\hat{A},\hat{R}_{2}{]}
&=&
-2(\hat{R}_{1}+\hat{S}_{1})\hat{C}
-2
{[}\hat{A},\hat{B}{]}
{[}\hat{A},\hat{C}{]}, \nonumber \\
{[}\hat{A},\hat{R}_{3}{]}
&=&
-3(\hat{R}_{2}+\hat{S}_{2})\hat{C}
-3
{[}\hat{A},\hat{B}^{2}{]}
{[}\hat{A},\hat{C}{]}
+
3(\hat{R}_{1}+\hat{S}_{1})\hat{A}, \nonumber \\
{[}\hat{A},\hat{R}_{4}{]}
&=&
-4(\hat{R}_{3}+\hat{S}_{3})\hat{C}
-4
{[}\hat{A},\hat{B}^{3}{]}
{[}\hat{A},\hat{C}{]}
+6(\hat{R}_{2}+\hat{S}_{2})\hat{A}
-4(\hat{R}_{1}+\hat{S}_{1})\hat{C} \nonumber \\
&&
-4
{[}\hat{A},\hat{B}{]}
{[}\hat{A},\hat{C}{]}, \nonumber \\
{[}\hat{A},\hat{R}_{5}{]}
&=&
-5(\hat{R}_{4}+\hat{S}_{4})\hat{C}
-5
{[}\hat{A},\hat{B}^{4}{]}
{[}\hat{A},\hat{C}{]}
+10(\hat{R}_{3}+\hat{S}_{3})\hat{A}
-10(\hat{R}_{2}+\hat{S}_{2})\hat{C} \nonumber \\
&&
-10
{[}\hat{A},\hat{B}^{2}{]}
{[}\hat{A},\hat{C}{]}
+5(\hat{R}_{1}+\hat{S}_{1})\hat{A}, \nonumber \\
{[}\hat{A},\hat{R}_{6}{]}
&=&
-6(\hat{R}_{5}+\hat{S}_{5})\hat{C}
-6
{[}\hat{A},\hat{B}^{5}{]}
{[}\hat{A},\hat{C}{]}
+15(\hat{R}_{4}+\hat{S}_{4})\hat{A}
-20(\hat{R}_{3}+\hat{S}_{3})\hat{C} \nonumber \\
&&
-20
{[}\hat{A},\hat{B}^{3}{]}
{[}\hat{A},\hat{C}{]}
+15(\hat{R}_{2}+\hat{S}_{2})\hat{A}
-6(\hat{R}_{1}+\hat{S}_{1})\hat{C}
-6
{[}\hat{A},\hat{B}{]}
{[}\hat{A},\hat{C}{]}, \nonumber \\
&&....
\end{eqnarray}
Thus,
we can rewrite the part including $\{[\hat{A},\hat{R}_{n}]\}$
in Eq.~(\ref{definition-3rd-order-correction-terms-1})
as
\begin{eqnarray}
\sum_{n=0}^{\infty}
g^{(n)}_{1}
[\hat{A},\hat{R}_{n}]
&=&
-
\sum_{n=0}^{\infty}
\left(
\begin{array}{c}
n+2 \\
1
\end{array}
\right)
g^{(n+2)}_{1}
(\hat{R}_{n+1}+\hat{S}_{n+1})\hat{C} \nonumber \\
&&
-
\sum_{n=0}^{\infty}
\left(
\begin{array}{c}
n+2 \\
1
\end{array}
\right)
g^{(n+2)}_{1}
[\hat{A},\hat{B}^{n+1}][\hat{A},\hat{C}] \nonumber \\
&&
+
\sum_{n=0}^{\infty}
\left(
\begin{array}{c}
n+3 \\
2
\end{array}
\right)
g^{(n+3)}_{1}
(\hat{R}_{n+1}+\hat{S}_{n+1})\hat{A} \nonumber \\
&&
-
\sum_{n=0}^{\infty}
\left(
\begin{array}{c}
n+4 \\
3
\end{array}
\right)
g^{(n+4)}_{1}
(\hat{R}_{n+1}+\hat{S}_{n+1})\hat{C} \nonumber \\
&&
-
\sum_{n=0}^{\infty}
\left(
\begin{array}{c}
n+4 \\
3
\end{array}
\right)
g^{(n+4)}_{1}
[\hat{A},\hat{B}^{n+1}][\hat{A},\hat{C}] \nonumber \\
&&
+
\sum_{n=0}^{\infty}
\left(
\begin{array}{c}
n+5 \\
4
\end{array}
\right)
g^{(n+5)}_{1}
(\hat{R}_{n+1}+\hat{S}_{n+1})\hat{A} \nonumber \\
&&
-
\sum_{n=0}^{\infty}
\left(
\begin{array}{c}
n+6 \\
5
\end{array}
\right)
g^{(n+6)}_{1}
(\hat{R}_{n+1}+\hat{S}_{n+1})\hat{C} \nonumber \\
&&
-
\sum_{n=0}^{\infty}
\left(
\begin{array}{c}
n+6 \\
5
\end{array}
\right)
g^{(n+6)}_{1}
(\hat{R}_{n+1}+\hat{S}_{n+1})[\hat{A},\hat{C}] \nonumber \\
&&
-
....
\label{sum-g-A-R-1}
\end{eqnarray}
In the following paragraphs,
we examine the terms in the right-hand side of Eq.~(\ref{sum-g-A-R-1}),
one by one.

Here,
we think about the first term of the right-hand side of Eq.~(\ref{sum-g-A-R-1}).
At first,
referring to Eq.~(\ref{defintion-Rn-Sn-ii}),
we calculate the part including $\{\hat{R}_{n+1}\}$
in the first term of the right-hand side of Eq.~(\ref{sum-g-A-R-1})
as
\begin{eqnarray}
&&
\sum_{n=0}^{\infty}
\left(
\begin{array}{c}
n+2 \\
1
\end{array}
\right)
g^{(n+2)}_{1}
\hat{R}_{n+1} \nonumber \\
&=&
-
\sum_{n=0}^{\infty}
\left(
\begin{array}{c}
n+2 \\
1
\end{array}
\right)
\left(
\begin{array}{c}
n+3 \\
1
\end{array}
\right)
g^{(n+3)}_{1}
[\hat{A},\hat{B}^{n+1}]\hat{C} \nonumber \\
&&
+
\sum_{n=0}^{\infty}
\left(
\begin{array}{c}
n+3 \\
2
\end{array}
\right)
\left(
\begin{array}{c}
n+4 \\
1
\end{array}
\right)
g^{(n+4)}_{1}
[\hat{A},\hat{B}^{n+1}]\hat{A} \nonumber \\
&&
-
\sum_{n=0}^{\infty}
\left(
\begin{array}{c}
n+4 \\
3
\end{array}
\right)
\left(
\begin{array}{c}
n+5 \\
1
\end{array}
\right)
g^{(n+5)}_{1}
[\hat{A},\hat{B}^{n+1}]\hat{C} \nonumber \\
&&
+
\sum_{n=0}^{\infty}
\left(
\begin{array}{c}
n+5 \\
4
\end{array}
\right)
\left(
\begin{array}{c}
n+6 \\
1
\end{array}
\right)
g^{(n+6)}_{1}
[\hat{A},\hat{B}^{n+1}]\hat{A} \nonumber \\
&&
-
....
\label{summation-gR-1}
\end{eqnarray}
We prepare the following formula:
\begin{eqnarray}
&&
\sum_{n=0}^{\infty}
\left(
\begin{array}{c}
n+m+1 \\
m
\end{array}
\right)
\left(
\begin{array}{c}
n+m+2 \\
1
\end{array}
\right)
g^{(n+m+2)}_{1}
x^{n+1} \nonumber \\
&=&
\frac{1}{m!}
\frac{d^{m}}{dx^{m}}
g'_{1}(x)
-
(m+1)g^{(m+1)}_{1}
\quad\quad
\mbox{for $m=1,2,3,...$}.
\end{eqnarray}
From Eq.~(\ref{commutation-relations-A-Bn-3}) and the above formula,
we can rewrite Eq.~(\ref{summation-gR-1}) as
\begin{eqnarray}
&&
\sum_{n=0}^{\infty}
\left(
\begin{array}{c}
n+2 \\
1
\end{array}
\right)
g^{(n+2)}_{1}
\hat{R}_{n+1} \nonumber \\
&=&
-
\Bigl(
\frac{d}{dx}g'_{1}(x)\bigg|_{x=\hat{B}-1}\hat{\mu}
-
\frac{d}{dx}g'_{1}(x)\bigg|_{x=\hat{B}+1}\hat{\nu}
-
\frac{d}{dx}g'_{1}(x)\bigg|_{x=\hat{B}}\hat{A}
\Bigr)
\hat{C} \nonumber \\
&&
+
\Bigl(
\frac{1}{2!}\frac{d^{2}}{dx^{2}}g'_{1}(x)\bigg|_{x=\hat{B}-1}\hat{\mu}
-
\frac{1}{2!}\frac{d^{2}}{dx^{2}}g'_{1}(x)\bigg|_{x=\hat{B}+1}\hat{\nu}
-
\frac{1}{2!}\frac{d^{2}}{dx^{2}}g'_{1}(x)\bigg|_{x=\hat{B}}\hat{A}
\Bigr)
\hat{A} \nonumber \\
&&
-
\Bigl(
\frac{1}{3!}\frac{d^{3}}{dx^{3}}g'_{1}(x)\bigg|_{x=\hat{B}-1}\hat{\mu}
-
\frac{1}{3!}\frac{d^{3}}{dx^{3}}g'_{1}(x)\bigg|_{x=\hat{B}+1}\hat{\nu}
-
\frac{1}{3!}\frac{d^{3}}{dx^{3}}g'_{1}(x)\bigg|_{x=\hat{B}}\hat{A}
\Bigr)
\hat{C} \nonumber \\
&&
+
\Bigl(
\frac{1}{4!}\frac{d^{4}}{dx^{4}}g'_{1}(x)\bigg|_{x=\hat{B}-1}\hat{\mu}
-
\frac{1}{4!}\frac{d^{4}}{dx^{4}}g'_{1}(x)\bigg|_{x=\hat{B}+1}\hat{\nu}
-
\frac{1}{4!}\frac{d^{4}}{dx^{4}}g'_{1}(x)\bigg|_{x=\hat{B}}\hat{A}
\Bigr)
\hat{A} \nonumber \\
&&
+... \nonumber \\
&=&
\Bigl(
e^{-d/dx}g'_{1}(x)\bigg|_{x=\hat{B}-1}
-
g'_{1}(\hat{B}-1)
-
e^{-d/dx}g'_{1}(x)\bigg|_{x=\hat{B}}
+
g'_{1}(\hat{B})
\Bigr)
\hat{\mu}^{2} \nonumber \\
&&
+
\Bigl(
-
e^{d/dx}g'_{1}(x)\bigg|_{x=\hat{B}-1}
+
g'_{1}(\hat{B}-1)
+
e^{d/dx}g'_{1}(x)\bigg|_{x=\hat{B}}
-
g'_{1}(\hat{B})
\Bigr)
\hat{\mu}\hat{\nu} \nonumber \\
&&
+
\Bigl(
-
e^{-d/dx}g'_{1}(x)\bigg|_{x=\hat{B}+1}
+
g'_{1}(\hat{B}+1)
+
e^{-d/dx}g'_{1}(x)\bigg|_{x=\hat{B}}
-
g'_{1}(\hat{B})
\Bigr)
\hat{\nu}\hat{\mu} \nonumber \\
&&
+
\Bigl(
e^{d/dx}g'_{1}(x)\bigg|_{x=\hat{B}+1}
-
g'_{1}(\hat{B}+1)
-
e^{d/dx}g'_{1}(x)\bigg|_{x=\hat{B}}
+
g'_{1}(\hat{B})
\Bigr)
\hat{\nu}^{2} \nonumber \\
&=&
\Bigl(
g'_{1}(\hat{B}-2)-2g'_{1}(\hat{B}-1)+g'_{1}(\hat{B})
\Bigr)
\hat{\mu}^{2} \nonumber \\
&&
+
\Bigl(
g'_{1}(\hat{B}-1)-2g'_{1}(\hat{B})+g'_{1}(\hat{B}+1)
\Bigr)
\hat{\mu}\hat{\nu} \nonumber \\
&&
+
\Bigl(
g'_{1}(\hat{B}-1)-2g'_{1}(\hat{B})+g'_{1}(\hat{B}+1)
\Bigr)
\hat{\nu}\hat{\mu} \nonumber \\
&&
+
\Bigl(
g'_{1}(\hat{B})-2g'_{1}(\hat{B}+1)+g'_{1}(\hat{B}+2)
\Bigr)
\hat{\nu}^{2}.
\label{summation-results-01}
\end{eqnarray}

Next,
from Eqs.~(\ref{formula-g-polynomial-1}) and (\ref{formula-Sn-1}),
we can compute
the part including $\{\hat{S}_{n+1}\}$
in the first term of the right-hand side of Eq.~(\ref{sum-g-A-R-1})
as
\begin{eqnarray}
&&
\sum_{n=0}^{\infty}
\left(
\begin{array}{c}
n+2 \\
1
\end{array}
\right)
g^{(n+2)}_{1}
\hat{S}_{n+1} \nonumber \\
&=&
-
\sum_{n=0}^{\infty}
\left(
\begin{array}{c}
n+2 \\
1
\end{array}
\right)
g^{(n+2)}_{1}
\Bigl(
(\hat{B}-1)^{n+1}-(\hat{B}+1)^{n+1}
\Bigr)
\hat{D} \nonumber \\
&=&
-
\Bigl(
g'_{1}(\hat{B}-1)-g'_{1}(\hat{B}+1)
\Bigr)
\hat{D}.
\label{summation-results-02}
\end{eqnarray}

From Eqs.~(\ref{commutation-relations-A-Bn-3})
and
(\ref{formula-g-polynomial-1}),
we calculate the second term in the right-hand-side of Eq.~(\ref{sum-g-A-R-1})
as
\begin{eqnarray}
&&
\sum_{n=0}^{\infty}
\left(
\begin{array}{c}
n+2 \\
1
\end{array}
\right)
g^{(n+2)}_{1}
{[}\hat{A},\hat{B}^{n+1}{]} \nonumber \\
&=&
\sum_{n=0}^{\infty}
\left(
\begin{array}{c}
n+2 \\
1
\end{array}
\right)
g^{(n+2)}_{1}
\Bigl(
(\hat{B}-1)^{n+1}\hat{\mu}-(\hat{B}+1)^{n+1}\hat{\nu}-\hat{B}^{n+1}\hat{A}
\Bigr) \nonumber \\
&=&
g'_{1}(\hat{B}-1)\hat{\mu}
-
g'_{1}(\hat{B}+1)\hat{\nu}
-
g'_{1}(\hat{B})\hat{A}.
\label{summation-results-03}
\end{eqnarray}

We consider the third term in the right-hand side of Eq.~(\ref{sum-g-A-R-1})
in the following manner.
At first,
referring to Eq.~(\ref{defintion-Rn-Sn-ii}),
we calculate the part including $\{\hat{R}_{n+1}\}$
in the third term of the right-hand side of Eq.~(\ref{sum-g-A-R-1})
as
\begin{eqnarray}
&&
\sum_{n=0}^{\infty}
\left(
\begin{array}{c}
n+3 \\
2
\end{array}
\right)
g^{(n+3)}_{1}
\hat{R}_{n+1} \nonumber \\
&=&
-
\sum_{n=0}^{\infty}
\left(
\begin{array}{c}
n+2 \\
1
\end{array}
\right)
\left(
\begin{array}{c}
n+4 \\
2
\end{array}
\right)
g^{(n+4)}_{1}
[\hat{A},\hat{B}^{n+1}]\hat{C} \nonumber \\
&&
+
\sum_{n=0}^{\infty}
\left(
\begin{array}{c}
n+3 \\
2
\end{array}
\right)
\left(
\begin{array}{c}
n+5 \\
2
\end{array}
\right)
g^{(n+5)}_{1}
[\hat{A},\hat{B}^{n+1}]\hat{A} \nonumber \\
&&
-
\sum_{n=0}^{\infty}
\left(
\begin{array}{c}
n+4 \\
3
\end{array}
\right)
\left(
\begin{array}{c}
n+6 \\
2
\end{array}
\right)
g^{(n+6)}_{1}
[\hat{A},\hat{B}^{n+1}]\hat{C} \nonumber \\
&&
+
\sum_{n=0}^{\infty}
\left(
\begin{array}{c}
n+5 \\
4
\end{array}
\right)
\left(
\begin{array}{c}
n+7 \\
2
\end{array}
\right)
g^{(n+7)}_{1}
[\hat{A},\hat{B}^{n+1}]\hat{A} \nonumber \\
&&
-
....
\label{summation-gR-2}
\end{eqnarray}
Here,
we prepare the following formula:
\begin{eqnarray}
&&
\sum_{n=0}^{\infty}
\left(
\begin{array}{c}
n+m+1 \\
m
\end{array}
\right)
\left(
\begin{array}{c}
n+m+3 \\
2
\end{array}
\right)
g^{(n+m+3)}_{1}
x^{n+1} \nonumber \\
&=&
\frac{1}{2}
\frac{1}{m!}
\frac{d^{m}}{dx^{m}}
g''_{1}(x)
-
\frac{(m+1)(m+2)}{2}g^{(m+2)}_{1}
\quad\quad
\mbox{for $m=1,2,3,...$}.
\end{eqnarray}
From Eq.~(\ref{commutation-relations-A-Bn-3})
and the above formula,
we can rewrite Eq.~(\ref{summation-gR-2}) as
\begin{eqnarray}
&&
\sum_{n=0}^{\infty}
\left(
\begin{array}{c}
n+3 \\
2
\end{array}
\right)
g^{(n+3)}_{1}
\hat{R}_{n+1} \nonumber \\
&=&
-
\frac{1}{2}
\Bigl(
\frac{d}{dx}g''_{1}(x)\bigg|_{x=\hat{B}-1}\hat{\mu}
-
\frac{d}{dx}g''_{1}(x)\bigg|_{x=\hat{B}+1}\hat{\nu}
-
\frac{d}{dx}g''_{1}(x)\bigg|_{x=\hat{B}}\hat{A}
\Bigr)
\hat{C} \nonumber \\
&&
+
\frac{1}{2}
\Bigl(
\frac{1}{2!}\frac{d^{2}}{dx^{2}}g''_{1}(x)\bigg|_{x=\hat{B}-1}\hat{\mu}
-
\frac{1}{2!}\frac{d^{2}}{dx^{2}}g''_{1}(x)\bigg|_{x=\hat{B}+1}\hat{\nu}
-
\frac{1}{2!}\frac{d^{2}}{dx^{2}}g''_{1}(x)\bigg|_{x=\hat{B}}\hat{A}
\Bigr)
\hat{A} \nonumber \\
&&
-
\frac{1}{2}
\Bigl(
\frac{1}{3!}\frac{d^{3}}{dx^{3}}g''_{1}(x)\bigg|_{x=\hat{B}-1}\hat{\mu}
-
\frac{1}{3!}\frac{d^{3}}{dx^{3}}g''_{1}(x)\bigg|_{x=\hat{B}+1}\hat{\nu}
-
\frac{1}{3!}\frac{d^{3}}{dx^{3}}g''_{1}(x)\bigg|_{x=\hat{B}}\hat{A}
\Bigr)
\hat{C} \nonumber \\
&&
+
\frac{1}{2}
\Bigl(
\frac{1}{4!}\frac{d^{4}}{dx^{4}}g''_{1}(x)\bigg|_{x=\hat{B}-1}\hat{\mu}
-
\frac{1}{4!}\frac{d^{4}}{dx^{4}}g''_{1}(x)\bigg|_{x=\hat{B}+1}\hat{\nu}
-
\frac{1}{4!}\frac{d^{4}}{dx^{4}}g''_{1}(x)\bigg|_{x=\hat{B}}\hat{A}
\Bigr)
\hat{A} \nonumber \\
&&
-... \nonumber \\
&=&
\frac{1}{2}
\Bigl(
e^{-d/dx}g''_{1}(x)\bigg|_{x=\hat{B}-1}
-
g''_{1}(\hat{B}-1)
-
e^{-d/dx}g''_{1}(x)\bigg|_{x=\hat{B}}
+
g''_{1}(\hat{B})
\Bigr)
\hat{\mu}^{2} \nonumber \\
&&
+
\frac{1}{2}
\Bigl(
-
e^{d/dx}g''_{1}(x)\bigg|_{x=\hat{B}-1}
+
g''_{1}(\hat{B}-1)
+
e^{d/dx}g''_{1}(x)\bigg|_{x=\hat{B}}
-
g''_{1}(\hat{B})
\Bigr)
\hat{\mu}\hat{\nu} \nonumber \\
&&
+
\frac{1}{2}
\Bigl(
-
e^{-d/dx}g''_{1}(x)\bigg|_{x=\hat{B}+1}
+
g''_{1}(\hat{B}+1)
+
e^{-d/dx}g''_{1}(x)\bigg|_{x=\hat{B}}
-
g''_{1}(\hat{B})
\Bigr)
\hat{\nu}\hat{\mu} \nonumber \\
&&
+
\frac{1}{2}
\Bigl(
e^{d/dx}g''_{1}(x)\bigg|_{x=\hat{B}+1}
-
g''_{1}(\hat{B}+1)
-
e^{d/dx}g''_{1}(x)\bigg|_{x=\hat{B}}
+
g''_{1}(\hat{B})
\Bigr)
\hat{\nu}^{2} \nonumber \\
&=&
\frac{1}{2}
\Bigl(
g''_{1}(\hat{B}-2)-2g''_{1}(\hat{B}-1)+g''_{1}(\hat{B})
\Bigr)
\hat{\mu}^{2} \nonumber \\
&&
+
\frac{1}{2}
\Bigl(
g''_{1}(\hat{B}-1)-2g''_{1}(\hat{B})+g''_{1}(\hat{B}+1)
\Bigr)
\hat{\mu}\hat{\nu} \nonumber \\
&&
+
\frac{1}{2}
\Bigl(
g''_{1}(\hat{B}-1)-2g''_{1}(\hat{B})+g''_{1}(\hat{B}+1)
\Bigr)
\hat{\nu}\hat{\mu} \nonumber \\
&&
+
\frac{1}{2}
\Bigl(
g''_{1}(\hat{B})-2g''_{1}(\hat{B}+1)+g''_{1}(\hat{B}+2)
\Bigr)
\hat{\nu}^{2}.
\label{summation-results-04}
\end{eqnarray}

Next,
using Eqs.~(\ref{formula-g-polynomial-1}) and (\ref{formula-Sn-1}),
we compute the part including $\{\hat{S}_{n+1}\}$
in the third term of the right-hand side of Eq.~(\ref{sum-g-A-R-1})
as
\begin{eqnarray}
&&
\sum_{n=0}^{\infty}
\left(
\begin{array}{c}
n+3 \\
2
\end{array}
\right)
g^{(n+3)}_{1}
\hat{S}_{n+1} \nonumber \\
&=&
-
\sum_{n=0}^{\infty}
\left(
\begin{array}{c}
n+3 \\
2
\end{array}
\right)
g^{(n+3)}_{1}
\Bigl(
(\hat{B}-1)^{n+1}-(\hat{B}+1)^{n+1}
\Bigr)
\hat{D} \nonumber \\
&=&
-
\frac{1}{2}
\Bigl(
g''_{1}(\hat{B}-1)
-
g''_{1}(\hat{B}+1)
\Bigr)
\hat{D}.
\label{summation-results-05}
\end{eqnarray}

Using Eqs.~(\ref{commutation-relations-A-Bn-3})
and
(\ref{formula-g-polynomial-1}),
we compute the fifth term in the right-hand side of Eq.~(\ref{sum-g-A-R-1})
as
\begin{eqnarray}
&&
\sum_{n=0}^{\infty}
\left(
\begin{array}{c}
n+4 \\
3
\end{array}
\right)
g^{(n+4)}_{1}
[\hat{A},\hat{B}^{n+1}] \nonumber \\
&=&
\sum_{n=0}^{\infty}
\left(
\begin{array}{c}
n+4 \\
3
\end{array}
\right)
g^{(n+4)}_{1}
\Bigl(
(\hat{B}-1)^{n+1}\hat{\mu}-(\hat{B}+1)^{n+1}\hat{\nu}-\hat{B}^{n+1}\hat{A}
\Bigr) \nonumber \\
&=&
\frac{1}{3!}
\Bigl(
\frac{d^{3}}{dx^{3}}g_{1}(x)\bigg|_{x=\hat{B}-1}\hat{\mu}
-
\frac{d^{3}}{dx^{3}}g_{1}(x)\bigg|_{x=\hat{B}+1}\hat{\nu}
-
\frac{d^{3}}{dx^{3}}g_{1}(x)\bigg|_{x=\hat{B}}\hat{A}
\Bigr).
\label{summation-results-06}
\end{eqnarray}

Second,
we examine the part which includes
$\{[\hat{A},\hat{S}_{n}]\}$ in Eq.~(\ref{definition-3rd-order-correction-terms-1}).
From Eq.~(\ref{formula-Sn-1}),
after slightly tough calculations,
we obtain
\begin{eqnarray}
{[}\hat{A},\hat{S}_{n}{]}
&=&
2
\left(
\begin{array}{c}
n \\
1
\end{array}
\right)
[\hat{A},\hat{B}^{n-1}]\hat{D}
+
2
\left(
\begin{array}{c}
n \\
3
\end{array}
\right)
[\hat{A},\hat{B}^{n-3}]\hat{D}
+
...
+
2
\left(
\begin{array}{c}
n \\
n-1
\end{array}
\right)
[\hat{A},\hat{B}]\hat{D} \nonumber \\
&&
+
\Bigl(
(\hat{B}+1)^{n}-(\hat{B}-1)^{n}
\Bigr)
[\hat{A},\hat{D}] \nonumber \\
&&
\mbox{for $n=0,2,4,...$ (even)}, \nonumber \\
{[}\hat{A},\hat{S}_{n}{]}
&=&
2
\left(
\begin{array}{c}
n \\
1
\end{array}
\right)
[\hat{A},\hat{B}^{n-1}]\hat{D}
+
2
\left(
\begin{array}{c}
n \\
3
\end{array}
\right)
[\hat{A},\hat{B}^{n-3}]\hat{D}
+
...
+
2
\left(
\begin{array}{c}
n \\
n-2
\end{array}
\right)
[\hat{A},\hat{B}^{2}]\hat{D} \nonumber \\
&&
+
\Bigl(
(\hat{B}+1)^{n}-(\hat{B}-1)^{n}
\Bigr)
[\hat{A},\hat{D}] \nonumber \\
&&
\mbox{for $n=1,3,5,...$ (odd)}.
\label{commutation-relation-A-Sn}
\end{eqnarray}

Here, we prepare the following formulae:
\begin{eqnarray}
&&
\sum_{n=0}^{\infty}
g^{(n+2m-1)}_{1}
\left(
\begin{array}{c}
n+2m-1 \\
2m-1
\end{array}
\right)
x^{n} \nonumber \\
&=&
\frac{1}{(2m-1)!}
\frac{d^{2m-1}}{dx^{2m-1}}g_{1}(x)
\quad\quad
\mbox{for $m=1,2,3,...$},
\label{summation-power-series-formula-1}
\end{eqnarray}
\begin{equation}
\frac{d}{dx}
+
\frac{1}{3!}
\frac{d^{3}}{dx^{3}}
+
\frac{1}{5!}
\frac{d^{5}}{dx^{5}}
+...
=
\frac{1}{2}(e^{d/dx}-e^{-d/dx}).
\label{differential-operator-series-formula-1}
\end{equation}
From Eqs.~(\ref{commutation-relations-A-Bn-3}),
(\ref{commutation-relation-A-Sn}),
(\ref{summation-power-series-formula-1})
and (\ref{differential-operator-series-formula-1}),
we obtain
\begin{eqnarray}
&&
\sum_{n=0}^{\infty}
g^{(n)}_{1}[\hat{A},\hat{S}_{n}] \nonumber \\
&=&
\sum_{n=0}^{\infty}
g^{(n)}_{1}
\Bigl(
(\hat{B}+1)^{n}-(\hat{B}-1)^{n}
\Bigr)
[\hat{A},\hat{D}] \nonumber \\
&&
+
2
\sum_{n=0}^{\infty}
g^{(n)}_{1}
\left(
\begin{array}{c}
n+1 \\
1
\end{array}
\right)
[\hat{A},\hat{B}^{n}]\hat{D} \nonumber \\
&&
+
2
\sum_{n=0}^{\infty}
g^{(n+3)}_{1}
\left(
\begin{array}{c}
n+3 \\
3
\end{array}
\right)
[\hat{A},\hat{B}^{n}]\hat{D} \nonumber \\
&&
+
2
\sum_{n=0}^{\infty}
g^{(n+5)}_{1}
\left(
\begin{array}{c}
n+5 \\
5
\end{array}
\right)
[\hat{A},\hat{B}^{n}]\hat{D} \nonumber \\
&&
+... \nonumber \\
&=&
\Bigl(
g_{1}(\hat{B}+1)-g_{1}(\hat{B}-1)
\Bigr)
[\hat{A},\hat{D}] \nonumber \\
&&
+
2
\Bigl(
\frac{d}{dx}g_{1}(x)\bigg|_{x=\hat{B}-1}\hat{\mu}
-
\frac{d}{dx}g_{1}(x)\bigg|_{x=\hat{B}+1}\hat{\nu}
-
\frac{d}{dx}g_{1}(x)\bigg|_{x=\hat{B}}\hat{A}
\Bigr)
\hat{D} \nonumber \\
&&
+
2
\Bigl(
\frac{1}{3!}
\frac{d^{3}}{dx^{3}}g_{1}(x)\bigg|_{x=\hat{B}-1}\hat{\mu}
-
\frac{1}{3!}
\frac{d^{3}}{dx^{3}}g_{1}(x)\bigg|_{x=\hat{B}+1}\hat{\nu}
-
\frac{1}{3!}
\frac{d^{3}}{dx^{3}}g_{1}(x)\bigg|_{x=\hat{B}}\hat{A}
\Bigr)
\hat{D} \nonumber \\
&&
+
2
\Bigl(
\frac{1}{5!}
\frac{d^{5}}{dx^{5}}g_{1}(x)\bigg|_{x=\hat{B}-1}\hat{\mu}
-
\frac{1}{5!}
\frac{d^{5}}{dx^{5}}g_{1}(x)\bigg|_{x=\hat{B}+1}\hat{\nu}
-
\frac{1}{5!}
\frac{d^{5}}{dx^{5}}g_{1}(x)\bigg|_{x=\hat{B}}\hat{A}
\Bigr)
\hat{D} \nonumber \\
&&
+... \nonumber \\
&=&
\Bigl(
g_{1}(\hat{B}+1)-g_{1}(\hat{B}-1)
\Bigr)
[\hat{A},\hat{D}] \nonumber \\
&&
+
\Bigl(
(e^{d/dx}-e^{-d/dx})g_{1}(x)\bigg|_{x=\hat{B}-1}\hat{\mu}
-
(e^{d/dx}-e^{-d/dx})g_{1}(x)\bigg|_{x=\hat{B}+1}\hat{\nu} \nonumber \\
&&
-
(e^{d/dx}-e^{-d/dx})g_{1}(x)\bigg|_{x=\hat{B}}\hat{A}
\Bigr)
\hat{D} \nonumber \\
&=&
\Bigl(
g_{1}(\hat{B}+1)-g_{1}(\hat{B}-1)
\Bigr)
[\hat{A},\hat{D}] \nonumber \\
&&
+
\Bigl(
(g_{1}(\hat{B})-g_{1}(\hat{B}-2))\hat{\mu}
-
(g_{1}(\hat{B+2})-g_{1}(\hat{B}))\hat{\nu} \nonumber \\
&&
-
(g_{1}(\hat{B}+1)-g_{1}(\hat{B}-1))\hat{A}
\Bigr)
\hat{D}.
\label{summation-results-07}
\end{eqnarray}

Putting together Eqs.~(\ref{sum-g-A-R-1}),
(\ref{summation-results-01}),
(\ref{summation-results-02}),
(\ref{summation-results-03}),
(\ref{summation-results-04}),
(\ref{summation-results-05}),
(\ref{summation-results-06})
and
(\ref{summation-results-07}),
we obtain
\begin{eqnarray}
&&
\sum_{n=0}^{\infty}
g^{(n)}_{1}
(
[\hat{A},\hat{R}_{n}]
+
[\hat{A},\hat{S}_{n}]
) \nonumber \\
&=&
\sum_{n=1}^{\infty}
\frac{(-1)^{n}}{n!}
\Bigl(
\frac{d^{n}}{dx^{n}}g_{1}(x)\bigg|_{x=\hat{B}-2}
-
2
\frac{d^{n}}{dx^{n}}g_{1}(x)\bigg|_{x=\hat{B}-1}
+
\frac{d^{n}}{dx^{n}}g_{1}(x)\bigg|_{x=\hat{B}}
\Bigr)
\hat{\mu}^{3} \nonumber \\
&&
-
\sum_{n=1}^{\infty}
\frac{1}{n!}
\Bigl(
\frac{d^{n}}{dx^{n}}g_{1}(x)\bigg|_{x=\hat{B}-2}
-
2
\frac{d^{n}}{dx^{n}}g_{1}(x)\bigg|_{x=\hat{B}-1}
+
\frac{d^{n}}{dx^{n}}g_{1}(x)\bigg|_{x=\hat{B}}
\Bigr)
\hat{\mu}^{2}\hat{\nu} \nonumber \\
&&
+
\sum_{n=1}^{\infty}
\frac{(-1)^{n}}{n!}
\Bigl(
\frac{d^{n}}{dx^{n}}g_{1}(x)\bigg|_{x=\hat{B}-1}
-
2
\frac{d^{n}}{dx^{n}}g_{1}(x)\bigg|_{x=\hat{B}}
+
\frac{d^{n}}{dx^{n}}g_{1}(x)\bigg|_{x=\hat{B}+1}
\Bigr)
\hat{\mu}\hat{\nu}\hat{\mu} \nonumber \\
&&
-
\sum_{n=1}^{\infty}
\frac{1}{n!}
\Bigl(
\frac{d^{n}}{dx^{n}}g_{1}(x)\bigg|_{x=\hat{B}-1}
-
2
\frac{d^{n}}{dx^{n}}g_{1}(x)\bigg|_{x=\hat{B}}
+
\frac{d^{n}}{dx^{n}}g_{1}(x)\bigg|_{x=\hat{B}+1}
\Bigr)
\hat{\mu}\hat{\nu}^{2} \nonumber \\
&&
+
\sum_{n=1}^{\infty}
\frac{(-1)^{n}}{n!}
\Bigl(
\frac{d^{n}}{dx^{n}}g_{1}(x)\bigg|_{x=\hat{B}-1}
-
2
\frac{d^{n}}{dx^{n}}g_{1}(x)\bigg|_{x=\hat{B}}
+
\frac{d^{n}}{dx^{n}}g_{1}(x)\bigg|_{x=\hat{B}+1}
\Bigr)
\hat{\nu}\hat{\mu}^{2} \nonumber \\
&&
-
\sum_{n=1}^{\infty}
\frac{1}{n!}
\Bigl(
\frac{d^{n}}{dx^{n}}g_{1}(x)\bigg|_{x=\hat{B}-1}
-
2
\frac{d^{n}}{dx^{n}}g_{1}(x)\bigg|_{x=\hat{B}}
+
\frac{d^{n}}{dx^{n}}g_{1}(x)\bigg|_{x=\hat{B}+1}
\Bigr)
\hat{\nu}\hat{\mu}\hat{\nu} \nonumber \\
&&
+
\sum_{n=1}^{\infty}
\frac{(-1)^{n}}{n!}
\Bigl(
\frac{d^{n}}{dx^{n}}g_{1}(x)\bigg|_{x=\hat{B}}
-
2
\frac{d^{n}}{dx^{n}}g_{1}(x)\bigg|_{x=\hat{B}+1}
+
\frac{d^{n}}{dx^{n}}g_{1}(x)\bigg|_{x=\hat{B}+2}
\Bigr)
\hat{\nu}^{2}\hat{\mu} \nonumber \\
&&
-
\sum_{n=1}^{\infty}
\frac{1}{n!}
\Bigl(
\frac{d^{n}}{dx^{n}}g_{1}(x)\bigg|_{x=\hat{B}}
-
2
\frac{d^{n}}{dx^{n}}g_{1}(x)\bigg|_{x=\hat{B}+1}
+
\frac{d^{n}}{dx^{n}}g_{1}(x)\bigg|_{x=\hat{B}+2}
\Bigr)
\hat{\nu}^{3} \nonumber \\
&&
-
\sum_{n=1}^{\infty}
\frac{(-1)^{n}}{n!}
\Bigl(
\frac{d^{n}}{dx^{n}}g_{1}(x)\bigg|_{x=\hat{B}-1}
-
\frac{d^{n}}{dx^{n}}g_{1}(x)\bigg|_{x=\hat{B}+1}
\Bigr)
\hat{D}\hat{\mu} \nonumber \\
&&
+
\sum_{n=1}^{\infty}
\frac{1}{n!}
\Bigl(
\frac{d^{n}}{dx^{n}}g_{1}(x)\bigg|_{x=\hat{B}-1}
-
\frac{d^{n}}{dx^{n}}g_{1}(x)\bigg|_{x=\hat{B}+1}
\Bigr)
\hat{D}\hat{\nu} \nonumber \\
&&
-
\Bigl(
\frac{d}{dx}g_{1}(x)\bigg|_{x=\hat{B}-1}\hat{\mu}
-
\frac{d}{dx}g_{1}(x)\bigg|_{x=\hat{B}+1}\hat{\nu}
-
\frac{d}{dx}g_{1}(x)\bigg|_{x=\hat{B}}\hat{A} \nonumber \\
&&
+
\frac{1}{3!}
\frac{d^{3}}{dx^{3}}g_{1}(x)\bigg|_{x=\hat{B}-1}\hat{\mu}
-
\frac{1}{3!}
\frac{d^{3}}{dx^{3}}g_{1}(x)\bigg|_{x=\hat{B}+1}\hat{\nu}
-
\frac{1}{3!}
\frac{d^{3}}{dx^{3}}g_{1}(x)\bigg|_{x=\hat{B}}\hat{A} \nonumber \\
&&
+
\frac{1}{5!}
\frac{d^{5}}{dx^{5}}g_{1}(x)\bigg|_{x=\hat{B}-1}\hat{\mu}
-
\frac{1}{5!}
\frac{d^{5}}{dx^{5}}g_{1}(x)\bigg|_{x=\hat{B}+1}\hat{\nu}
-
\frac{1}{5!}
\frac{d^{5}}{dx^{5}}g_{1}(x)\bigg|_{x=\hat{B}}\hat{A} \nonumber \\
&&
+...
\Bigr)
[\hat{A},\hat{C}] \nonumber \\
&&
+
\Bigl(
g_{1}(\hat{B}+1)-g_{1}(\hat{B}-1)
\Bigr)
[\hat{A},\hat{D}] \nonumber \\
&&
+
[
\Bigl(
g_{1}(\hat{B})-g_{1}(\hat{B}-2)
\Bigr)
\hat{\mu}
-
\Bigl(
g_{1}(\hat{B}+2)-g_{1}(\hat{B})
\Bigr)
\hat{\nu} \nonumber \\
&&
-
\Bigl(
g_{1}(\hat{B}+1)-g_{1}(\hat{B}-1)
\Bigr)
\hat{A}
]
\hat{D} \nonumber \\
&=&
\Bigl(
e^{-d/dx}g_{1}(\hat{B}-2)-g_{1}(\hat{B}-2)
-
2e^{-d/dx}g_{1}(\hat{B}-1)+2g_{1}(\hat{B}-1) \nonumber \\
&&
+
e^{-d/dx}g_{1}(\hat{B})-g_{1}(\hat{B})
\Bigr)
\hat{\mu}^{3} \nonumber \\
&&
+
\Bigl(
-e^{d/dx}g_{1}(\hat{B}-2)+g_{1}(\hat{B}-2)
+
2e^{d/dx}g_{1}(\hat{B}-1)-2g_{1}(\hat{B}-1) \nonumber \\
&&
-
e^{d/dx}g_{1}(\hat{B})+g_{1}(\hat{B})
\Bigr)
\hat{\mu}^{2}\hat{\nu} \nonumber \\
&&
+
\Bigl(
e^{-d/dx}g_{1}(\hat{B}-1)-g_{1}(\hat{B}-1)
-
2e^{-d/dx}g_{1}(\hat{B})+2g_{1}(\hat{B}) \nonumber \\
&&
+
e^{-d/dx}g_{1}(\hat{B}+1)-g_{1}(\hat{B}+1)
\Bigr)
\hat{\mu}\hat{\nu}\hat{\mu} \nonumber \\
&&
+
\Bigl(
-e^{d/dx}g_{1}(\hat{B}-1)+g_{1}(\hat{B}-1)
+
2e^{d/dx}g_{1}(\hat{B})-2g_{1}(\hat{B}) \nonumber \\
&&
-
e^{d/dx}g_{1}(\hat{B}+1)+g_{1}(\hat{B}+1)
\Bigr)
\hat{\mu}\hat{\nu}^{2} \nonumber \\
&&
+
\Bigl(
e^{-d/dx}g_{1}(\hat{B}-1)-g_{1}(\hat{B}-1)
-
2e^{-d/dx}g_{1}(\hat{B})+2g_{1}(\hat{B}) \nonumber \\
&&
+
e^{-d/dx}g_{1}(\hat{B}+1)-g_{1}(\hat{B}+1)
\Bigr)
\hat{\nu}\hat{\mu}^{2} \nonumber \\
&&
+
\Bigl(
-e^{d/dx}g_{1}(\hat{B}-1)+g_{1}(\hat{B}-1)
+
2e^{d/dx}g_{1}(\hat{B})-2g_{1}(\hat{B}) \nonumber \\
&&
-
e^{d/dx}g_{1}(\hat{B}+1)+g_{1}(\hat{B}+1)
\Bigr)
\hat{\nu}\hat{\mu}\hat{\nu} \nonumber \\
&&
+
\Bigl(
e^{-d/dx}g_{1}(\hat{B})-g_{1}(\hat{B})
-
2e^{-d/dx}g_{1}(\hat{B}+1)+2g_{1}(\hat{B}+1) \nonumber \\
&&
+
e^{-d/dx}g_{1}(\hat{B}+2)-g_{1}(\hat{B}+2)
\Bigr)
\hat{\nu}^{2}\hat{\mu} \nonumber \\
&&
+
\Bigl(
-e^{d/dx}g_{1}(\hat{B})+g_{1}(\hat{B})
+
2e^{d/dx}g_{1}(\hat{B}+1)-2g_{1}(\hat{B}+1) \nonumber \\
&&
-
e^{d/dx}g_{1}(\hat{B}+2)+g_{1}(\hat{B}+2)
\Bigr)
\hat{\nu}^{3} \nonumber \\
&&
+
\Bigl(
-e^{-d/dx}g_{1}(\hat{B}-1)+g_{1}(\hat{B}-1)
+
e^{-d/dx}g_{1}(\hat{B}+1)-g_{1}(\hat{B}+1)
\Bigr)
\hat{D}\hat{\mu} \nonumber \\
&&
+
\Bigl(
e^{d/dx}g_{1}(\hat{B}-1)-g_{1}(\hat{B}-1)
-
e^{d/dx}g_{1}(\hat{B}+1)+g_{1}(\hat{B}+1)
\Bigr)
\hat{D}\hat{\nu} \nonumber \\
&&
-
\Bigl(
\frac{1}{2}(e^{d/dx}-e^{-d/dx})g_{1}(\hat{B}-1)\hat{\mu}
-
\frac{1}{2}(e^{d/dx}-e^{-d/dx})g_{1}(\hat{B}+1)\hat{\nu} \nonumber \\
&&
-
\frac{1}{2}(e^{d/dx}-e^{-d/dx})g_{1}(\hat{B})\hat{A}
\Bigr)
[\hat{A},\hat{C}] \nonumber \\
&&
+
\Bigl(
g_{1}(\hat{B}+1)-g_{1}(\hat{B}-1)
\Bigr)
[\hat{A},\hat{D}] \nonumber \\
&&
+
[
\Bigl(
g_{1}(\hat{B})-g_{1}(\hat{B}-2)
\Bigr)
\hat{\mu}
-
\Bigl(
g_{1}(\hat{B}+2)-g_{1}(\hat{B})
\Bigr)
\hat{\nu} \nonumber \\
&&
-
\Bigl(
g_{1}(\hat{B}+1)-g_{1}(\hat{B}-1)
\Bigr)
\hat{A}
]
\hat{D} \nonumber \\
&=&
\Bigl(
-g_{1}(\hat{B})+3g_{1}(\hat{B}-1)-3g_{1}(\hat{B}-2)+g_{1}(\hat{B}-3)
\Bigr)
\hat{\mu}^{3} \nonumber \\
&&
+
\Bigl(
-g_{1}(\hat{B}+1)+3g_{1}(\hat{B})-3g_{1}(\hat{B}-1)+g_{1}(\hat{B}-2)
\Bigr)
\hat{\mu}^{2}\hat{\nu} \nonumber \\
&&
+
\Bigl(
-g_{1}(\hat{B}+1)+3g_{1}(\hat{B})-3g_{1}(\hat{B}-1)+g_{1}(\hat{B}-2)
\Bigr)
\hat{\mu}\hat{\nu}\hat{\mu} \nonumber \\
&&
+
\Bigl(
-g_{1}(\hat{B}+2)+3g_{1}(\hat{B}+1)-3g_{1}(\hat{B})+g_{1}(\hat{B}-1)
\Bigr)
\hat{\mu}\hat{\nu}^{2} \nonumber \\
&&
+
\Bigl(
-g_{1}(\hat{B}+1)+3g_{1}(\hat{B})-3g_{1}(\hat{B}-1)+g_{1}(\hat{B}-2)
\Bigr)
\hat{\nu}\hat{\mu}^{2} \nonumber \\
&&
+
\Bigl(
-g_{1}(\hat{B}+2)+3g_{1}(\hat{B}+1)-3g_{1}(\hat{B})+g_{1}(\hat{B}-1)
\Bigr)
\hat{\nu}\hat{\mu}\hat{\nu} \nonumber \\
&&
+
\Bigl(
-g_{1}(\hat{B}+2)+3g_{1}(\hat{B}+1)-3g_{1}(\hat{B})+g_{1}(\hat{B}-1)
\Bigr)
\hat{\nu}^{2}\hat{\mu} \nonumber \\
&&
+
\Bigl(
-g_{1}(\hat{B}+3)+3g_{1}(\hat{B}+2)-3g_{1}(\hat{B}+1)+g_{1}(\hat{B})
\Bigr)
\hat{\nu}^{3} \nonumber \\
&&
+
\Bigl(
-g_{1}(\hat{B}-2)+g_{1}(\hat{B})+g_{1}(\hat{B}-1)-g_{1}(\hat{B}+1)
\Bigr)
\hat{D}\hat{\mu} \nonumber \\
&&
+
\Bigl(
g_{1}(\hat{B})-g_{1}(\hat{B}+2)-g_{1}(\hat{B}-1)+g_{1}(\hat{B}+1)
\Bigr)
\hat{D}\hat{\nu} \nonumber \\
&&
+
2
\Bigl(
g_{1}(\hat{B})-g_{1}(\hat{B}-2)
\Bigr)
\hat{\mu}\hat{D} \nonumber \\
&&
-
2
\Bigl(
g_{1}(\hat{B}+2)-g_{1}(\hat{B})
\Bigr)
\hat{\nu}\hat{D} \nonumber \\
&&
-
\Bigl(
g_{1}(\hat{B}+1)-g_{1}(\hat{B}-1)
\Bigr)
\hat{D}\hat{A} \nonumber \\
&&
-
\Bigl(
g_{1}(\hat{B}+1)-g_{1}(\hat{B}-1)
\Bigr)
\hat{A}\hat{D}.
\label{3rd-order-correction-terms-2}
\end{eqnarray}

Here,
to compute the right-hand side of Eq.~(\ref{3rd-order-correction-terms-2}),
we arrange $\hat{\mu}$ in the left side of the product of operators
and $\hat{\nu}$ in the right side of the product of operators.
For the arrangement of operators,
we carry out the following calculations:
\begin{eqnarray}
\hat{B}^{n}\hat{\mu}^{3}
&=&
\hat{\mu}^{3}(\hat{B}+3)^{n}, \nonumber \\
\hat{B}^{n}\hat{\mu}^{2}\hat{\nu}
&=&
\hat{\mu}^{2}(\hat{B}+2)^{n}\hat{\nu}, \nonumber \\
\hat{B}^{n}\hat{\mu}\hat{\nu}\hat{\mu}
&=&
\hat{\mu}^{2}(\hat{B}+2)^{n}\hat{\nu}
+
\hat{\mu}(\hat{B}+1)^{n}\hat{D}, \nonumber \\
\hat{B}^{n}\hat{\mu}\hat{\nu}^{2}
&=&
\hat{\mu}(\hat{B}+1)^{n}\hat{\nu}^{2}, \nonumber \\
\hat{B}^{n}\hat{\nu}\hat{\mu}^{2}
&=&
\hat{\mu}^{2}(\hat{B}+2)^{n}\hat{\nu}
+
2\hat{\mu}(\hat{B}+1)^{n}(\hat{D}+1), \nonumber \\
\hat{B}^{n}\hat{\nu}\hat{\mu}\hat{\nu}
&=&
\hat{\mu}(\hat{B}+1)^{n}\hat{\nu}^{2}
+
\hat{B}^{n}\hat{D}\hat{\nu}, \nonumber \\
\hat{B}^{n}\hat{\nu}^{2}\hat{\mu}
&=&
\hat{\mu}(\hat{B}+1)^{n}\hat{\nu}^{2}
+
2\hat{B}^{n}(\hat{D}+1)\hat{\nu}, \nonumber \\
\hat{B}^{n}\hat{D}\hat{\mu}
&=&
\hat{\mu}(\hat{B}+1)^{n}(\hat{D}+2), \nonumber \\
\hat{B}^{n}\hat{\mu}
&=&
\hat{\mu}(\hat{B}+1)^{n}, \nonumber \\
\hat{B}^{n}\hat{A}\hat{D}
&=&
\hat{\mu}(\hat{B}+1)^{n}\hat{D}
-
\hat{B}^{n}(\hat{D}+2)\hat{\nu}, \nonumber \\
\hat{B}^{n}\hat{D}\hat{A}
&=&
\hat{\mu}(\hat{B}+1)^{n}(\hat{D}+2)
-
\hat{B}^{n}\hat{D}\hat{\nu} \nonumber \\
&&
\mbox{for $n=1,2,3,...$},
\label{mu-nu-arrangement-formulae-1}
\end{eqnarray}
\begin{equation}
\hat{\nu}\hat{D}
=
(\hat{D}+2)\hat{\nu}.
\label{nu-D-arrangement-formula}
\end{equation}

With making use of Eqs.~(\ref{g-B-D-foemula})
and (\ref{definition-3rd-order-correction-terms-1}),
substitution of Eqs.~(\ref{mu-nu-arrangement-formulae-1})
and (\ref{nu-D-arrangement-formula})
into Eq.~(\ref{3rd-order-correction-terms-2})
yields
\begin{eqnarray}
P^{(3)}_{g,1}(t)
&=&
-
\langle\alpha|\langle\tilde{\alpha}|
\sum_{n=0}^{\infty}
g_{1}^{(n)}
([\hat{A},\hat{R}_{n}]+[\hat{A},\hat{S}_{n}])
|\alpha\rangle|\tilde{\alpha}\rangle \nonumber \\
&=&
-
8\alpha^{6}
\langle\alpha|\langle\tilde{\alpha}|
[
-g_{1}(\hat{B}+3)
+3g_{1}(\hat{B}+2)
-3g_{1}(\hat{B}+1)
+g_{1}(\hat{B})
]
|\alpha\rangle|\tilde{\alpha}\rangle \nonumber \\
&&
-
2\alpha^{2}
\langle\alpha|\langle\tilde{\alpha}|
[
-g_{1}(\hat{B}+2)
+3g_{1}(\hat{B}+1)
-3g_{1}(\hat{B})
+g_{1}(\hat{B}-1)
]
\hat{D}
|\alpha\rangle|\tilde{\alpha}\rangle \nonumber \\
&&
-
4\alpha^{2}
\langle\alpha|\langle\tilde{\alpha}|
[
-g_{1}(\hat{B}+2)
+3g_{1}(\hat{B}+1)
-3g_{1}(\hat{B})
+g_{1}(\hat{B}-1)
]
(\hat{D}+1)
|\alpha\rangle|\tilde{\alpha}\rangle \nonumber \\
&&
-
\alpha^{2}
\langle\alpha|\langle\tilde{\alpha}|
[
-g_{1}(\hat{B}-1)
+g_{1}(\hat{B}+1)
+g_{1}(\hat{B})
-g_{1}(\hat{B}+2)
]
(\hat{D}+2)
|\alpha\rangle|\tilde{\alpha}\rangle \nonumber \\
&&
-
\alpha^{2}
\langle\alpha|\langle\tilde{\alpha}|
[
g_{1}(\hat{B})
-g_{1}(\hat{B}+2)
-g_{1}(\hat{B}-1)
+g_{1}(\hat{B}+1)
]
\hat{D}
|\alpha\rangle|\tilde{\alpha}\rangle \nonumber \\
&&
-2
\alpha^{2}
\langle\alpha|\langle\tilde{\alpha}|
[
g_{1}(\hat{B}+1)
-g_{1}(\hat{B}-1)
]
\hat{D}
|\alpha\rangle|\tilde{\alpha}\rangle \nonumber \\
&&
+2
\alpha^{2}
\langle\alpha|\langle\tilde{\alpha}|
[
g_{1}(\hat{B}+2)
-g_{1}(\hat{B})
]
(\hat{D}+2)
|\alpha\rangle|\tilde{\alpha}\rangle \nonumber \\
&&
+
\alpha^{2}
\langle\alpha|\langle\tilde{\alpha}|
[
g_{1}(\hat{B}+2)
-g_{1}(\hat{B})
]
\hat{D}
|\alpha\rangle|\tilde{\alpha}\rangle \nonumber \\
&&
-
\alpha^{2}
\langle\alpha|\langle\tilde{\alpha}|
[
g_{1}(\hat{B}+1)
-g_{1}(\hat{B}-1)
]
(\hat{D}+2)
|\alpha\rangle|\tilde{\alpha}\rangle \nonumber \\
&&
+
\alpha^{2}
\langle\alpha|\langle\tilde{\alpha}|
[
g_{1}(\hat{B}+2)
-g_{1}(\hat{B})
]
(\hat{D}+2)
|\alpha\rangle|\tilde{\alpha}\rangle \nonumber \\
&&
-
\alpha^{2}
\langle\alpha|\langle\tilde{\alpha}|
[
g_{1}(\hat{B}+1)
-g_{1}(\hat{B}-1)
]
\hat{D}
|\alpha\rangle|\tilde{\alpha}\rangle \nonumber \\
&=&
4\alpha^{4}(2\alpha^{2}+3)
Q_{1}^{(3)}(t) \nonumber \\
&&
-
12\alpha^{2}(2\alpha^{4}+\alpha^{2}-2)
Q_{1}^{(2)}(t) \nonumber \\
&&
+
4\alpha^{2}(6\alpha^{4}-3\alpha^{2}-10)
Q_{1}^{(1)}(t) \nonumber \\
&&
-
4\alpha^{2}(2\alpha^{4}-3\alpha^{2}-4)
Q_{1}^{(0)}(t).
\label{3rd-order-perturbation-term-g1}
\end{eqnarray}
Similarly,
we obtain $P^{(3)}_{g,2}(t)$ as
\begin{eqnarray}
P^{(3)}_{g,2}(t)
&=&
4\alpha^{4}(2\alpha^{2}+3)
Q_{2}^{(3)}(t) \nonumber \\
&&
-
12\alpha^{2}(2\alpha^{4}+\alpha^{2}-2)
Q_{2}^{(2)}(t) \nonumber \\
&&
+
4\alpha^{2}(6\alpha^{4}-3\alpha^{2}-10)
Q_{2}^{(1)}(t) \nonumber \\
&&
-
4\alpha^{2}(2\alpha^{4}-3\alpha^{2}-4)
Q_{2}^{(0)}(t).
\label{3rd-order-perturbation-term-g2}
\end{eqnarray}

Here, we pay attention to the following facts:
In Eqs.~(\ref{3rd-order-perturbation-term-g1}) and (\ref{3rd-order-perturbation-term-g2}),
replacing
$Q_{1}^{(n)}(t)$ and $Q_{2}^{(n)}(t)$ for $n\in\{0,1,2,3\}$
with a certain constant that is not equal to zero,
we can rewrite both of $P^{(3)}_{g,1}(t)$ and $P^{(3)}_{g,2}(t)$
as polynomials of $\alpha$.
Because all of the terms cancel each other out in these polynomials,
we can finally confirm
that they are equal to zero.
This fact can be found in any order corrections of the perturbative expansion,
such as the first, second and third order corrections.

\end{document}